\documentclass[]{aa}
\usepackage{natbib}
\usepackage{amssymb,verbatim}
\usepackage{graphicx}
\usepackage{appendix}
\newcommand{\mincir}{\raise -2.truept\hbox{\rlap{\hbox{$\sim$}}\raise5.truept
\hbox{$<$}\ }}
\newcommand{\magcir}{\raise -2.truept\hbox{\rlap{\hbox{$\sim$}}\raise5.truept
\hbox{$>$}\ }}
\newcommand{\siml}{\raise -2.truept\hbox{\rlap{\hbox{$\sim$}}\raise5.truept
\hbox{$<$}\ }}
\newcommand{\simg}{\raise -2.truept\hbox{\rlap{\hbox{$\sim$}}\raise5.truept
\hbox{$>$}\ }}
\usepackage[normalem]{ulem}

\begin{document}

\titlerunning{Growth and disruption in the Lyra complex}
\authorrunning{S. Clavico, S. et al.}
\title{Growth and disruption in the Lyra complex}

\author{S. Clavico\inst{1,2}, S. De Grandi\inst{1}, S. Ghizzardi\inst{3}, M. Rossetti\inst{3}, S. Molendi\inst{3}, F. Gastaldello\inst{3}, M. Girardi\inst{4,5}, W. Boschin\inst{6,7}, A. Botteon\inst{8,9,10}, R. Cassano\inst{9}, M. Br\"uggen\inst{11}, G. Brunetti\inst{9}, D. Dallacasa\inst{8,9}, D. Eckert\inst{12}, S. Ettori\inst{13,14}, M. Gaspari\inst{15}, M. Sereno\inst{13,14}, T. Shimwell\inst{10,16} and R. J. van Weeren \inst{10}.   
}

%1
\institute{INAF - Osservatorio Astronomico di Brera, 
     via E. Bianchi 46, 23807 Merate, Italy \\
\email{sara.clavico@inaf.it, sabrina.degrandi@inaf.it}
%2
\and Universit\`a degli Studi di Milano-Bicocca, Piazza della Scienza, 3, I-20126 Milano, Italy 
%3
\and INAF - IASF-Milano, Via E. Bassini 15, I-20133 Milano, Italy
%4
\and Dipartimento di Fisica dell'Universit\`a degli Studi di Trieste, Italy
%5
\and INAF - Osservatorio Astronomico di Trieste, via Tiepolo 11, I-34143 Trieste, Italy
%6
\and Fundaci\'on Galileo Galilei - INAF (Telescopio Nazionale Galileo),    Rambla Jos\'e Ana Fern\'andez Perez 7, E-38712 Bre\~na Baja (La Palma), Canary Islands, Spain
%7
\and Instituto de Astrof\'{\i}sica de Canarias, C/V\'{\i}a L\'actea s/n, E-38205 La Laguna (Tenerife), Canary Islands, Spain
%8
\and Dipartimento di Fisica e Astronomia, Universit\`a di Bologna, via P. Gobetti 93/2, I-40129 Bologna, Italy
%9
\and INAF - IRA, via P.~Gobetti 101, I-40129 Bologna, Italy
%10
\and Leiden Observatory, Leiden University, PO Box 9513, NL-2300 RA Leiden, The Netherlands
%11
\and Hamburger Sternwarte, Universit\"at Hamburg, Gojenbergsweg 112, D-21029 Hamburg, Germany
%12
\and Department of Astronomy, University of Geneva, ch. d'Ecogia 16, 1290 Versoix, Switzerland
%13
\and INAF, Osservatorio di Astrofisica e Scienza dello Spazio, via Pietro Gobetti 93/3, I-40129 Bologna, Italy 
%14
\and INFN, Sezione di Bologna, viale Berti Pichat 6/2, I-40127 Bologna, Italy
%15
\and Department of Astrophysical Sciences, Princeton University, 4 Ivy Lane, Princeton, NJ 08544-1001, USA - Lyman Spitzer Jr. Fellow
%16
\and ASTRON, the Netherlands Institute for Radio Astronomy, Postbus 2, NL-7990 AA Dwingeloo, The Netherlands}

 \abstract
% Context % context heading (optional) leave it empty if necessary  
 {Nearby clusters of galaxies, $z\lesssim 0.1$, are cosmic structures still under formation. Understanding the thermodynamic properties of merging clusters can provide crucial information on how they grow in the local universe.}
% Aims
 {A detailed study of the intra-cluster medium (ICM) properties of un-relaxed systems is essential to understand the fate of in-falling structures  and, more generally,  the  virialization process.}
% Methods
 {We analyzed a mosaic of XMM-Newton observations (240 ks)  of the Lyra system ($z\sim 0.067$) that shows a complex dynamical state. 
 }
% Results
 {We find the main cluster RXC J1825.3+3026 to be in a late merger phase, whereas its companion CIZA J1824.1+3029 is a relaxed cool-core cluster. We estimate  a mass ratio of $\sim 1:2$ for the pair. No diffuse X-ray emission is found in the region between them, indicating that these clusters are in a pre-merger phase. 
 We found evidence of a galaxy group infalling on RXC J1825.3+3026 in an advanced state of disruption. The SG, one of the brightest galaxies in the Lyra complex, was very likely at the center of the infalling group. This galaxy has a gaseous "corona" indicating that it was able to retain some of its gas after the ram-pressure stripping of the intra-group medium. In this scenario the diffuse emission excess observed southwest of RXC J1825.3+3026 could be due to gas once belonging to the group and/or to cluster ICM dislocated by the passage of the group. Finally, we identified three high-velocity galaxies aligned between RXC J1825.3+3026 and the SG, two of these showing evidence of gas stripped from them during infall. We estimate them to be currently falling onto the main cluster at an infall velocity of $\sim 3000$ km/s. }
% conclusions  
 {Our study of the Lyra complex provides important clues about the processes presiding over the virialization of massive clusters in the local Universe.
 }
\maketitle

\section{Introduction}
\label{sec:intro}
The currently favored cosmological model predicts that structure formation proceeds hierarchically with the more massive dark matter halos growing by the accretion of smaller halos driven by gravity. Galaxy clusters are the most massive gravitationally bound objects found at the present epoch and have been forming relatively recently, doubling their mass on average since  $z\sim0.5$ \citep{boylan09,gao12}. They are formed at the nodes of the filamentary network which constitutes the cosmic web, the structure of the Universe on the largest scales as described in the $\Lambda$CDM model (for a review see \citealt{kravtsov12_rev}).

Although major mergers of equal-size halos can be the most spectacular and energetic events since the Big Bang and leave substantial imprints on the observational appearance of clusters, they are rare events (e.g., \citealt{berrier09,mcgee09,fakhouri10}). A continuous and more gentle accretion of group-scale systems is the other important contributor to the growth of clusters.
Observations show that accretion of group-size systems is responsible for at least half of the accreted cluster mass from $z\sim 0.2$ to the present day, whereas the other half is likely derived from smooth accretion of unbound matter within halos \citep[e.g.,][]{haines18_locuss}.
Recent mergers have left an imprint on the distribution of the intra-cluster medium (ICM), which constitutes most of the baryonic mass of clusters and emits via bremsstrahlung in the soft X-ray band. Indeed, X-ray observations find that a high fraction ($> 40\%$) of the clusters in a representative sample of the nearby Universe ($z < 0.5$) display clear disturbance \cite[e.g.,][]{rossetti16,lovisari17}.

Large-scale cosmological hydrodynamic simulations show that an important region for studying the mechanisms of the growth of substructures in clusters is found between their R$_{500}$\footnote{For a given over-density $\Delta$, $R_\Delta$ is the radius for which $M_\Delta/(4/3\pi R_\Delta^3)=\Delta\rho_c$.} and R$_{200}$ \cite[e.g.,][]{walker19_rev}. 
In fact, in this region it is possible to observe the connections of the cosmic filaments with the innermost and already virialized part of the cluster \citep[e.g.,][]{eckert15_nature}. 

In addition to information on the mass growth of clusters, the study of accreting sub-structures at large radii provides an opportunity to study the physical properties of the ICM in the outskirts.
During the infall of groups in clusters, the ram pressure applied by the ambient ICM is responsible for stripping their gas and heating it up, leading to the virialization of the gas in the main dark-matter halo. The dynamical scales involved in this process indicate that the thermal conductivity of the gas must be highly suppressed at these large distances from the cluster core  \citep[e.g.,][]{eckert14_a2142,eckert17_a2142,degrandi16_hydraa}.

\begin{figure}
    \centering{
\includegraphics[angle=0,width=9.cm]{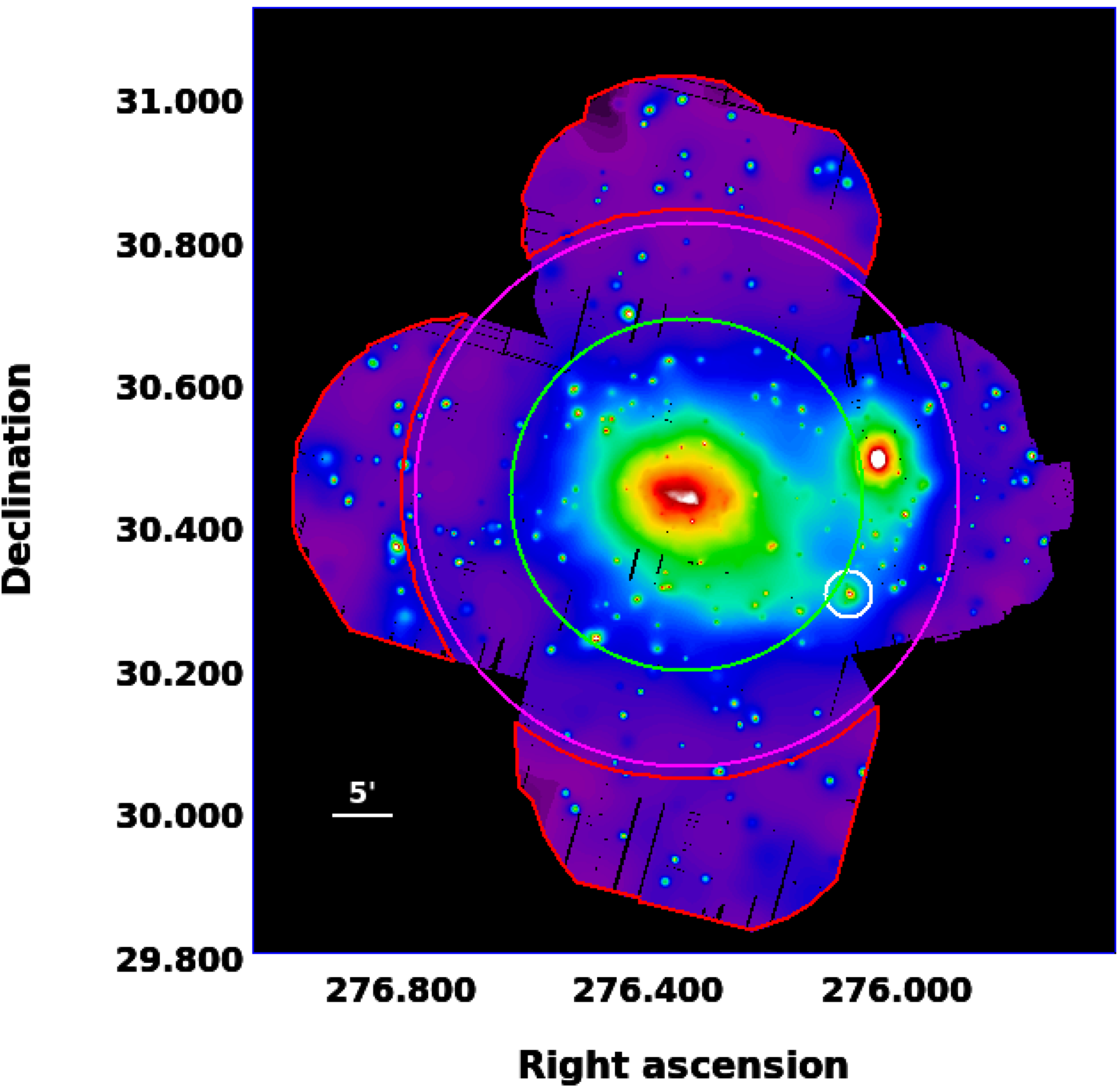}}
\caption{XMM-Newton mosaic image of the Lyra cluster complex in units of counts pixel$^{-1}$ in the [0.7–1.2] keV energy band. The cluster RXCJ1825 is in the center of the image with the  green and magenta circles representing the location of its R$_{500}$ and R$_{200}$, respectively. The cluster CIZAJ1824 is west of RXCJ1825, whereas the white circle is centered on the Southern Galaxy (SG). The image is corrected for the particle background (NXB).  The red regions are the ones chosen for the estimate of the local sky background (see Sect.\ref{sec:datax}).}
\label{fig:sbx}
\end{figure}

\begin{table*}
        \caption[]{XMM-Newton observations of the Lyra complex (RXCJ1825-CIZAJ1824) complex: field name, archival observation identification number, pointing position, total nominal exposure time and effective exposure time after soft-protons cleaning of the three EPIC detectors and ratio of the events IN and OUT of the MOS1 field of view.}
         \label{tab:obsx}
              $$ 
           \begin{array}{c c c c c c c c}
            \hline
            \noalign{\smallskip}
            \hline
            \noalign{\smallskip}
\mathrm{Pointing} &\mathrm{Obs.ID.} & \mathrm{RA}~\mathrm{DEC}~(\mathrm{J}2000) & \mathrm{Nominal} & \mathrm{MOS1} & \mathrm{MOS2} & pn & \mathrm{IN/OUT} \\ 
 &   & \mathrm{[deg]} &  \mathrm{[ksec]}  & \mathrm{[ksec]}  & \mathrm{[ksec]}  & \mathrm{[ksec]}  &  \\
            \hline
            \noalign {\smallskip}  
\text{Center1} &0744413501 & 276.3450, +30.4422  &   52.0 & 47.3 & 47.6 & 34.4 & 1.265\\      
\text{Center2} &0744414101 & 276.3450, +30.4422  &   32.9 & 20.6 & 21.7 & 9.4  & 2.568 \\      
\text{East}    &0744413901 & 276.7014, +30.4422  &   44.3 & 31.7 & 32.0 & 14.2 & 1.307\\      
\text{North}   &0744413601 & 276.3385, +30,7921  &   32.3 & 19.7 & 19.9 & 12.4 & 1.273\\      
\text{West}    &0744413701 & 275.9885, +30.4422  &   43.9 & 40.9 & 40.3 & 31.7 & 1.130\\      
\text{South}   &0744414801 & 276.3385, +30.0923  &   34.6 & 25.1 & 29.2 & 10.8 & 1.291 \\      
                        \noalign{\smallskip}                        
            \hline                                          
            \noalign{\smallskip}                            
            \hline                                          
         \end{array}
        $$
\end{table*}

In this work, we study the very special case of a system showing several substructures in different dynamical states (both pre- and post-merger) and different mass scales,  at about R$_{500}$ of the main cluster.
This is the Lyra complex formed by the galaxy clusters pair RXC J1825.3+3026 and CIZA J$1824.1+3029$ (Fig.~\ref{fig:sbx}).
RXC J1825.3+3026 (RXCJ1825 hereafter) was discovered by ROSAT in the X-rays ($z=0.065$, \citealt{ebeling02}). Because of its relatively low galactic latitude of $b=18.5$ degrees, it has been almost ignored until recently, when it was found to be one of the strongest (S/N$>$12) sources of Sunyaev-Zeldovich (SZ) signal in the Planck all-sky cluster survey \citep{planck14} and became part of the XMM-Newton Cluster Outskirts Project (X-COP), an SZ-selected sample of 13 massive clusters observed in the X-rays by XMM-Newton up to the virial radius \citep{eckert17_xcop}.  RXCJ1825 is a massive cluster with a total mass of $\sim 6\times10^{14}$ M$_{\odot}$ within R$_{200}$ \citep{ettori19_xcop_mass}, and an irregular west-east X-ray morphology that suggests an un-relaxed dynamical status. Interestingly, its X-ray morphology shows also a clear extension to the southwest culminating in a bright elliptical galaxy (called the Southern Galaxy (SG)). 
At $\sim 16$ arcmin northwest of RXCJ1825, there is another, smaller cluster: CIZA J$1824.1+3029$ (also known as NPM1G$+30.0$, and named CIZAJ1824 hereafter). Unlike RXCJ1825, CIZAJ1824 appears very regular in the X-rays.
The only optical observation available in the literature for this cluster is the  redshift of its brightest cluster galaxy (BCG) $z=0.072$ \citep{kocevski07_ciza}.
Recently, \cite{girardi19} presented spectroscopic observations of the Lyra complex confirming that the two clusters and the SG are very close in the redshifts space. 
These latter authors performed a dynamical analysis of the system and found that all these objects are likely to be gravitationally bound at a mean redshift of $z=0.067$ and that their redshift differences are of kinematic nature.
Their new estimates for the redshifts of RXCJ1825 and CIZAJ1824 are $z=0.0645$ and $z=0.0708$, respectively. 
Here we focus our study on the analysis of the X-COP XMM-Newton mosaic centered on RXCJ1825 (240 ks in total) to determine the thermodynamic properties and the dynamic state of the whole Lyra system.

This paper is structured as follows: in Sect. 2 we describe our XMM-Newton  mosaic observations together with their respective data reduction, imaging (Sect. 2.1), and spectral analysis (Sect. 2.2) techniques. In Sect. 3, we report the results for the whole Lyra system obtained from our data analysis, first focusing on its X-ray surface brightness features (Sect. 3.1), then on the thermodynamic profiles of the two clusters RXCJ1825 and CIZAJ1824 (Sect. 3.2), and finally on the 2D analysis of the spectral properties of several interesting regions (Sect. 3.3).  In Sect. 4, we interpret and discuss our results. We summarize our main results in Sect. 5.

Throughout the paper, we assume a $\Lambda$CDM cosmology with $H_0=70$ km s$^{-1}$, $\Omega_m=0.3$ and $\Omega_\Lambda=0.7$. At the redshift of the Lyra complex, $z=0.067$ \citep{girardi19}, 1 arcmin corresponds to $\sim$ 77.5 kpc. Metal abundances are in units of [X/H] normalized by the solar abundances of \cite{asplund09}. All the quoted errors hereafter are at the $1\sigma$ confidence level.

\begin{figure}[ht]
    \centering{
   \includegraphics[angle=0,width=8.3cm]{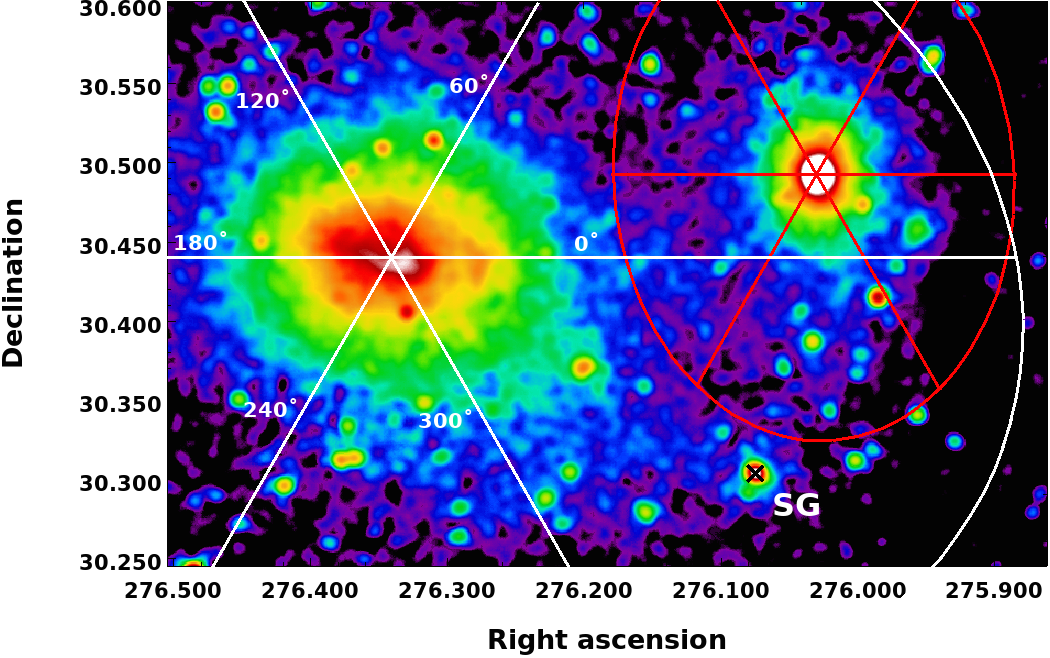}
   \includegraphics[angle=0,width=8.4cm]{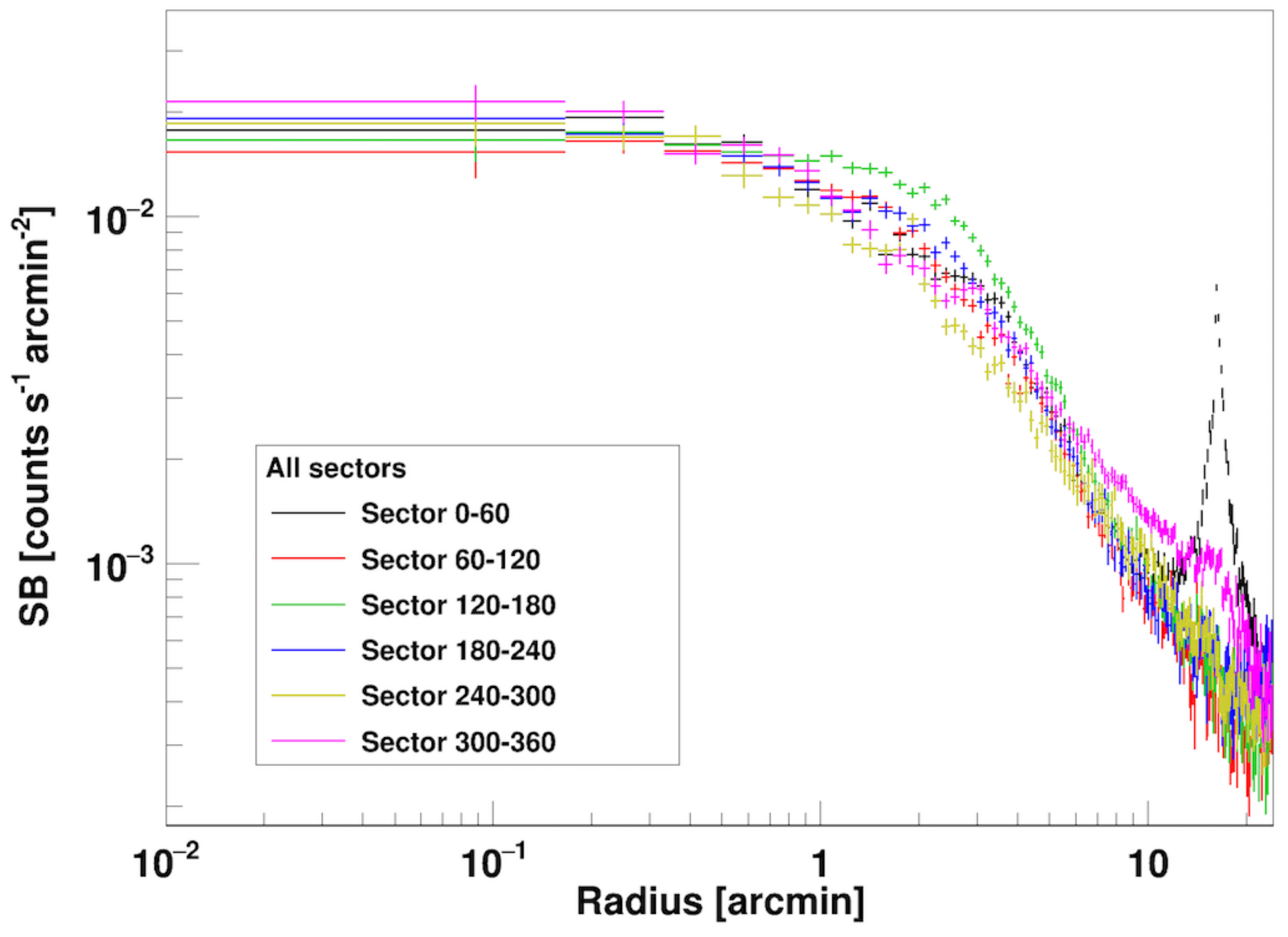}  
   \includegraphics[angle=0,width=8.7cm]{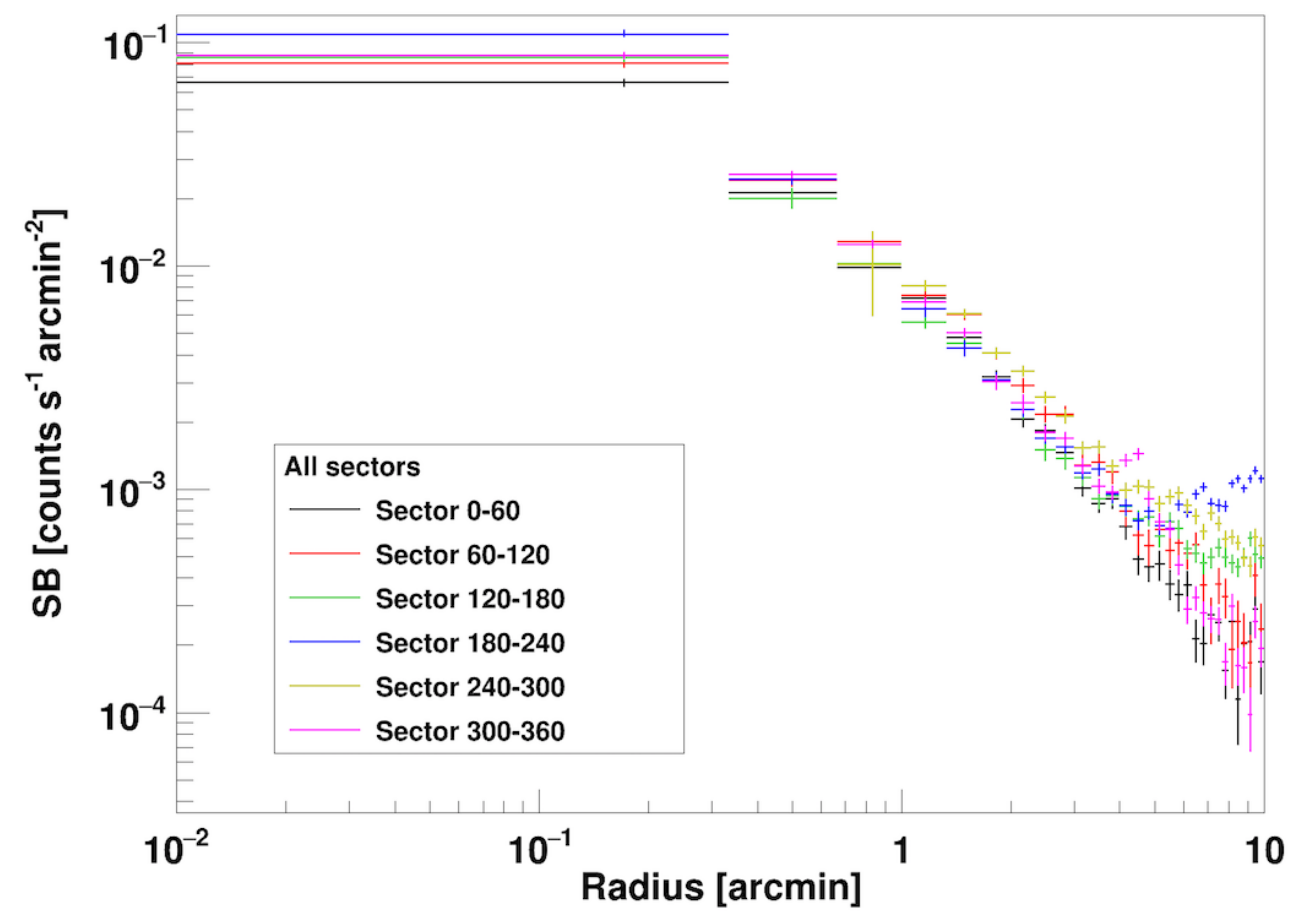}
}
 \caption{{\it Top panel}: Elliptical sectors (60 degrees wide) used for the extraction of surface brightness profiles, overplotted on the XMM-Newton mosaic image (Fig.~\ref{fig:sbx}). The orientation angle is calculated counterclockwise from the Right Ascension axis. {\it Middle panel} and {\it bottom panel} show the elliptical surface brightness profiles of RXCJ1825 and CIZAJ1824 extracted in the sectors, respectively.  In the x-axis we plot the distances along the major axis of the ellipses. Point-like sources are removed. The black cross shows the position of the SG. The features seen in the profiles are described in Sect.\ref{sec:sectors}. }
 \label{fig:allsectors}
\end{figure}

\section{Observations and data reduction}
\label{sec:datax}
The Lyra complex was observed by XMM-Newton in 2014 with a mosaic of two central and four offset pointings. We did not consider one of the two central observations (Obs.ID. 0744414101) as we found that it was highly contaminated by quiescent soft protons (see Sect.~\ref{sec:specx}). 

Table~\ref{tab:obsx} contains information regarding the observations, such as the observation ID, the total and clean exposure times (after applying the ESAS tasks {\it mos-filter} and {\it pn-filter}) and the level of soft-proton contamination obtained by comparing the measured count rate in a hard spectral band ($10-12$ keV) in the exposed and unexposed part of the field of view (IN/OUT, \citealt{leccardi08_t}). The analysis of the five remaining observations allowed us to explore the entire azimuth of the cluster out to $\sim$ R$_{500}$ and its close companion CIZAJ1824. The data were processed using the XMM-Newton Scientific Analysis System (XMM-SAS) v.16.1.

\subsection{Imaging}
\label{sec:image}
We produced an image in the [0.7–1.2] keV band for all three EPIC detectors (MOS1, MOS2, and pn) using the Extended Source Analysis Software package (ESAS, \citealt{snowden08_esas}). We used this band as it maximizes the source-to-background ratio for galaxy clusters \citep{ettori10, ettori11}, while maintaining a large effective area of the XMM-Newton telescopes. We obtained the count rate map for each EPIC instrument by dividing the raw count image by the exposure maps (tool {\it eexpmap}), accounting for vignetting. 
The total image, which is shown in Fig.~\ref{fig:sbx}, was subtracted by the nonX-ray background (NXB) image produced by ESAS (tools {\it mos-spectra, pn-spectra, mos-back} and {\it pn-back}), and by the soft-proton contribution as measured within X-COP \citep{ghirardini18}. Point sources were detected down to a fixed flux threshold with the XMM-SAS tool {\it ewavelet} and excluded using the ESAS task {\it cheese}. The detailed imaging procedure is described in \cite{ghirardini19}.

\subsection{Spectral analysis}
\label{sec:specx}
We performed a spectral analysis of the interesting regions following the procedure developed for the X-COP clusters and described in \cite{ghirardini19}. Here we report the main steps. 

Spectra and response files for each region were extracted using the ESAS tasks {\it mos-spectra} and {\it pn-spectra} and were fitted using XSPEC v$12.9.1$ after  grouping to ensure a minimum of 20 counts per spectral channel in the energy band [0.5-10] keV.
Point sources detected in the field were always removed before the extraction of the spectra.
 
We used a detailed modeling of all the various background components to obtain reliable measurements of the physical parameters in regions where the surface brightness barely exceeds the background level. The spectral components for the background are the following:
(1)the NXB was estimated for each region from closed-filter observations following the procedure given by \cite{snowden08_esas}. We left the normalization of the NXB component and those of the prominent background lines free to vary during the fitting procedure, which allows for possible systematic variations of the NXB level.  Since all five useful observations were very weakly contaminated by soft proton flares (IN over OUT ratio always $ \leq 1.3$), we did not include any component to model residual soft protons. 
(2) {\it The sky background and foreground components}: we considered the sum of an absorbed power law with a photon index fixed to $1.46$, which describes the residual cosmic X-ray background  \citep[CXB;][]{deluca04}, an absorbed APEC thermal plasma model with a temperature allowed to vary in the range $[0.15-0.6]$ keV, which models the Galactic halo emission \citep{mccammon02}, and an unabsorbed APEC model with a temperature fixed to $0.11$ keV, which represents the local hot bubble. 
To estimate the parameters of these sky background components we used a joint analysis of the spectra extracted from the North, East, and South observations, in regions at radii larger than R$_{200}$ ($\sim 1.7$ Mpc) from the emission peak of RXCJ1825. These regions are shown in red in Fig.~\ref{fig:sbx}. 
(3) Finally, the source is modeled using the thin-plasma emission code {\it apec} \citep{smith01_apec}, with temperature, metal abundance, and normalization as free parameters, and fixed redshift and Galactic column density absorption NH ({\it phabs} model). We discuss the NH variation in the field of view in Sect.~\ref{sec:profrxc}.

In the particular case of the detailed spectral study of the temperature and abundance maps in the core regions of RXCJ1825 (Sect.\ref{sec:mapx}), we subtracted the background using stacked blank-sky fields following the procedure given in \cite{ghizzardi14}, and then we extracted the total (source plus background) spectra starting from the soft photon cleaned files produced with ESAS. We preferred to use this method since in this case we decided to extract spectra from polygonal regions (which is not possible with the ESAS software), to better follow the surface brightness fluctuations present in the core of RXCJ1825. This procedure is fully justified as in the core regions the source is dominant over the background.

\begin{figure*}
    \centering{
    \includegraphics[angle=0,width=9.1cm]{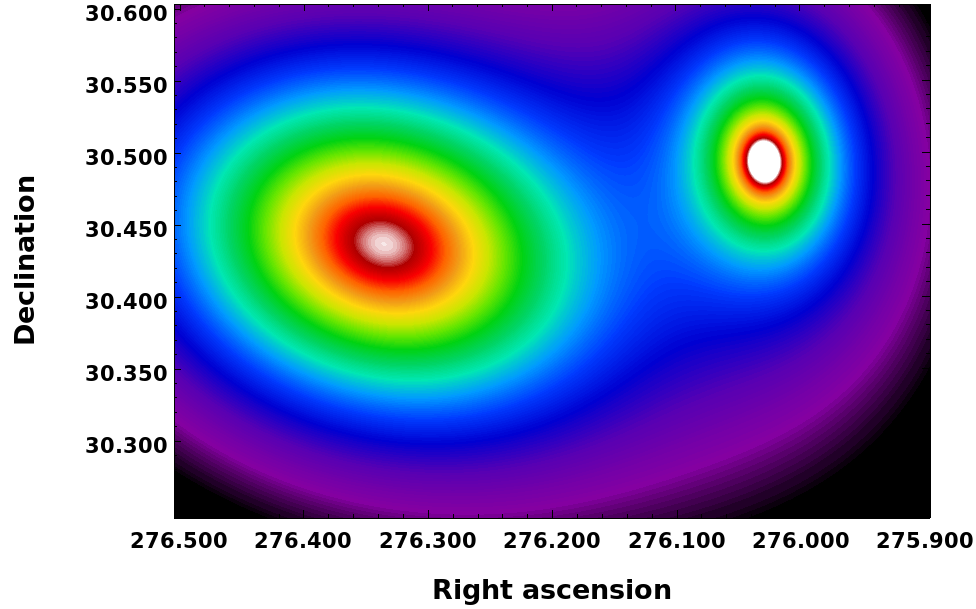}
    \includegraphics[angle=0,width=9.1cm]{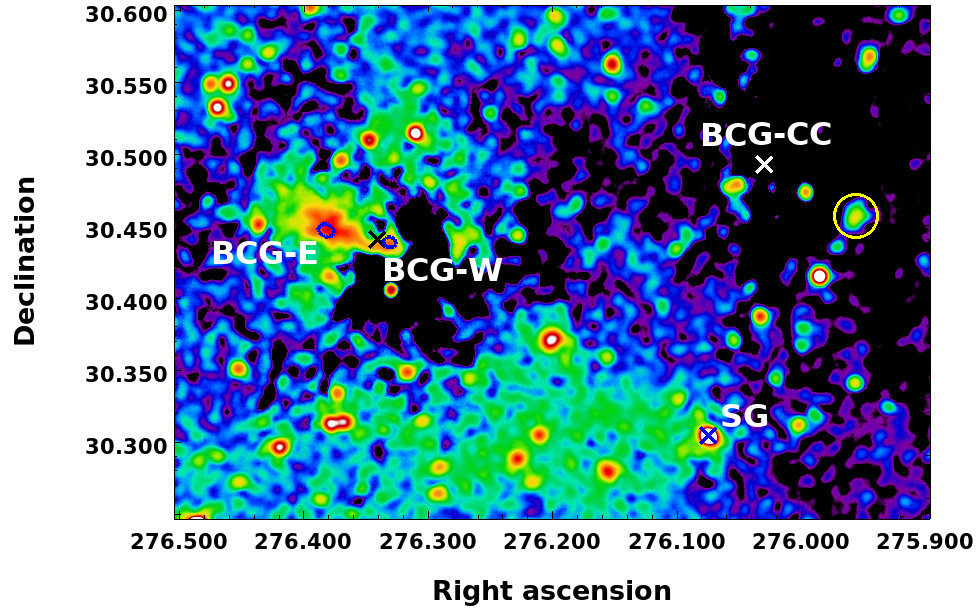}}
\caption{{\it Left panel}: Surface brightness model that includes both elliptical single $\beta$-models of RXCJ1825 and CIZAJ1824 clusters. 
{\it Right panel}: Residual image after subtracting the model from the X-ray image.  Blue small ellipses show the position of the two BCGs of RXCJ1825, labeled BCG-W and BGC-E, and the black cross is the X-ray centroid of the cluster. The white and blue crosses show the position of the BCG of CIZAJ1825, BCG-CC, and of the SG, respectively. }
\label{fig:residual}
\end{figure*}

\section{X-Ray analysis}
\label{sec:dataxres}
In this section we present the results of the X-ray analysis of the extended sources  detected in the mosaic image centered on RXCJ1825, including a study of the surface brightness features, temperature and abundance profiles, as well as quantities derived from these profiles. Maps of various thermodynamic quantities for the core of RXCJ1825 and the SG region are also presented. 

\subsection{X-ray surface brightness}
\label{sec:sbfeat}

\subsubsection{Radial profiles in sectors}
\label{sec:sectors}
We first calculated the vignetting-corrected NXB subtracted surface brightness profiles of RXCJ1825 and CIZAJ1824 with the Proffit v.1.5 software \citep{eckert16}. We extracted the profiles from a series of concentric elliptical annuli in bins of 10 arcsec ($\sim 12.5$ kpc) along the minor axis, centered on the centroid of the clusters (R.A. = 18h:25m:21.8s, Dec = +30d:26m:25.34s J2000.0 for RXCJ1825, R.A. = 18h:24m:7.1s, Dec= +30d:29m:34.7s for CIZAJ1824). The centroids were determined by the shape of the isophotes on large scales (7 arcmin and 5 arcmin aperture for RXCJ1825 and CIZAJ1824, respectively). Point-like sources detected in the field were always removed before the extraction of the profiles.

Figure~\ref{fig:allsectors} shows the surface brightness profiles extracted in six sectors, corresponding to an opening angle of 60 deg (with position angle calculated counterclockwise from the R.A. axis), overplotted on top of each other.

The profiles of RXCJ1825 (Fig.~\ref{fig:allsectors}, middle panel) show large differences between one another; most notably there is an excess surface brightness in the sector with position angle 120-180 deg (green data points), corresponding to a clear asymmetry in the northeast direction for radii smaller that 6 arcmin, whereas sectors 240-300 deg (yellow data) and 300-360 (magenta data) show a clear excess beyond 6 arcmin towards south and southwest, in particular where the SG  is located. The surface brightness peak due to CIZAJ1824 is clearly visible in sectors 0-60 deg (black data) and partially in sectors 300-360 deg.

Conversely, the profiles of CIZAJ1824 (Fig.~\ref{fig:allsectors}, bottom panel) are remarkably similar up to $\sim 4-5$ arcmin. Beyond this radius the scatter increases showing an excess in the three sectors towards RXCJ1825, that is, sectors 120-180 deg (green), 180-240 deg (blue), and 240-300 deg (yellow). We note than in the 300-360 deg sector (magenta data) there is a peak at $\sim 4$ arcmin due to a small extended source close to CIZAJ1824 that is further discussed in Sect.\ref{sec:ciza_d}.

\begin{figure*}
    \centering{
    \includegraphics[angle=0,width=9.1cm]{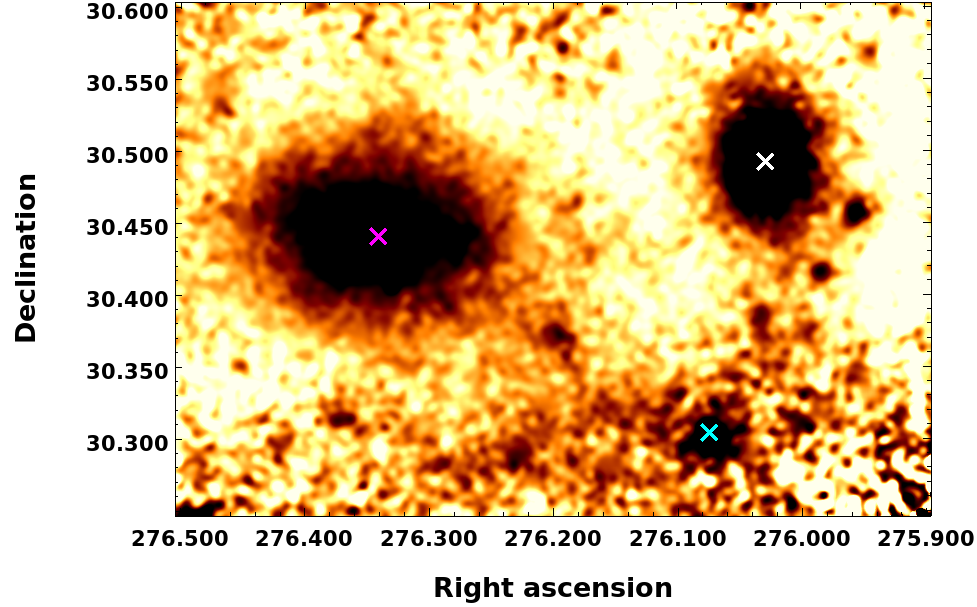}
    \includegraphics[angle=0,width=9.1cm]{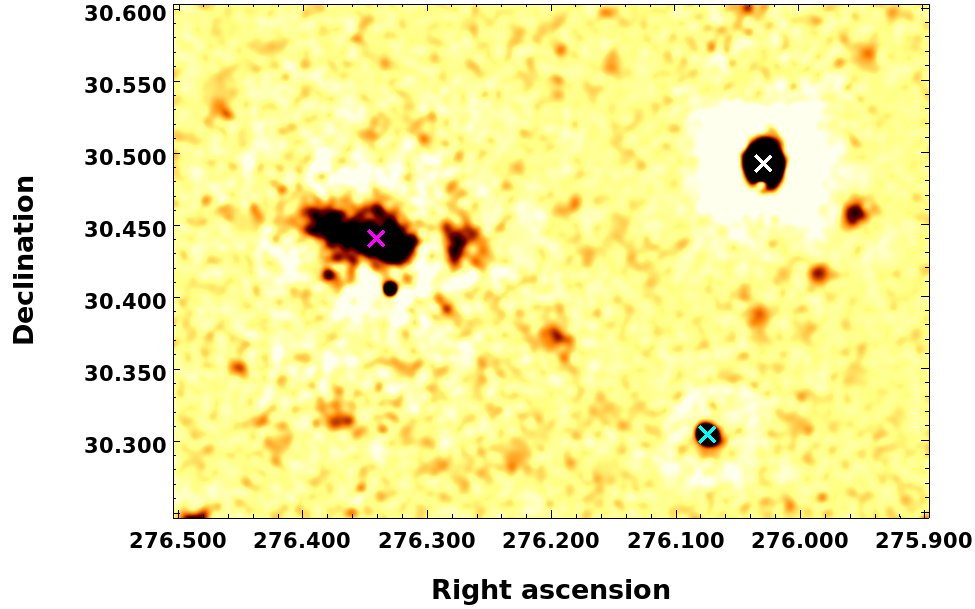}}
 \caption{{\it Left panel}: Unsharp-mask image using a 20$\sigma$ and a 800$\sigma$  Gaussian to enlighten large-scale, low-surface brightness features. Point-like sources were removed. Crosses show the position of the X-ray centroid of RXCJ1824 (magenta), BCG-CC (white), and SG (cyan). {\it Right panel}: Unsharp-mask image using  20$\sigma$ and a 240$\sigma$ Gaussian to search for small-scale features.}
\label{fig:unsharp}
\end{figure*}
 
\subsubsection{Residual and unsharp-mask images}
\label{sec:residual}
We extracted also the azimuthally averaged (0-360 deg) surface brightness profile for both clusters using concentric elliptical annuli: position angle and ellipticity (i.e., major over minor axis ratio) for the two clusters were determined using the {\it proffit - ellipticity} command at  $R<7$ arcmin for RXCJ1825 and $R<5$  arcmin for CIZA (P.A.=165 deg $\epsilon$=0.25 for RXCJ1825, and P.A.=94 deg, $\epsilon$=0.24 for CIZAJ1824). To avoid contamination by emission due to the companion cluster, when extracting the profile of one cluster we appropriately masked the other. 
Each profile was then fitted with a single-$\beta$ model profile \citep{cavaliere76}: 
\begin{equation}
 S(r) = S_0 [(1 + (r/{r{_c}})^2)]^{(-3\beta+0.5)} + const 
,\end{equation}
with free parameters normalization $S_0$, $\beta$, core radius $r_c$, and constant sky background level. A double-$\beta$ model did not significantly improve the fit (based on an F- test) in either cluster. The profiles with their best-fit models overlapped and the best-fit parameters are shown in Appendix A.
We used these best-fit profiles to produce a surface brightness model image of the two clusters that we subtracted pixel-by-pixel from the observed image to obtain a residual map (in units of $\sigma$, as we also divide by the poissonian error on each pixel).
Both images are shown in Fig.~\ref{fig:residual}.

The residual image is a useful tool to look for regions where there are significant gradients in the surface brightness: a candidate excess south and southwest of RXCJ1825 towards the SG is clearly visible in the image. Moreover, a complex structure emerges in the core of RXCJ1825, with another excess to the northeast along the direction traced by the position of the two BCGs (BCG-E and BCG-W in Fig.~\ref{fig:residual}).

 \begin{figure*}
    \centering{
    \includegraphics[angle=0,width=6.cm]{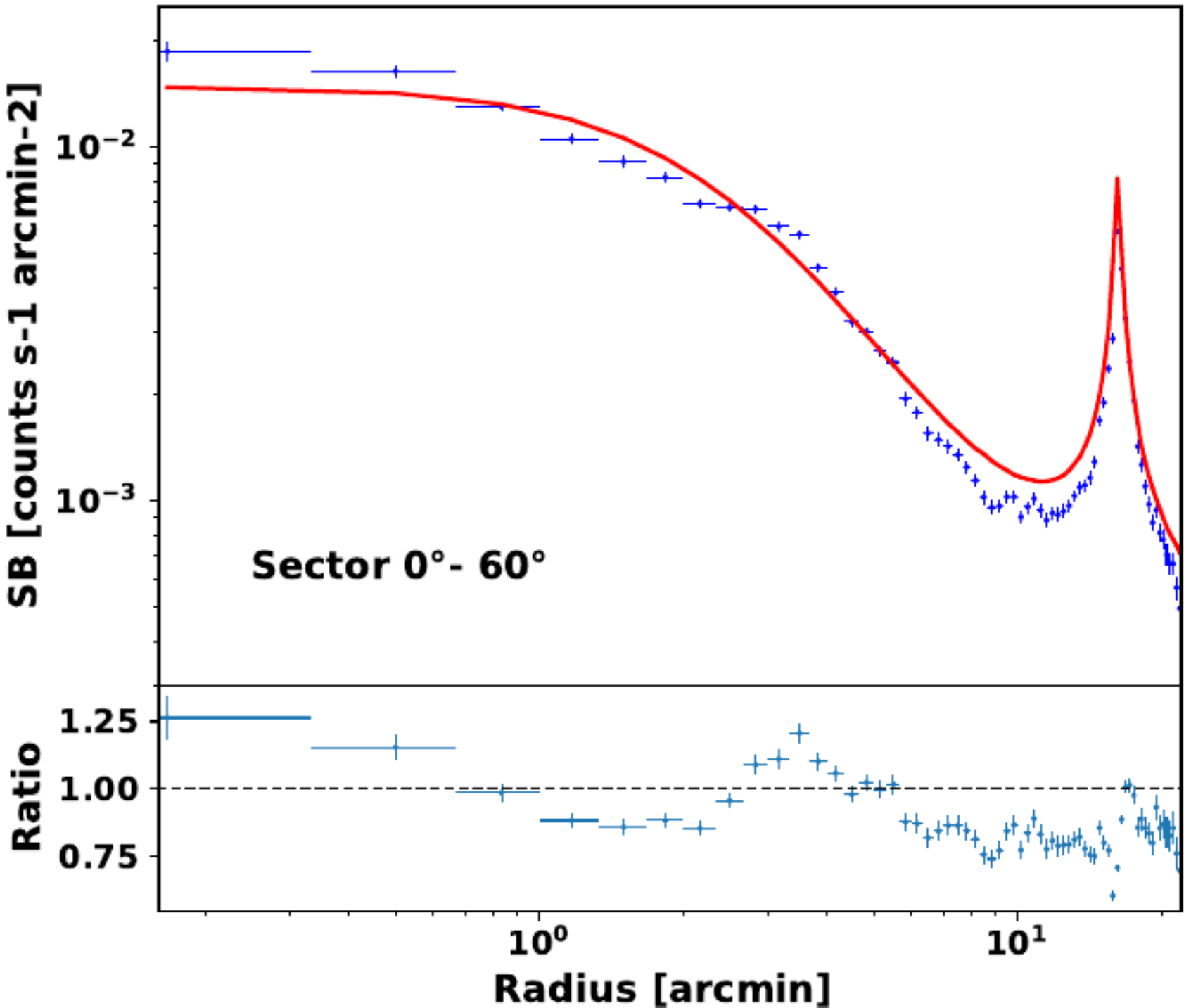}
    \includegraphics[angle=0,width=6.cm]{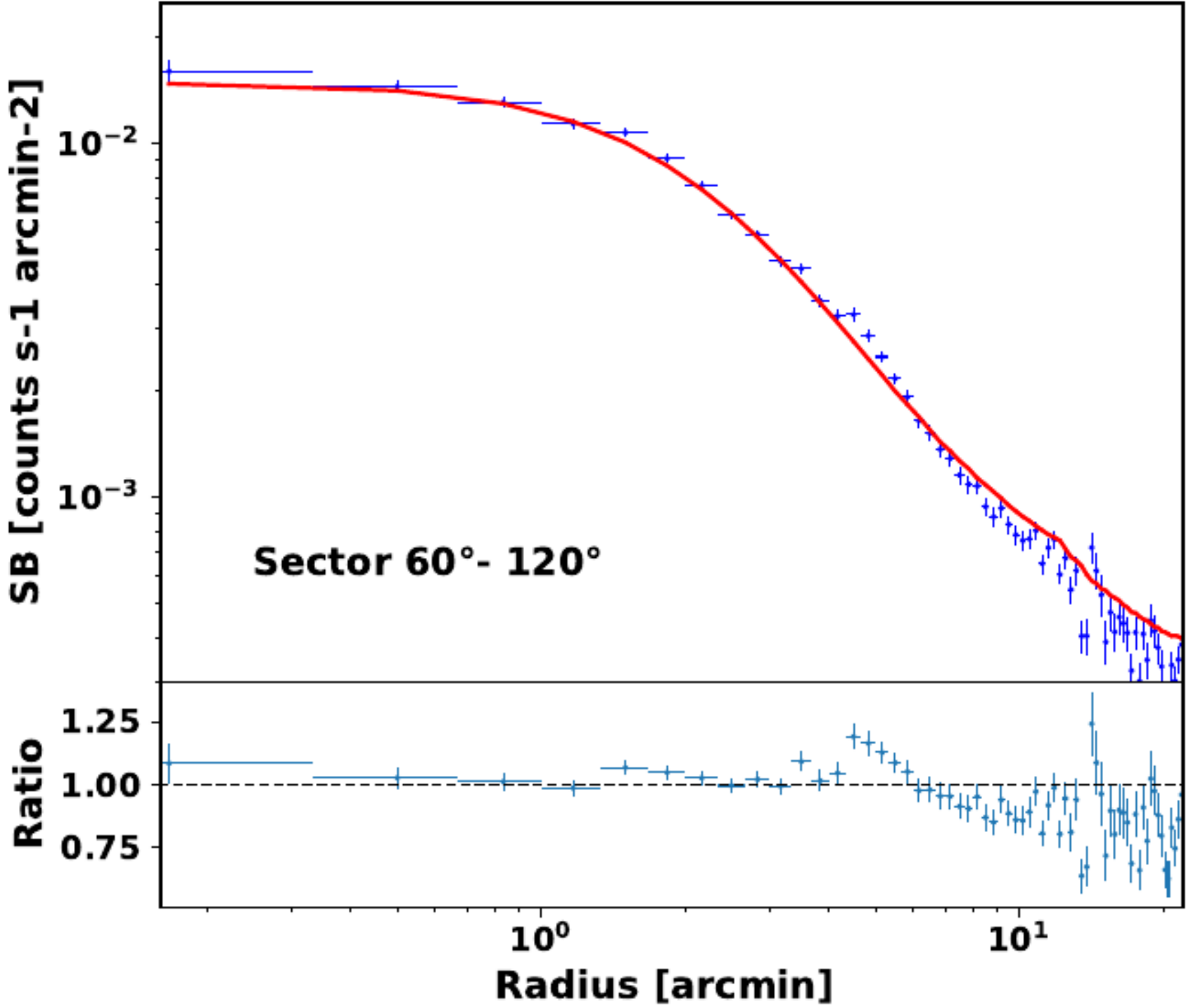}
    \includegraphics[angle=0,width=6.cm]{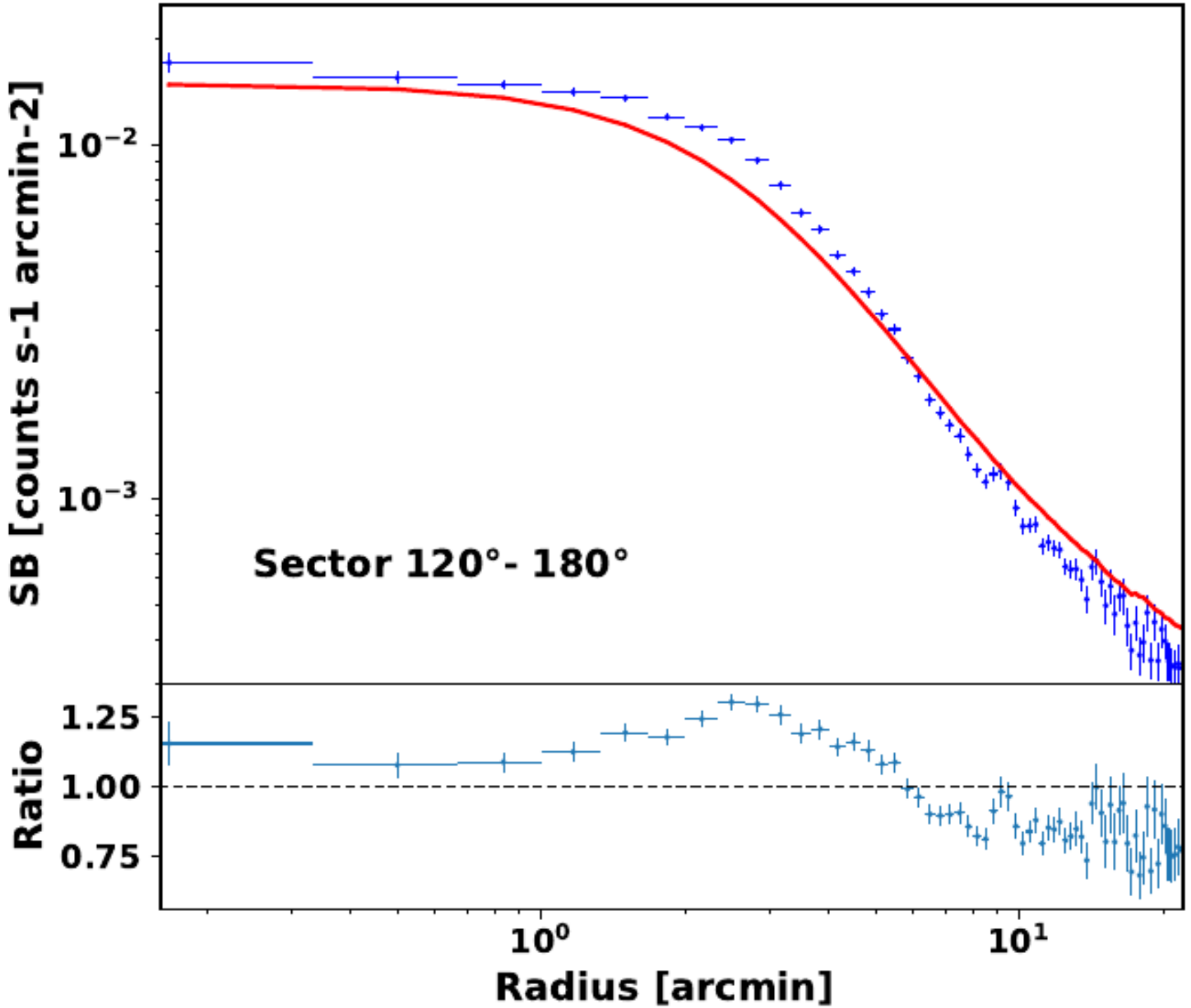}
    \vskip 0.3cm
    \includegraphics[angle=0,width=6.cm]{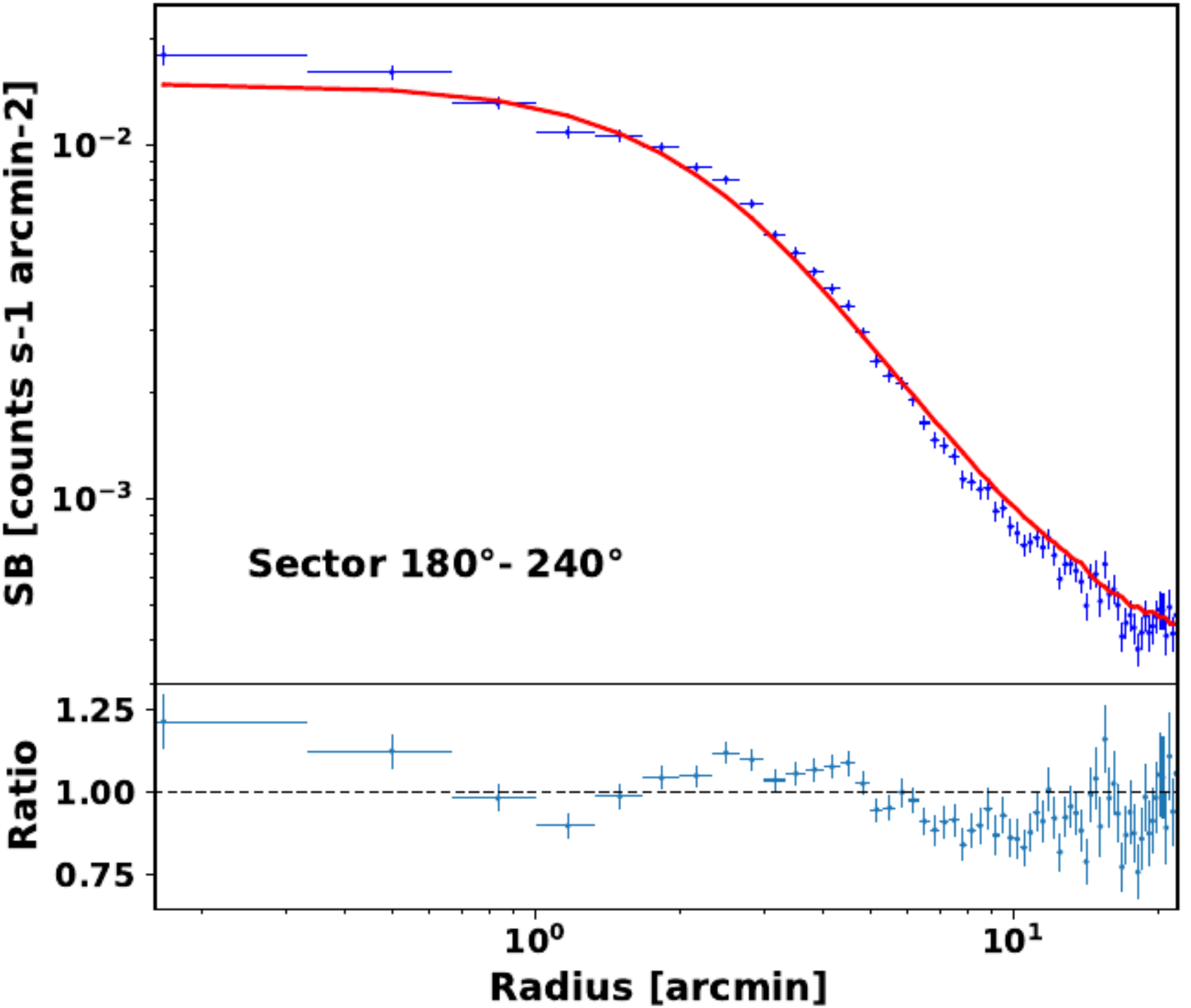}   
    \includegraphics[angle=0,width=6.cm]{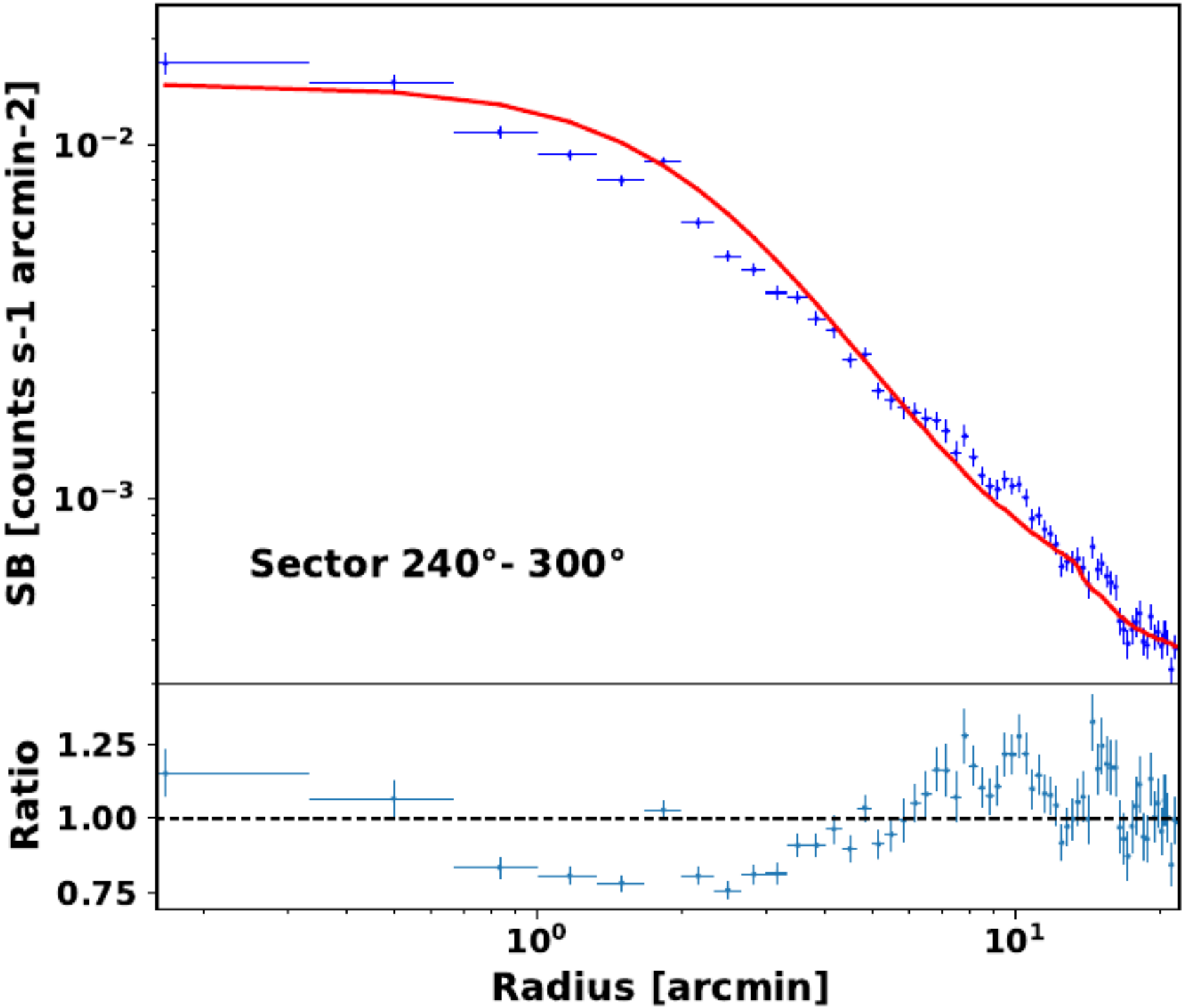}   
    \includegraphics[angle=0,width=6.cm]{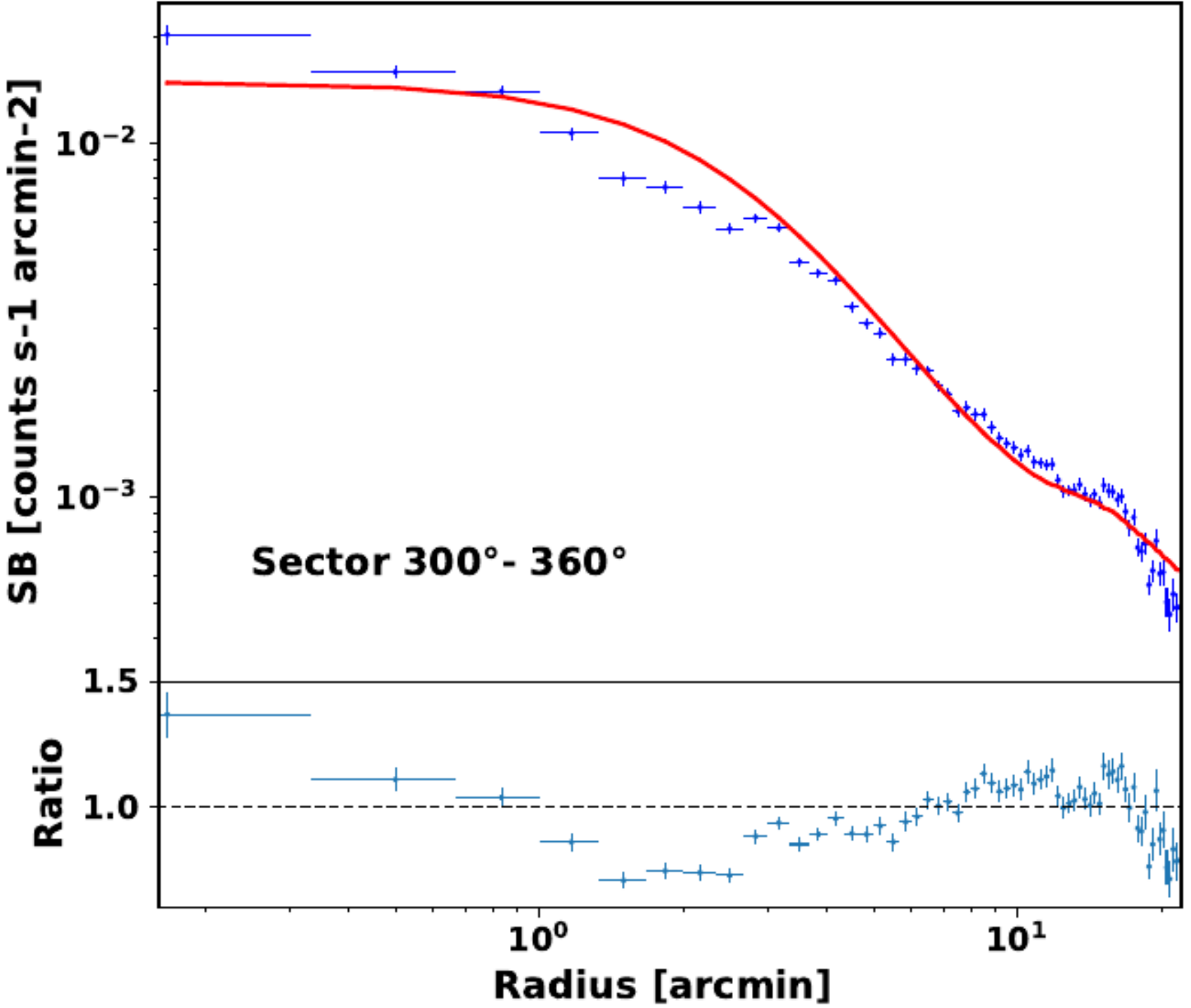}}
\caption{Surface brightness profiles of RXCJ1825 (in blue) extracted in sectors of 60 deg starting from the R.A. axis and moving counterclockwise. Point-like sources were removed from the image. The red lines show the model profiles extracted in the same sectors as in Fig. \ref{fig:allsectors}. Bottom panels report the data over model ratio.}
\label{fig:prof60deg}
\end{figure*}
 
\begin{figure*}[ht]
    \centering{
\includegraphics[angle=0,width=9.1cm]{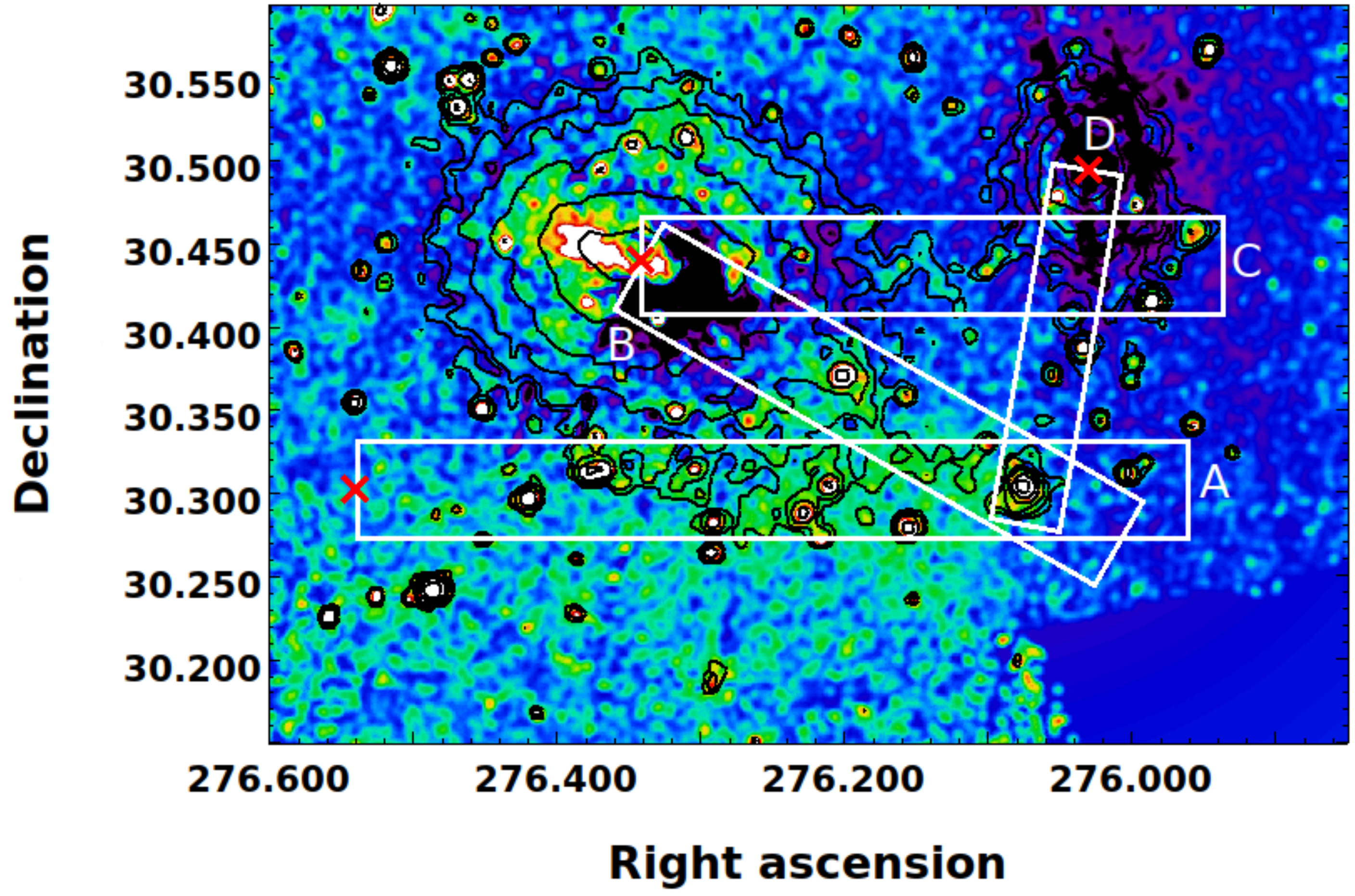}
    }
\caption{Regions used for the extractions of the surface brightness profiles shown in Fig.~\ref{fig:sector-box} plotted on the residual map. Each region is identified by a letter, and the zero point along the length is indicated by a red cross. In black are the surface brightness contours.}
\label{fig:box}
\vskip 1.cm
    \centering{
    \includegraphics[angle=0,width=8.cm]{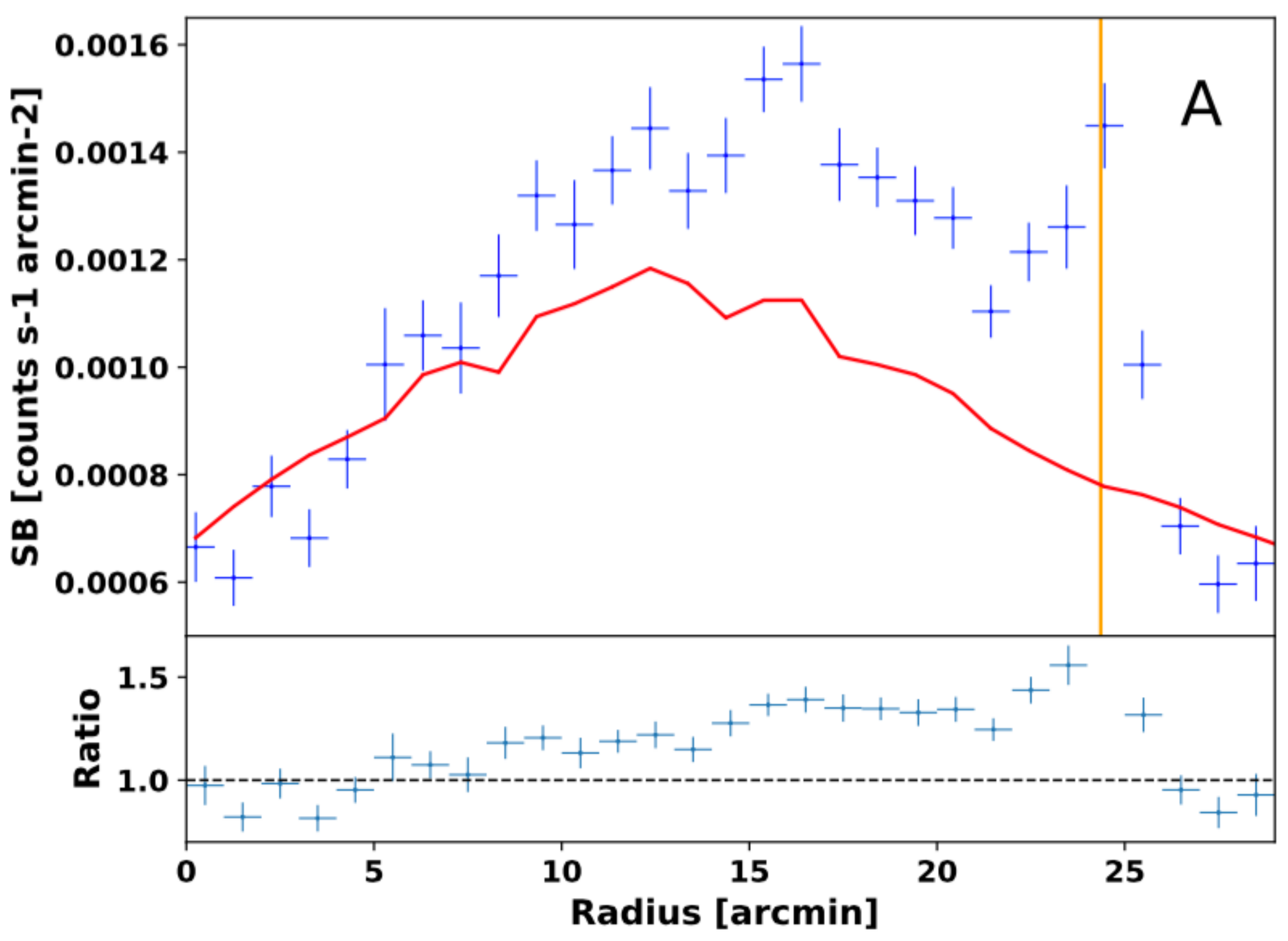}
 \hskip 1cm
    \includegraphics[angle=0,width=8.cm]{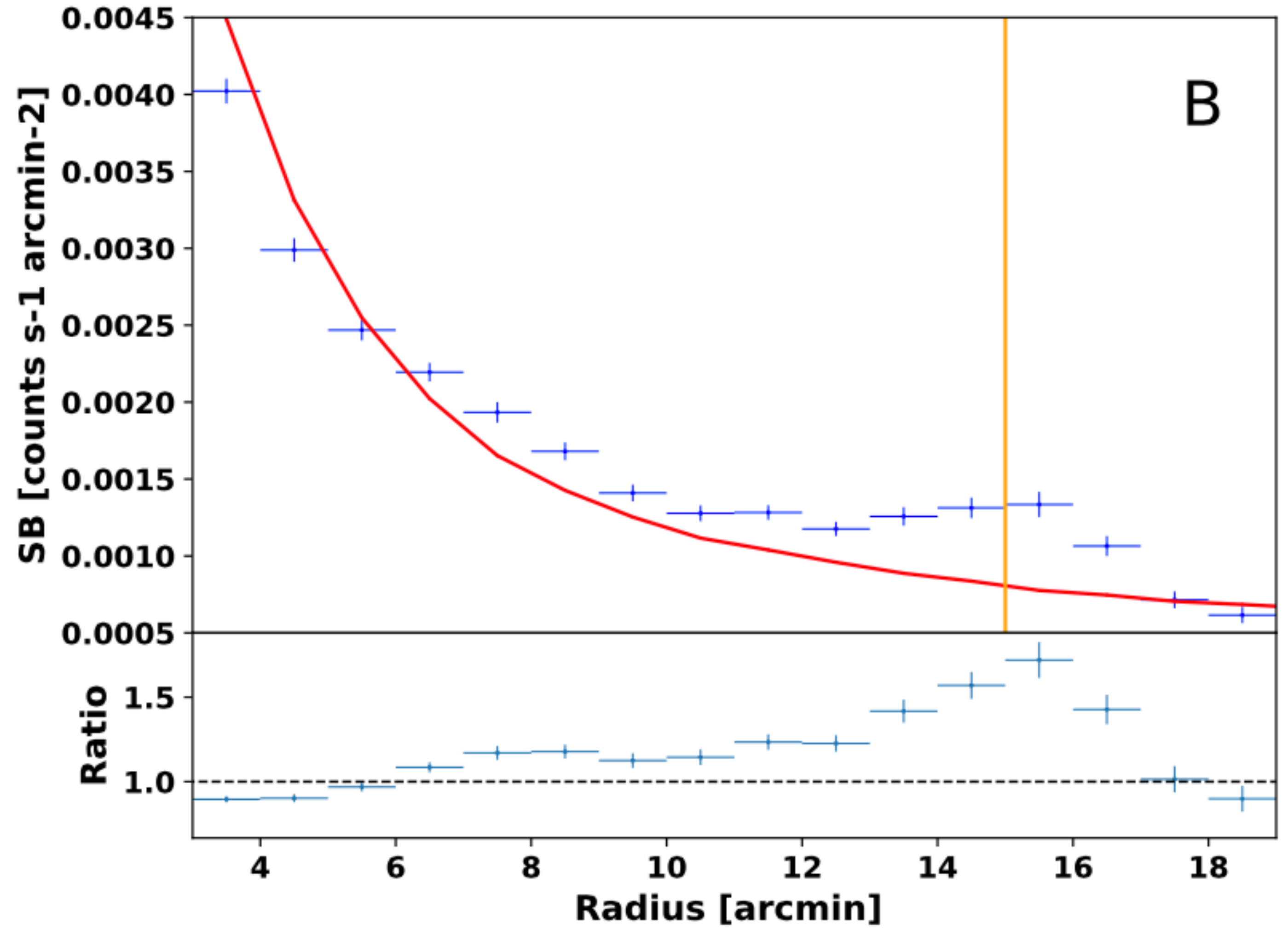}
    \includegraphics[angle=0,width=8.cm]{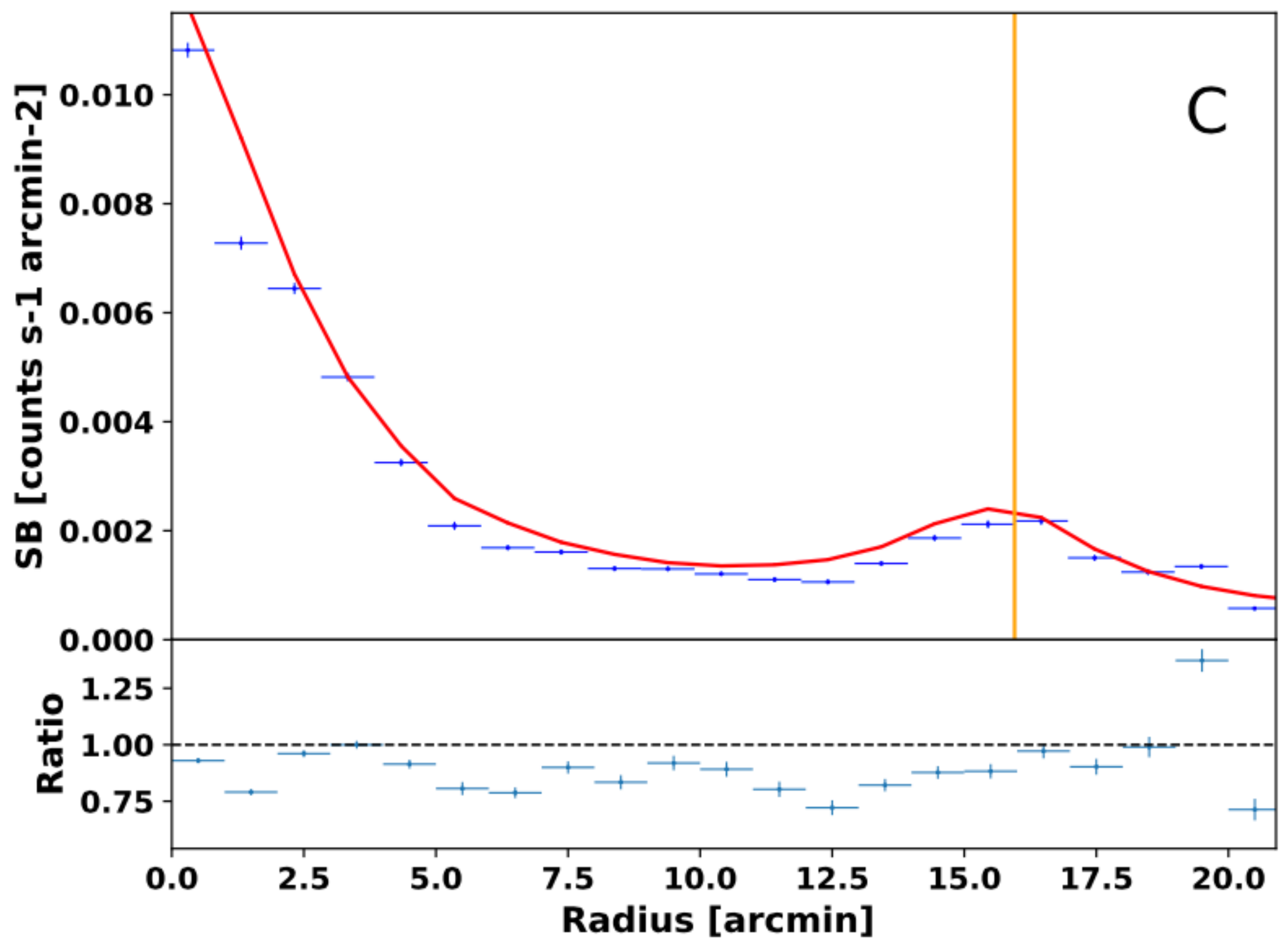}
 \hskip 1cm
    \includegraphics[angle=0,width=8.cm]{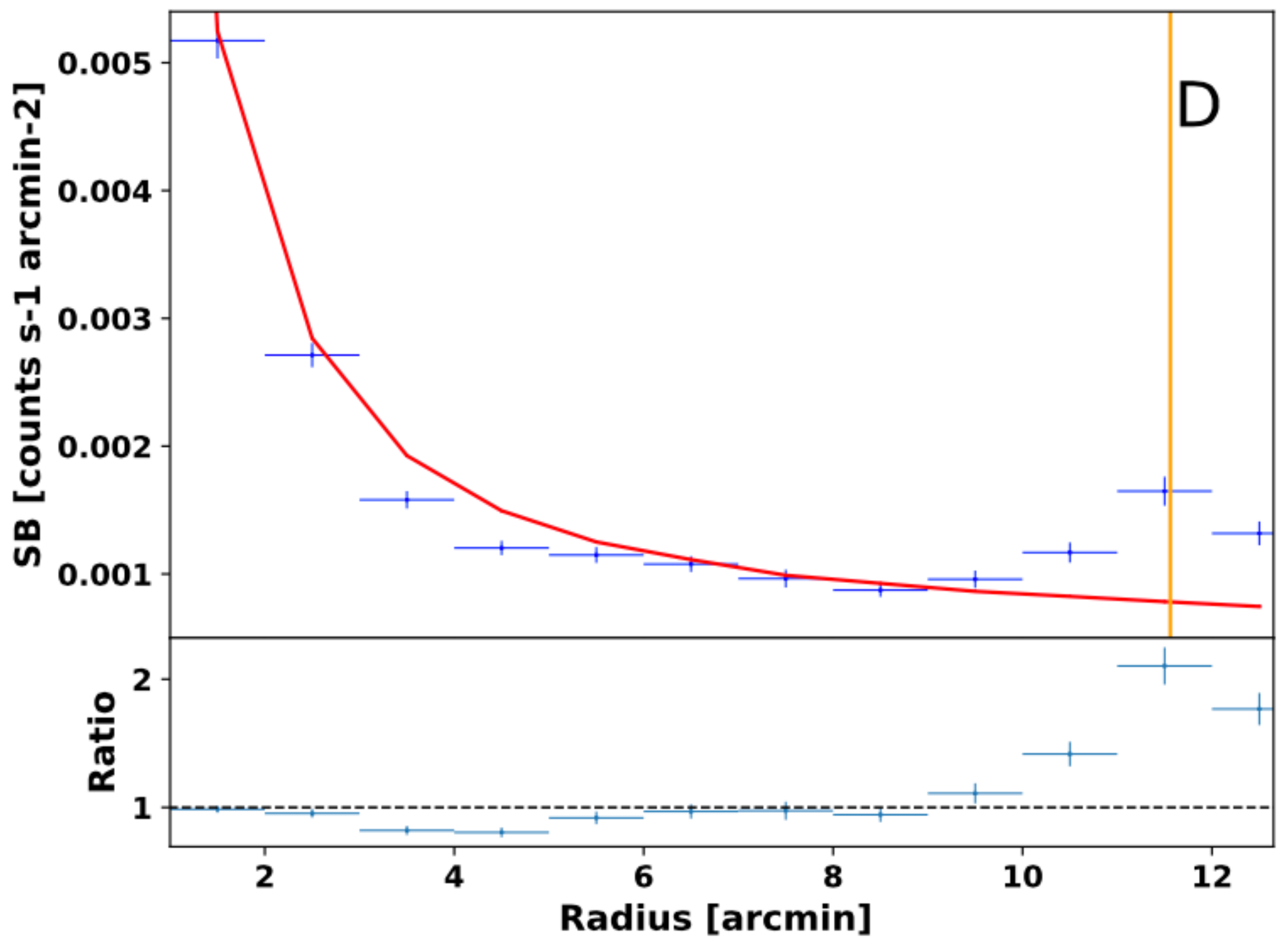}
    }
\caption{Surface brightness profiles extracted from the boxes shown in Fig.~\ref{fig:box}. Data are plotted in blue and the model in red. Point-like sources are always excluded (the point-like source associated to the SG is also excluded; see the discussion in Sect.~\ref{sec:sg_d}). Bottom panels report the data over model ratio. Orange vertical lines show the position of the SG (boxes A, B and D) and of CIZAJ1824 (box C). 
}
\label{fig:sector-box}
\end{figure*}

\begin{figure}
    \centering{
    \includegraphics[angle=0,width=9.5cm]{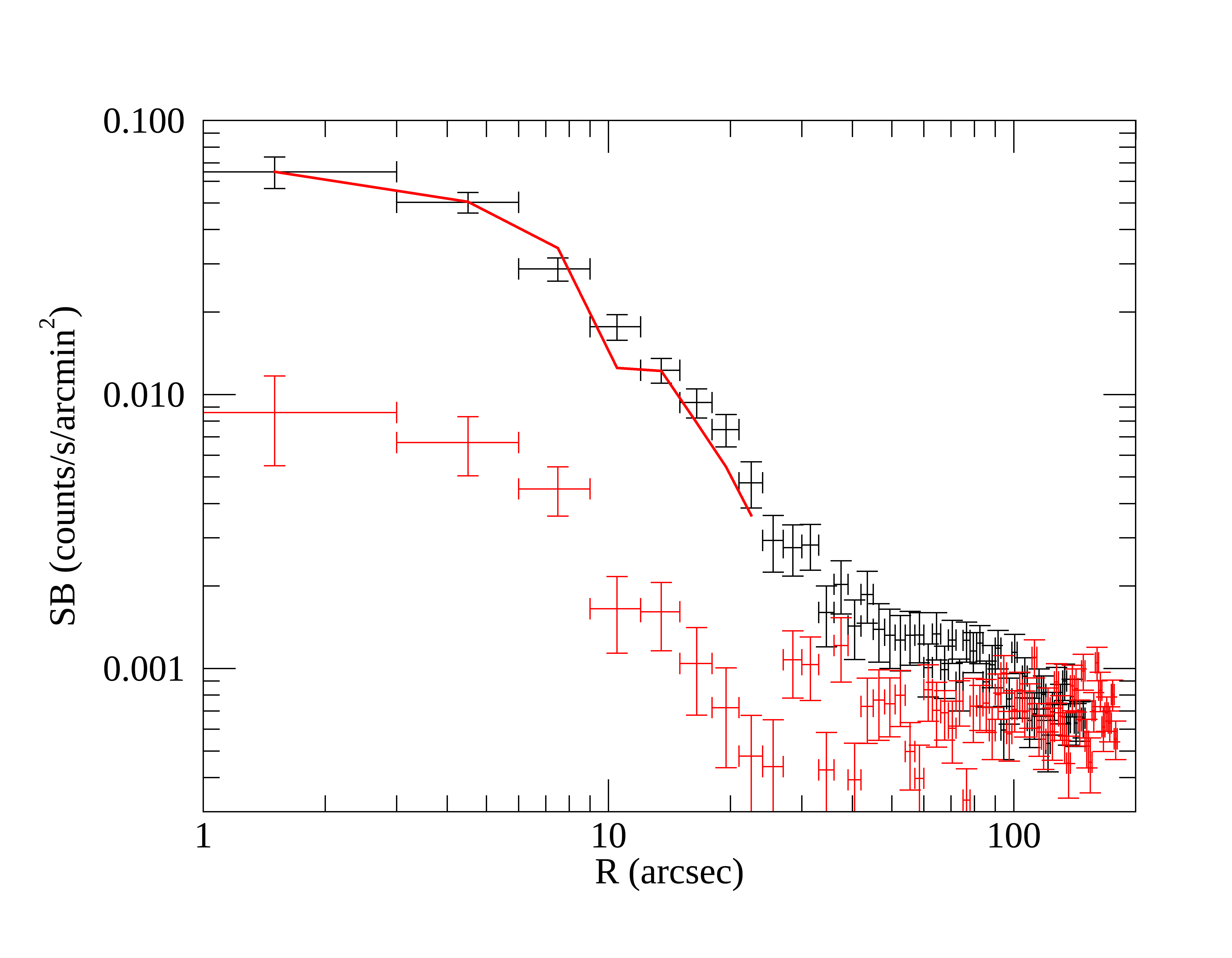}
}
 \caption{XMM-Newton surface brightness profile of the SG (black crosses) as compared to the profile of the nearby comparison point source (red crosses). The red continuous profile shows the surface brightness profile of the point source rescaled to match the intensity of the SG to simplify the comparison.}
        \label{fig:sbcorona}
\end{figure}

\begin{figure}
    \centering{
    \includegraphics[angle=0,width=9.cm]{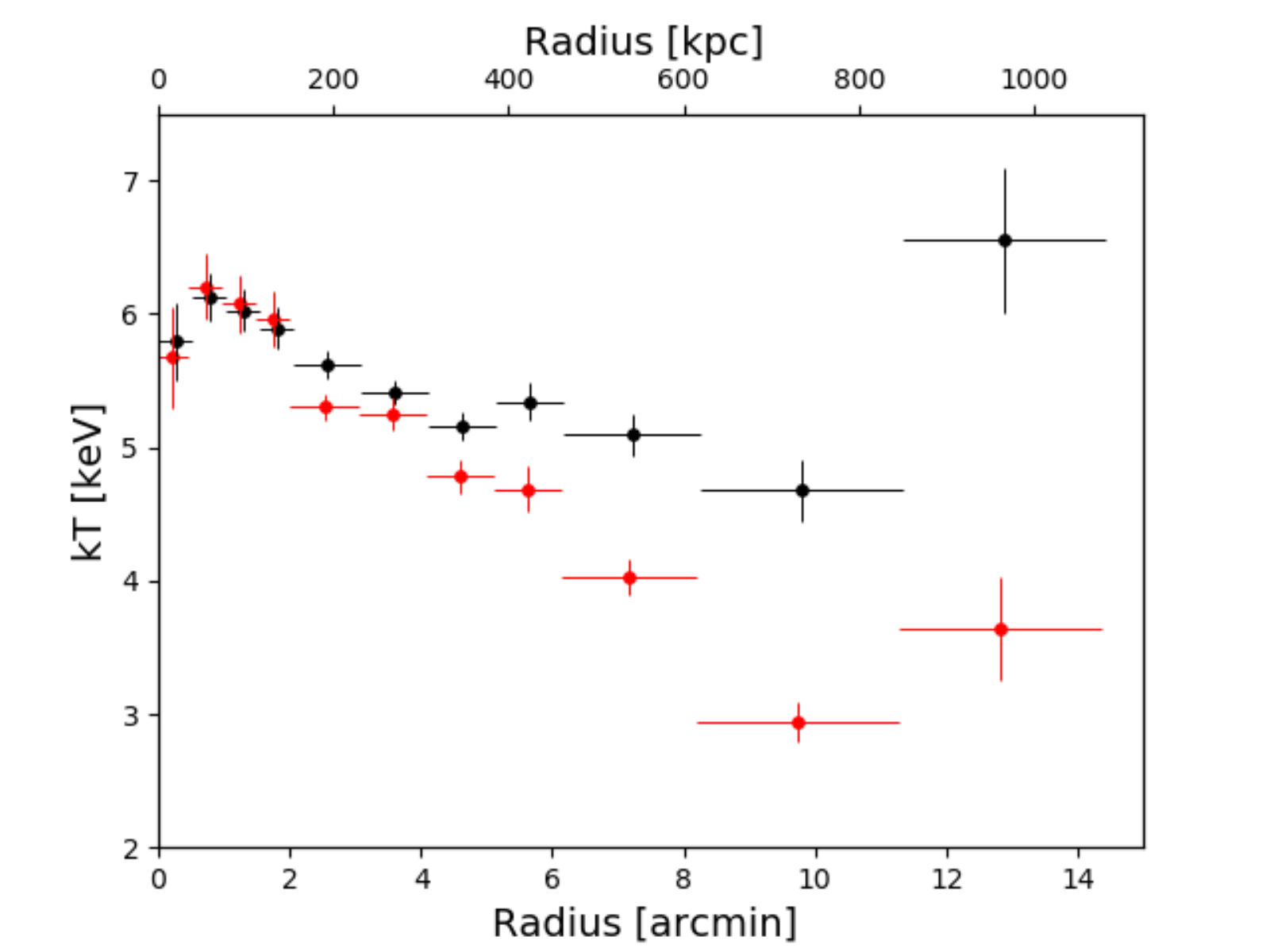}
    \includegraphics[angle=0,width=9.cm]{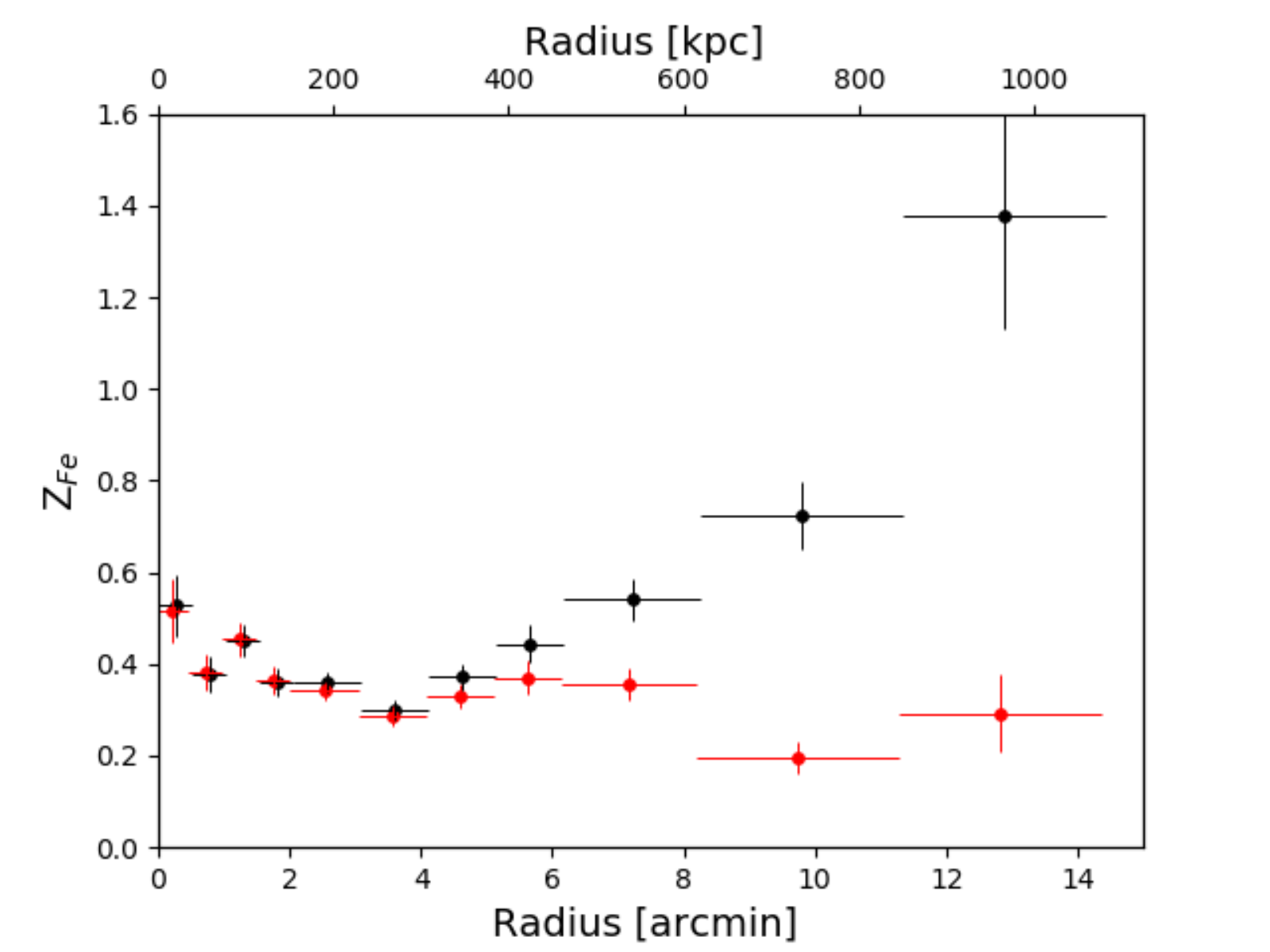}
    \includegraphics[angle=0,width=9.cm]{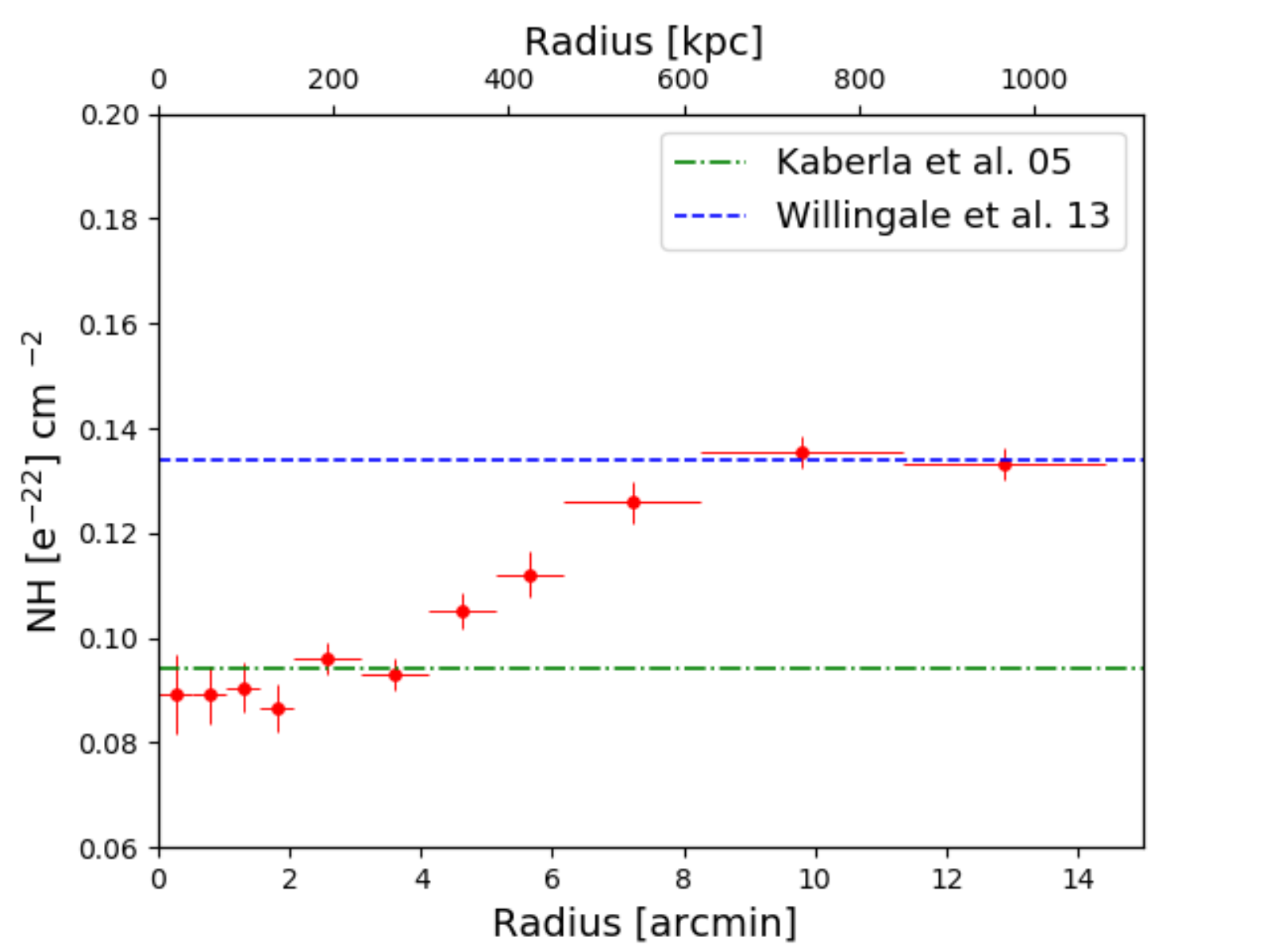}}
\caption{Temperature ({\it upper panel}) and metal abundance ({\it middle panel}) profiles of RXCJ1825. Black points show the results from the standard analysis with NH fixed to the LAB value and energy band [0.5-10.] keV. Red points show the results with NH left free to vary in the fit (the points in each bin have been shifted slightly for visual clarity). { \it Lower panel}: Galactic hydrogen column density profile of RXCJ1825. The green dot-dashed line is the NH LAB value \citep{kaberla05}, and the blue dashed line is the total NH (NHI + NHII) value taken from \cite{willingale13}.
}
\label{fig:profrxc}
\end{figure}

Another way to investigate the presence of features such as sharp discontinuities in the surface brightness map is through the unsharp-masking method. Unsharp-masked images are the ratio between the difference of two Gaussian-smoothed images (having different smoothing lengths) and their sum. Different choices of smoothing lengths can highlight features on different scales. In Fig. ~\ref{fig:unsharp} we show the unsharp-masks for the Lyra complex obtained with Gaussian smoothing lengths $\sigma=20$ and $800$ arcsec (left) and with Gaussian smoothing lengths  $\sigma=20$ and $240$ arcsec (right), chosen as they maximize the contrast at the large and small scales, respectively. Point sources were removed before smoothing and were replaced with an average value computed from nearby surrounding pixels. 
The image in the left panel is more sensitive to large-scale features and highlights the presence of a patchy bridge of surface brightness between RXCJ1825 and the SG at R.A.=18h:24m:17.7s, DEC=+30d:18m:15s (J2000.0), with clear traces of extended emission around the SG position. No sharp discontinuities are visible in this image. 
On the contrary, the image in the right panel reveals structures on small scales in the core region of RXCJ1825.
The image shows a long structure connecting the two BCGs of RXCJ1825, BCG-E and BCG-W, that is sharper in the south and less defined in the north. Another structure, with a triangular shape, is clearly visible west of the BCG-W. Small-scale emission is also present in the core of CIZAJ1824 and at the position of the SG.
These features are discussed further in Sect.\ref{sec:disc}.

\subsubsection{Search for surface brightness excesses and density discontinuities}
\label{sec:search}
We compared the surface brightness profiles of RXCJ1825 with the surface brightness of the model image in the six sectors of Fig.~\ref{fig:allsectors}. The profiles are shown in Fig.~\ref{fig:prof60deg}.
We find a statistically significant (at $\sim 3\sigma$) excess at radii smaller than 6 arcmin in sector 120-180 deg only; this excess corresponds to the elongated emission between the two BCGs and extends beyond in the northeast direction. At larger radii ($>$ 6 arcmin), excesses are present in the 240-300 deg and 300-360 deg sectors toward the direction of the SG.

We also looked for statistically significant surface brightness excesses in other directions using profiles extracted in ad-hoc boxes. We chose the boxes guided by the candidate excesses visible in the residual map. An example of interesting regions and boxes is
shown in Fig.~\ref{fig:box}, and the profiles extracted from the respective boxes are shown in Fig.~\ref{fig:sector-box}.
We found that the surface brightness profile between RXCJ1825 and CIZAJ1824 (box C in Fig.~\ref{fig:sector-box}) does not show any excess from the model (the same is true for a box directly connecting  the centers of the two clusters, not reported here, and in the 0-60 deg sector in Fig. \ref{fig:prof60deg}); indeed the emission is consistent with being produced only by the overlap of the two cluster halos. 
No excess in the surface brightness is present from CIZAJ1824 towards the SG (box D in Fig.~\ref{fig:sector-box});  some diffuse emission occurs at the SG position. 
Conversely, we found that the extended excess south and southwest of RXCJ1825 towards the SG (A and B boxes in Fig.~\ref{fig:sector-box}) is statistically significant (at more than $3\sigma$), suggesting some physical connection between RXCJ1825 and the SG.
We note that the excess emission in the two boxes A and B is rather irregular, suggesting the presence of gas inhomogeneities. 

From visual inspection of the two unsharp-mask images (Fig.~\ref{fig:unsharp}) and the radial profiles in Fig.~\ref{fig:prof60deg}, we find no significant discontinuities in the surface brightness that could be indicators of cold fronts or shocks.
We tried to verify the presence of surface brightness jumps in various regions of the cluster, not shown here for clarity, without finding any trace.
This is somewhat surprising as RXCJ1825 shows strong evidences of recent interactions and is discussed is Sect.~\ref{sec:disc}. 

\subsubsection{The Southern Galaxy} 
\label{sec:sbsg}
Investigation of the right panel of Fig.~\ref{fig:unsharp} shows evidence of an excess coincident with the position of the SG on small scales. 
We extracted the surface brightness profile of the X-ray source corresponding to the SG using R.A.=18h24m17.7s Dec=+30d18m15s (J2000.0)  as central coordinates and 3 arcsec for the bin width. The profile is shown in black in Fig.~\ref{fig:sbcorona}: it decreases rapidly from the center up to about 20 arcsec, then smoothly to reach a constant background level at radii $> 100$ arcsec. To assess if the source we detected is extended or point-like, we compared its surface brightness profile with a similar profile extracted from a nearby point source ($276.02003, 30.34340$) located at the same distance from the RXCJ1825 cluster center as the SG. We used this profile to estimate an ``effective'' point-spread function (PSF) close to the location of the SG, since this effect cannot be estimated in a more precise way given the mosaicing of multiple observations in the X-COP observation strategy. The profile of the nearby point source is also shown in Fig.~\ref{fig:sbcorona}. The shape of the profile in the central part of the SG ($r<20$ arcsec) is consistent with the profile of the point source, suggesting the presence of a source not spatially resolved by XMM-Newton. The fit of the central part of the SG profile with a Gaussian returns $\sigma=9.3\pm0.7$ arcsec, indicating that the central source must be smaller than $\sim 12-13$ kpc in physical units.
The behavior of the profiles differs in the external regions: while the point source immediately reaches the background level at $r>20$ arcsec, the source corresponding to the SG is embedded in a diffuse emission (see Fig.~\ref{fig:unsharp} left). 
 
\subsection{Temperature and metal abundance profiles}
\label{sec:profx}
We measured the radial profiles of the temperature and metal abundance of RXCJ1825 and CIZAJ1824 from the X-ray centroid up to their respective R$_{500}$ (i.e., $\sim 1100$ kpc for RXCJ1825 and $\sim 900$ kpc for CIZAJ1824, Sect. \ref{sec:profrxc} and \ref{sec:profciza}). For RXCJ1825 we selected circular annuli up to 5 arcmin and sectors from 5 to 14 arcmin to avoid contamination from CIZAJ1824 (we excluded the sector within 300-390 deg). In the case of CIZAJ1824 we extracted spectra from circular annuli up to 5 arcmin and excluded the contribution of RXCJ1825 selecting sectors of above 5 arcmin up to 10 arcmin (the excluded sector is in this case 130-270 deg).

\subsubsection{Profiles of RXCJ1825}
\label{sec:profrxc}
We fitted the spectra twice: the first time we fixed the redshift and the Galactic gas column density, NH, to the value of the 21 cm measurement \citep{kaberla05}, and the second time we left NH free to vary. The resulting temperature, metal abundance, and NH profiles of RXCJ1825 are shown in Fig.~\ref{fig:profrxc}.
We find that leaving NH free to vary has a significant impact on both temperature and metal abundance profiles, especially in the outer regions (r$\gtrsim$500 kpc) where the temperature profile decreases and the metal abundance profile becomes flatter. 
Indeed, the overall value of NH changes significantly with the distance from the center, varying from a central value consistent with the pure neutral hydrogen estimate (NH=$9.43\times 10^{20}$  cm$^{-2}$) to a larger value, which is consistent with the estimate of the column density corrected for the contribution of molecular hydrogen (NH=$1.33\times 10^{21}$  cm$^{-2}$, \citealt{willingale13}).
 
Our results point towards a variation of the Galactic absorption in the field of the Lyra complex. To test the robustness of our NH estimates, we looked at the IR emission of the Galactic dust in the same field of view by considering the IRAS 100 $\mu $m cleaned map of \cite{schlegel98}, and computed the expected NH using the conversion from IR flux into total (hydrogen+molecular) NH given by \cite{boulanger96} (see their Table 1). We found that the expected NH ranges between $\sim$1.2-1.6$\times 10^{21}$ cm$^{-2}$ in good agreement with our values. The IRAS map also shows a radial variation of the dust emission, lower in the center of RXCJ1825 and gradually higher towards the outside, again in good agreement with the results of our spectral fits.

The resulting mean temperature and metal abundance within R$_{500}$ of RXCJ1825 are $4.86\pm0.05$ keV and $0.22\pm0.01$, respectively.

We deprojected the temperature and the surface brightness profiles of this cluster to derive the gas density and deprojected temperature profiles. After a slight smoothing of the data in order to reduce nonphysical fluctuations that would be further enhanced by the deprojection process, we adopted the standard onion-skin technique to deproject data \citep{kriss83,ettori02_a1795}. We included a correction factor to account for the emission of the cluster beyond the outermost bin (see \citealt{ghizzardi04} for details).
Total and gas masses were then computed under the hypothesis of hydrostatic equilibrium and, by assuming a power-law behaviour for the mass in the outskirts to extrapolate the mass profile at larger radii, we estimated M$_{500}=3.47^{ +0.61}_{-0.56} \times 10^{14}$ M$_\odot$ within R$_{500}=1047^{+56}_{-59}$  kpc and M$_{200}=7.32^{ +1.99}_{-1.90} \times 10^{14}$ M$_\odot$ within R$_{200}=1822^{+156}_{-173}$ kpc.
Our value derived for M$_{200}$ agrees within $1\sigma$ c.l. with that reported in \cite{ettori19_xcop_mass}, M$_{200} = 6.15 \pm 0.56 \times 10^{14}$ M$_\odot$.  The uncertainties in the gravitating mass given by \cite{ettori19_xcop_mass} are smaller because their measurement relies on the X-ray temperature only in the cluster center while for large radii they used temperatures derived from the Sunyaev-Zel'dovich pressure profile (details are given in \citealt{ghirardini19}). Finally, we estimated the luminosity  L$_{500}=2.42^{+0.02}_{-0.02} \times 10^{44}$ erg/s within R$_{500}$.

\begin{figure}
    \centering{
    \includegraphics[angle=0,width=9.cm]{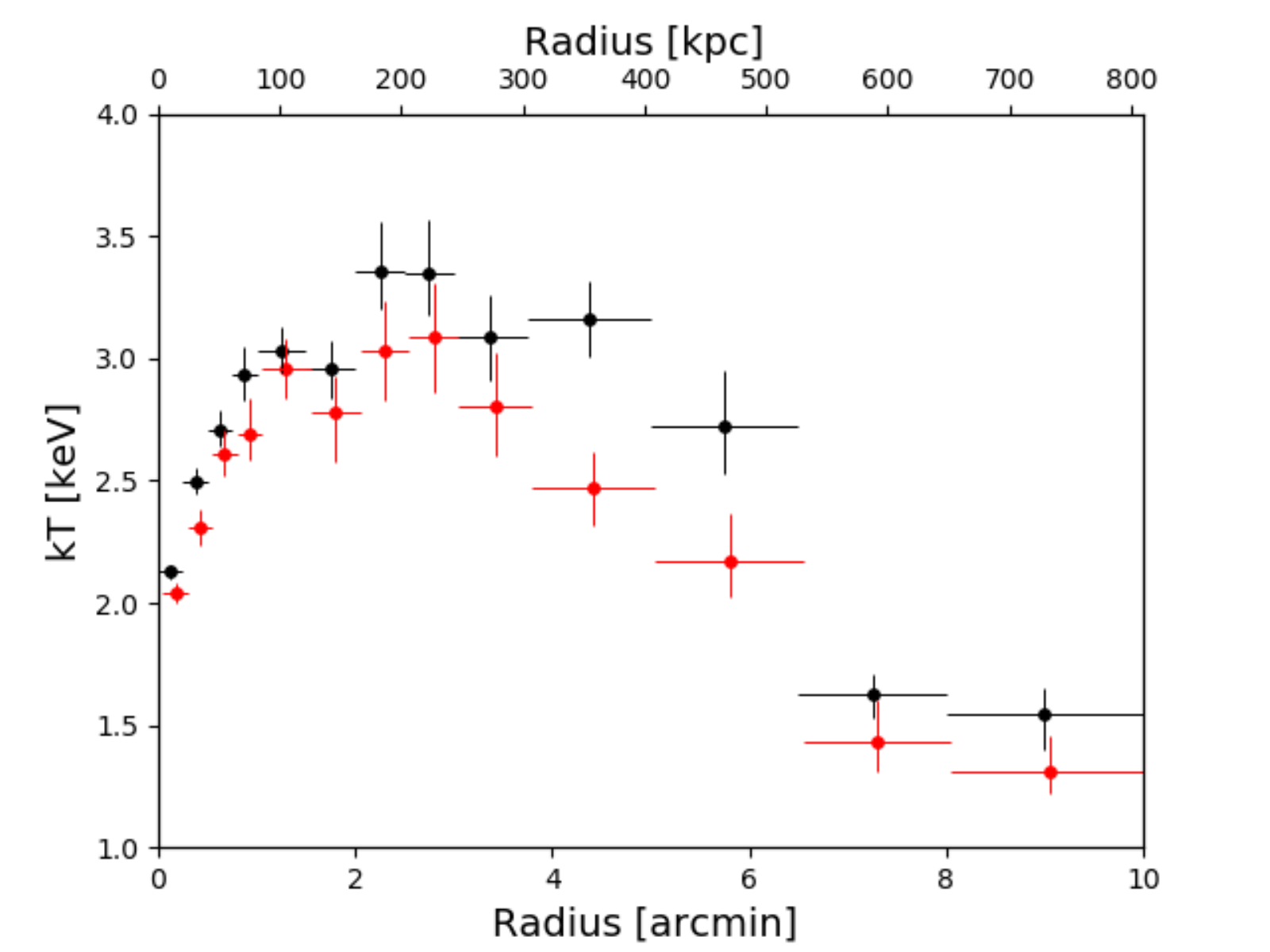}
    \includegraphics[angle=0,width=9.cm]{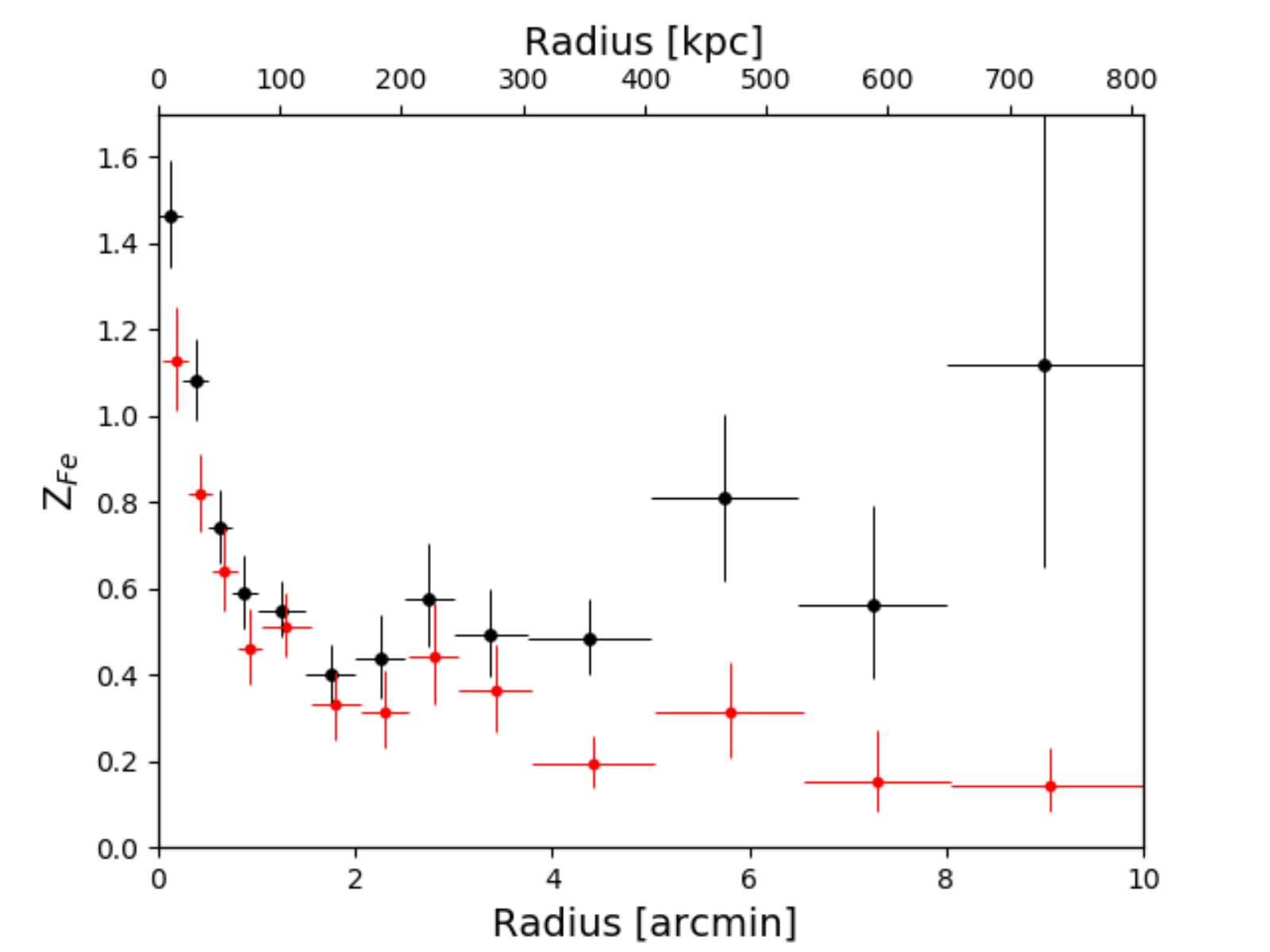}
    \includegraphics[angle=0,width=9.cm]{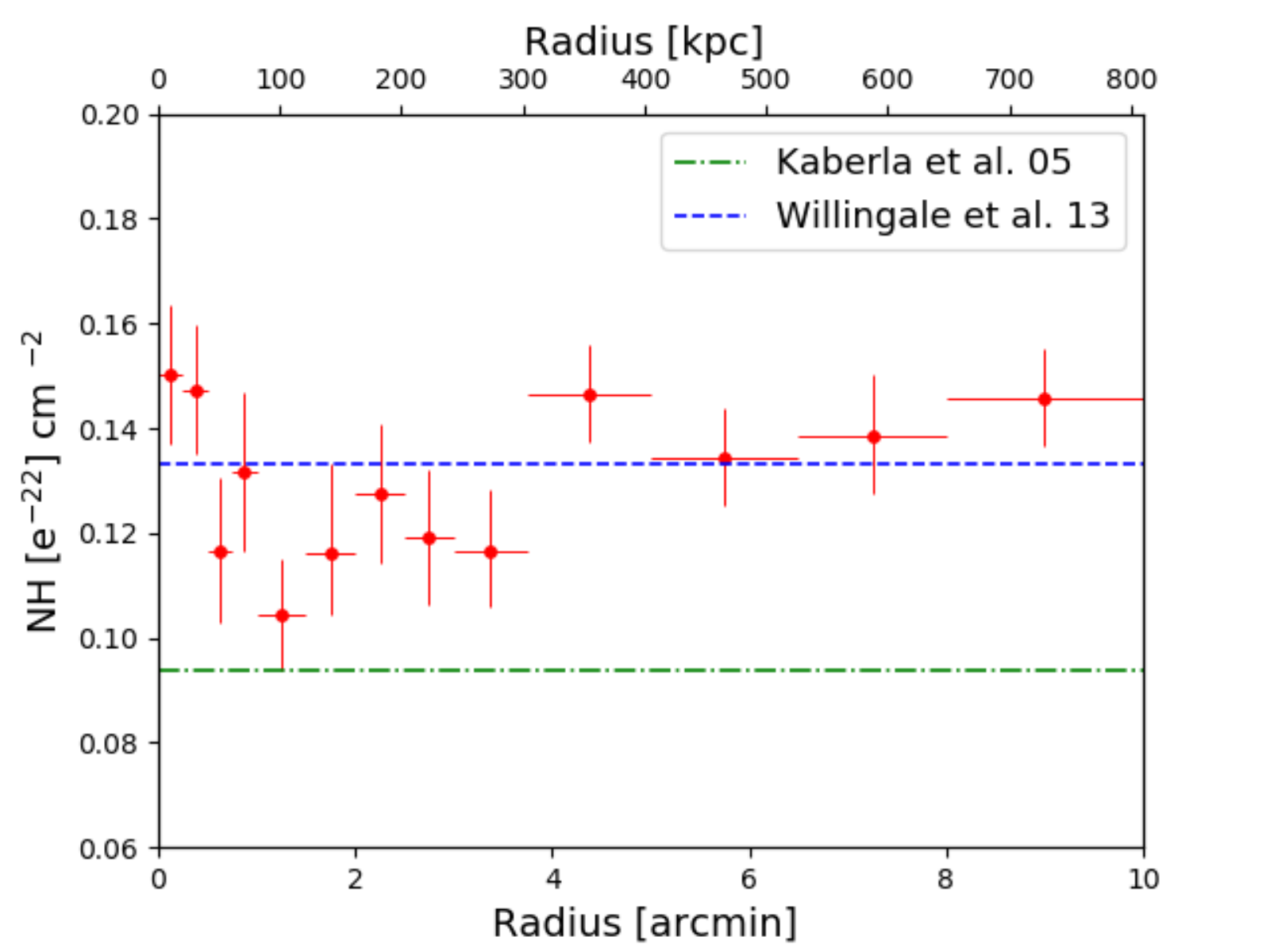}}   
\caption{Radial temperature ({\it upper panel}) and metal abundance ({\it middle panel}) profiles of CIZAJ1824. Black points show the results from the standard analysis with NH fixed to the LAB value and energy band [0.5-10.] keV and red points show the results with NH left free to vary in the fit (the points of each bin have been shifted slightly for visual clarity). {\it Lower panel}: Galactic hydrogen column density profile of CIZAJ1824. The green dot-dashed line is the NH LAB value \citep{kaberla05}, and the blue dashed line is the total NH (NHI + NHII) value taken from \cite{willingale13}.
}
\label{fig:profciza}
\end{figure}

\begin{figure*}
\centering{
    \includegraphics[angle=0,width=9.05cm]{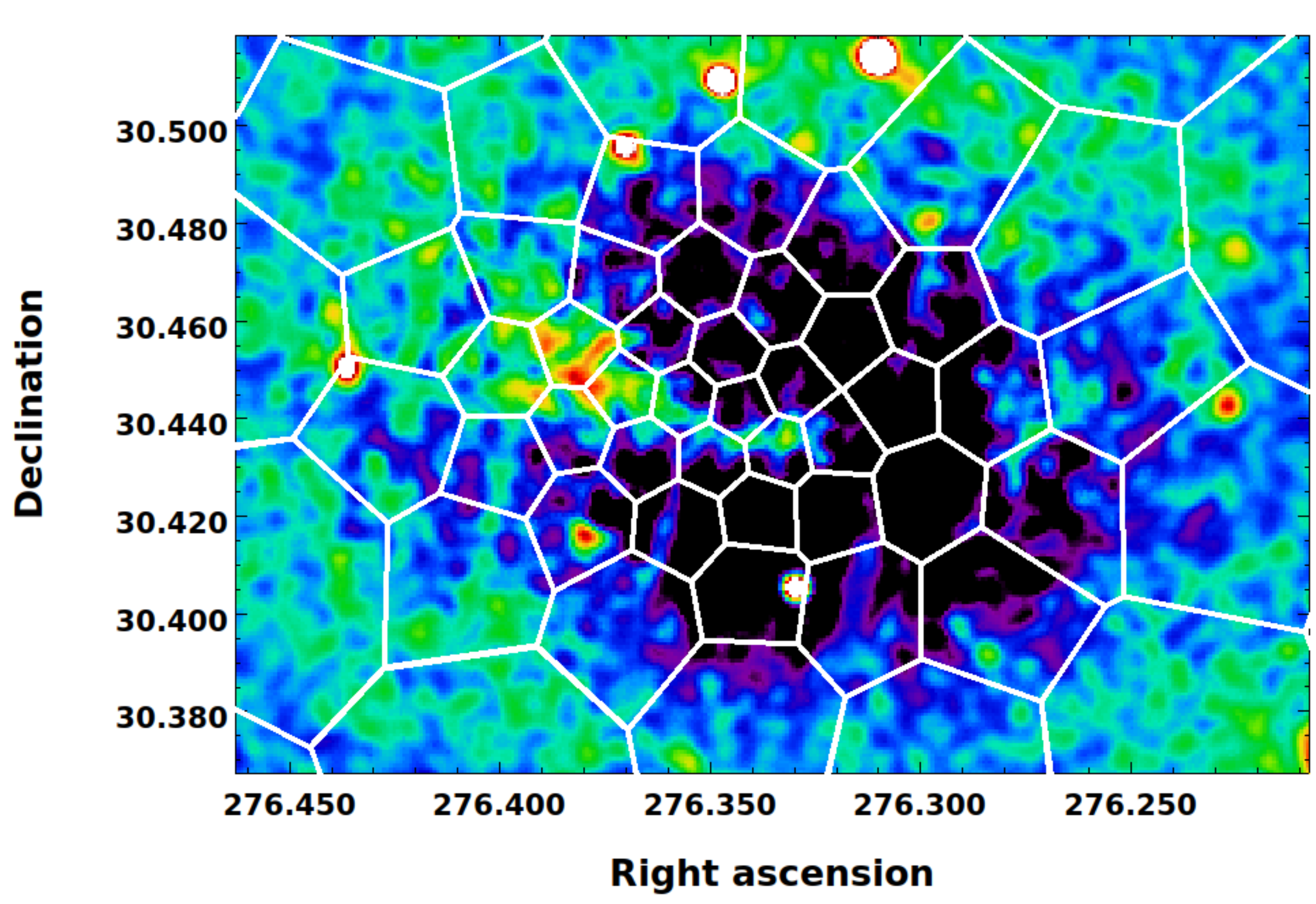}
    \includegraphics[angle=0,width=9.1cm]{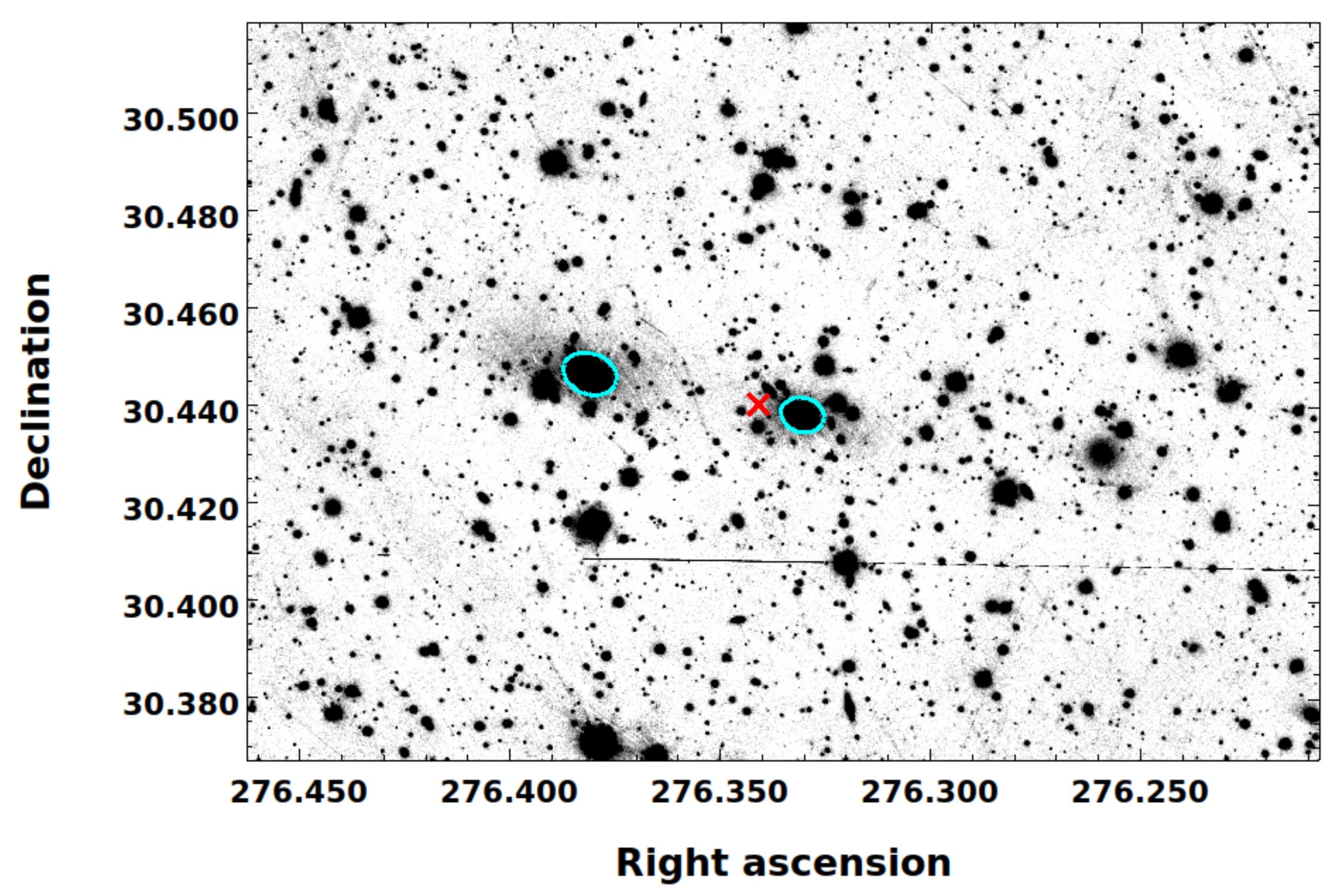}}
    \caption{{\it Left panel:} Residual map of the central regions of RXCJ1825. The regions chosen to perform the spectral analysis, selected using a S/N > 20 with source-to-background ratio I > 0.6 (\citealt{leccardi08_t}), are shown in white.  {\it Right panel:} Pan-STARRs r-band image of the core of RXCJ1825.}
\label{fig:map20}
\end{figure*}

\begin{figure*}
\centering{
    \includegraphics[angle=0,width=9.1cm]{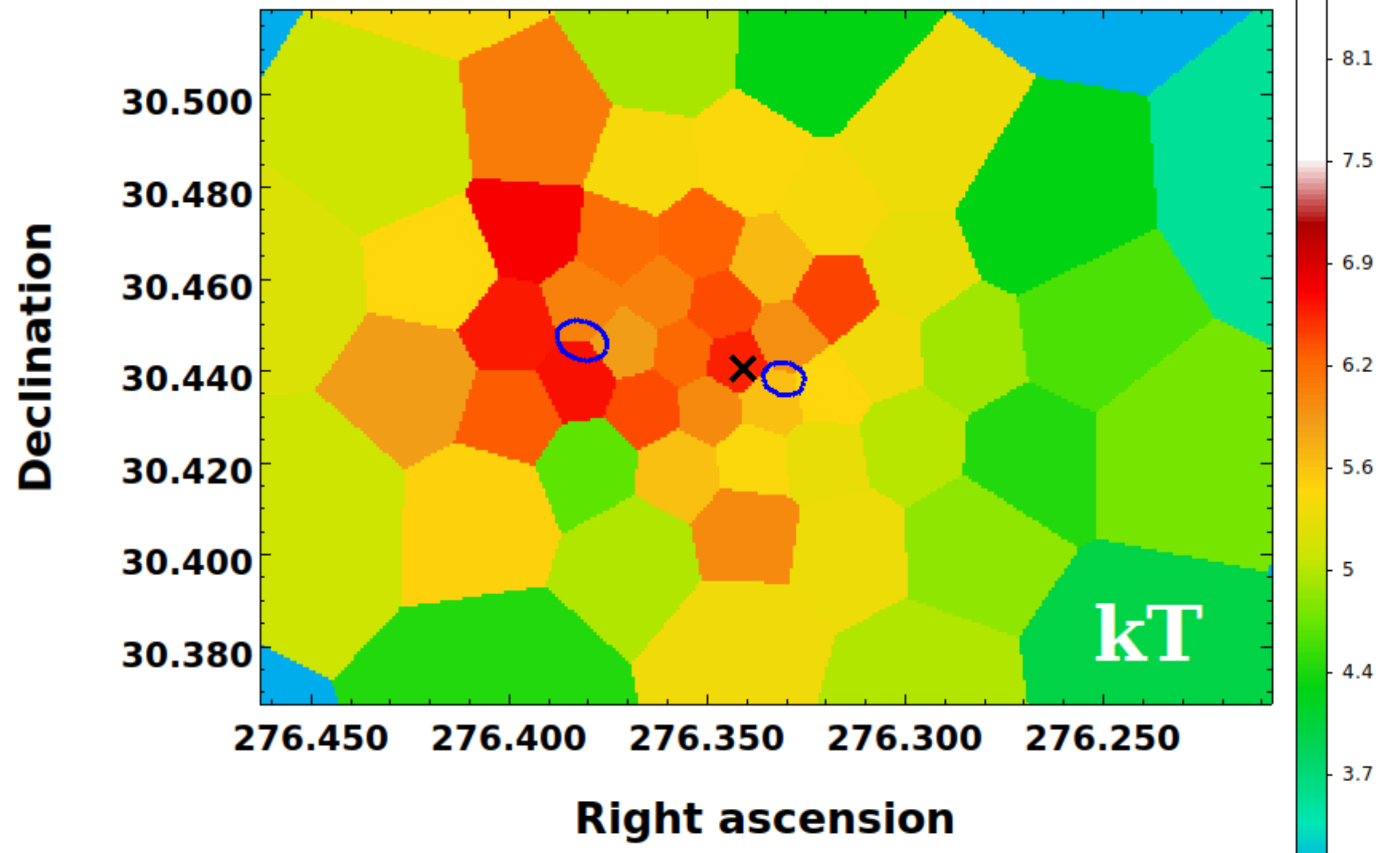}
    \includegraphics[angle=0,width=9.1cm]{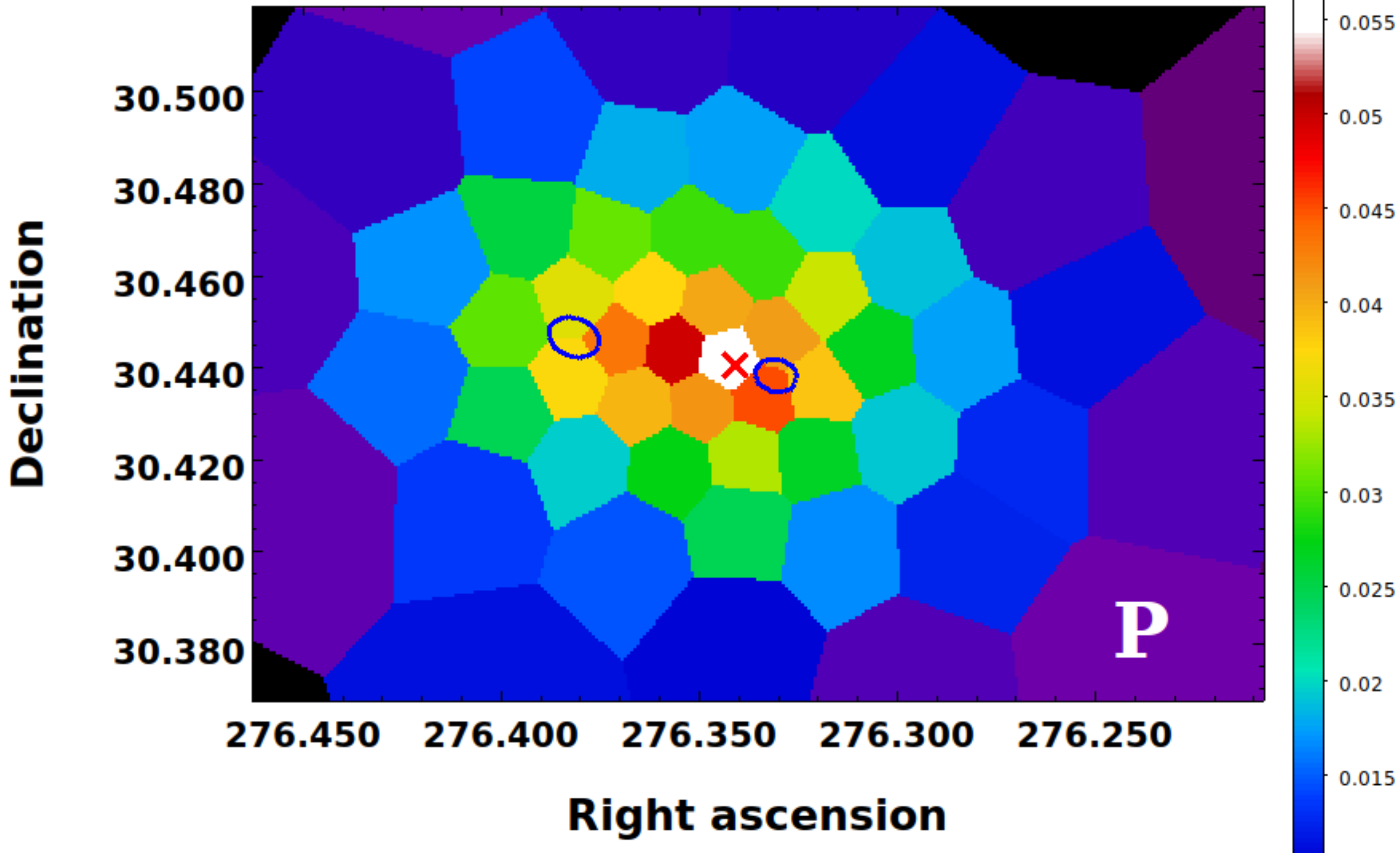}
    \includegraphics[angle=0,width=9.1cm]{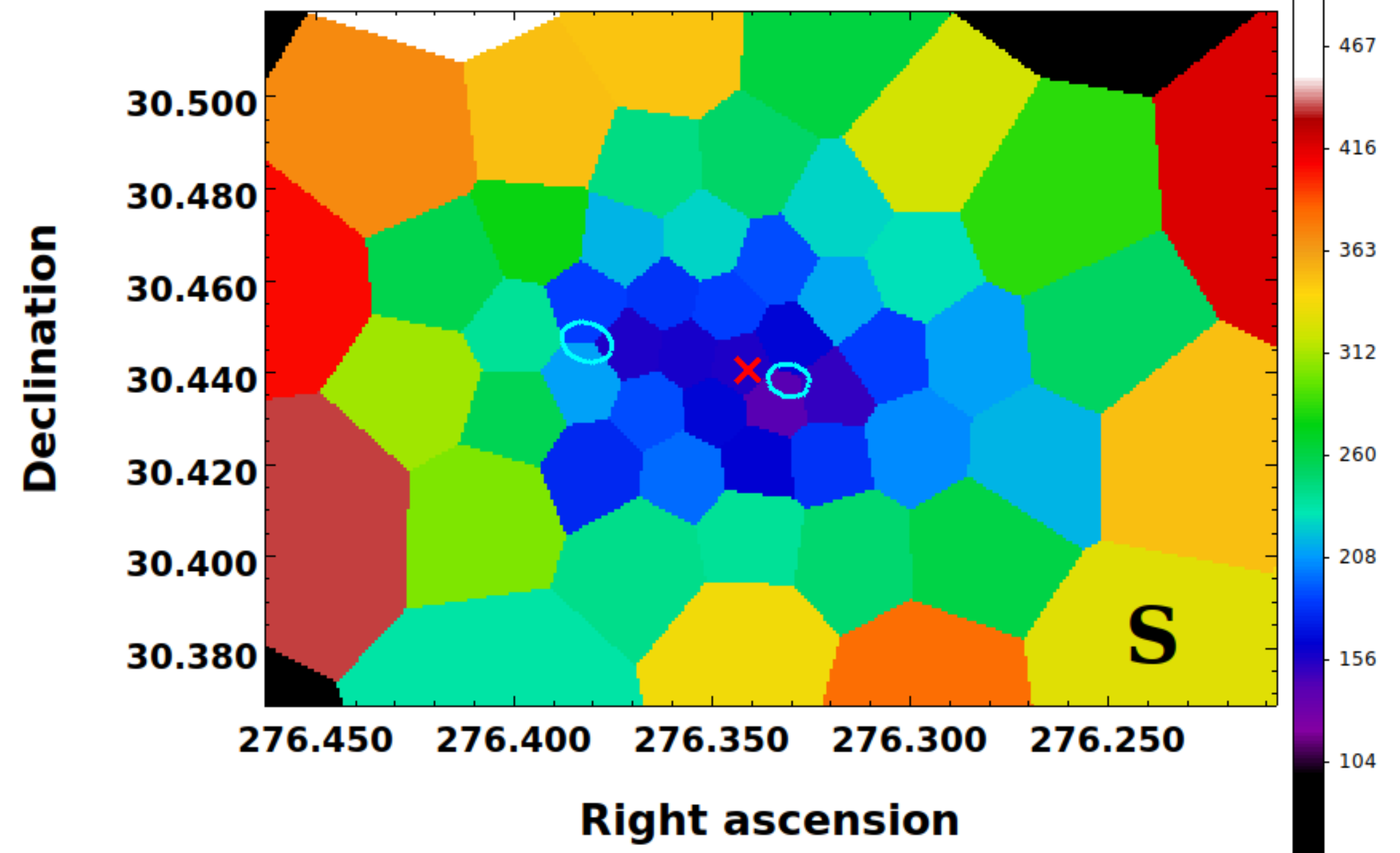}
    \includegraphics[angle=0,width=9.1cm]{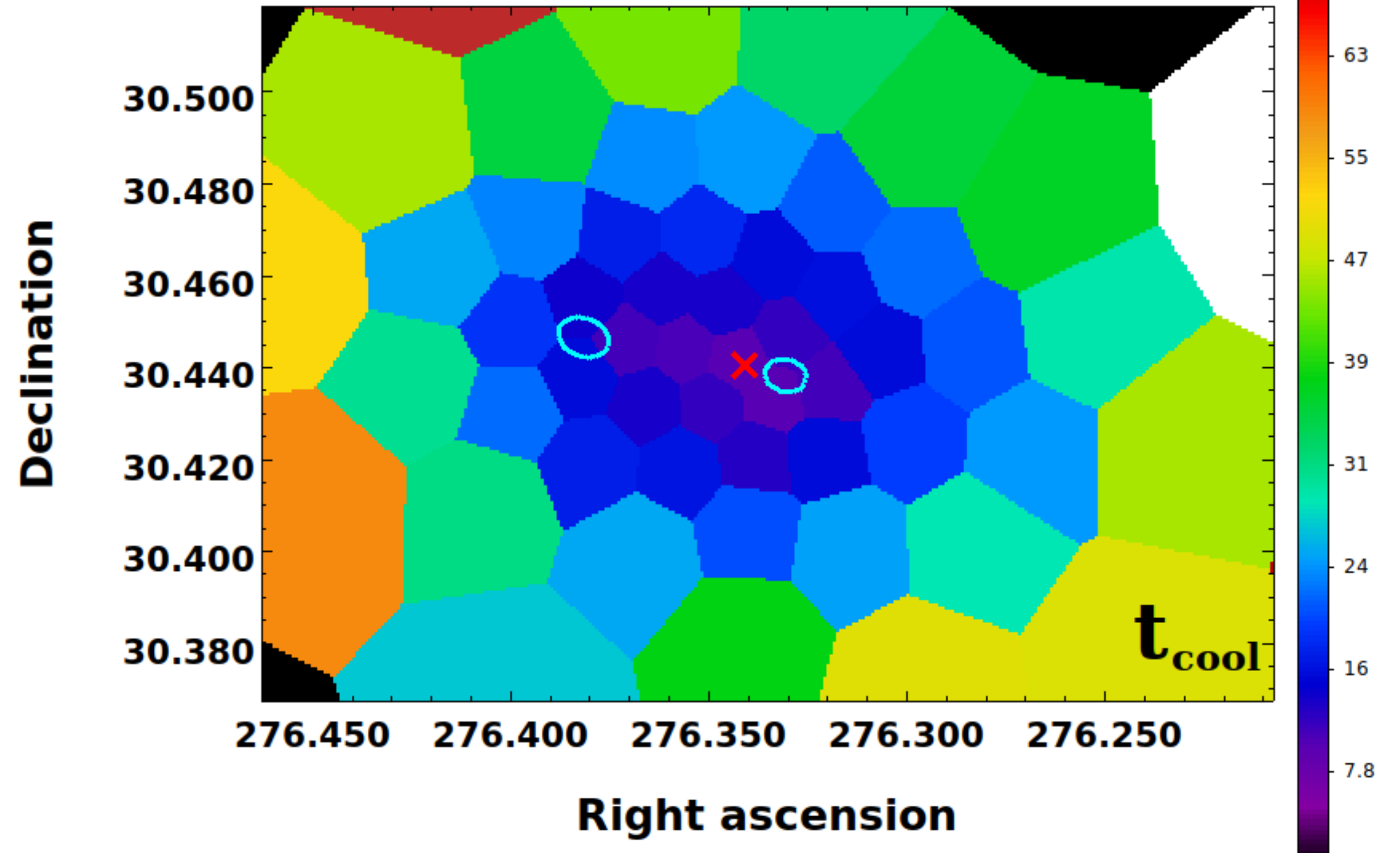}}
    \caption{Temperature  (keV), pressure  (keV cm$^{-3}$), entropy  (keV cm$^2$), and  cooling time (Gyr) maps of the center of RXCJ1825. For the uncertainties on the physical quantities see Fig.~\ref{fig:profmaprxc}. Small ellipses show the position of the two BCGs, the cross shows the position of the cluster centroid determined by the shape of the isophotes on large scales.}
\label{fig:map20rxc}
\end{figure*}

\subsubsection{Profiles of CIZAJ1824}
\label{sec:profciza}
We applied the first two spectral models described in the previous section to CIZAJ1824 and we show our results in Fig.~\ref{fig:profciza}.
Leaving NH free has a  smaller effect than in RXCJ1825, but the effect is qualitatively similar. In the core we find a temperature decrease and a metal abundance enhancement which are the hallmark of cool-core clusters. In the central bin ($r\lesssim21$ kpc) the entropy is $16.1\pm0.3$ keV cm$^{2}$, which is also in line with cool-core cluster central values \citep{cavagnolo09}.

Further out we observe a temperature decline and a metal abundance flattening which are typical of most clusters. 
By excising the central 100 kpc containing the cool-core we find a mean temperature of $2.14\pm0.05$ keV and a mean metal abundance of $0.28\pm0.03$. 

Following the same procedure described in the previous section, 
for CIZAJ1824 we estimated a total mass of M$_{500}=2.46^{ +0.70}_{-0.56} \times 10^{14}$ M$_\odot$ and a luminosity of L$_{500}=7.43^{+0.15}_{-0.15} \times 10^{43}$ erg/s within R$_{500}=932^{+80}_{-78}$  kpc and a total mass of M$_{200}=4.18^{ +1.71}_{-1.38} \times 10^{14}$ M$_\odot$ within R$_{200}=1512^{+186}_{-189}$ kpc.

\begin{figure*}
\centering{
  \includegraphics[angle=0,width=9.1cm]{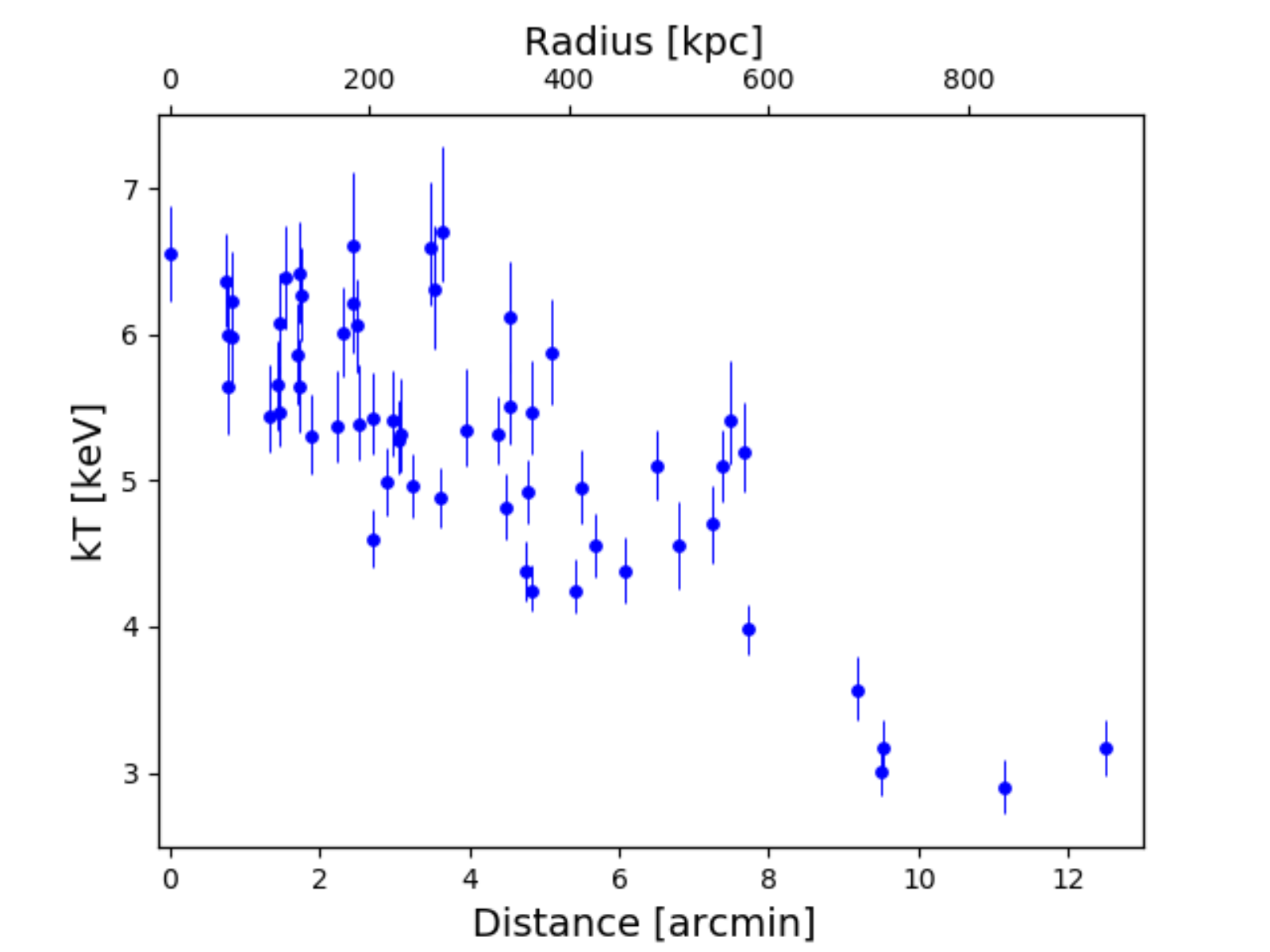}
  \includegraphics[angle=0,width=9.1cm]{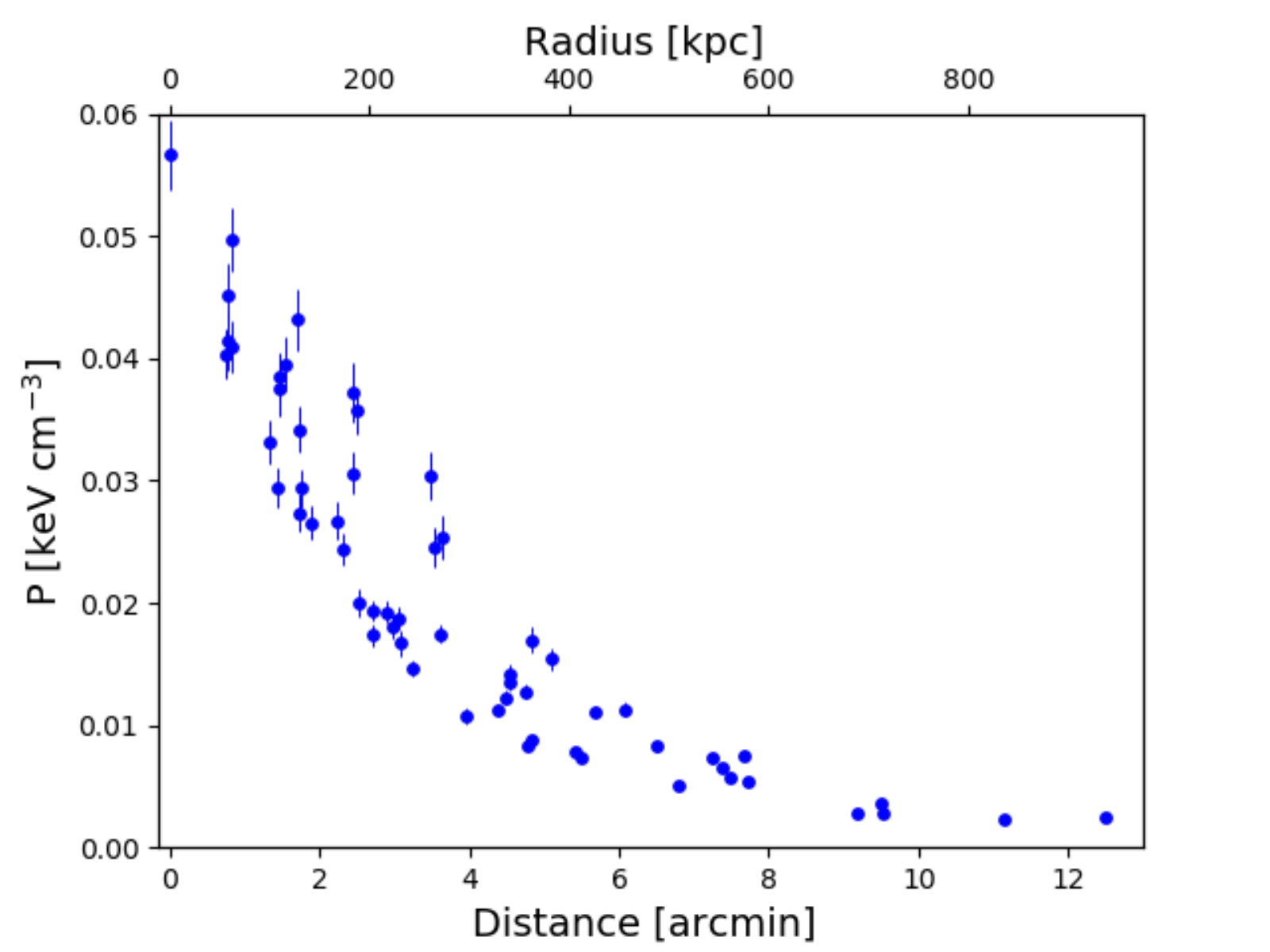}
  \includegraphics[angle=0,width=9.1cm]{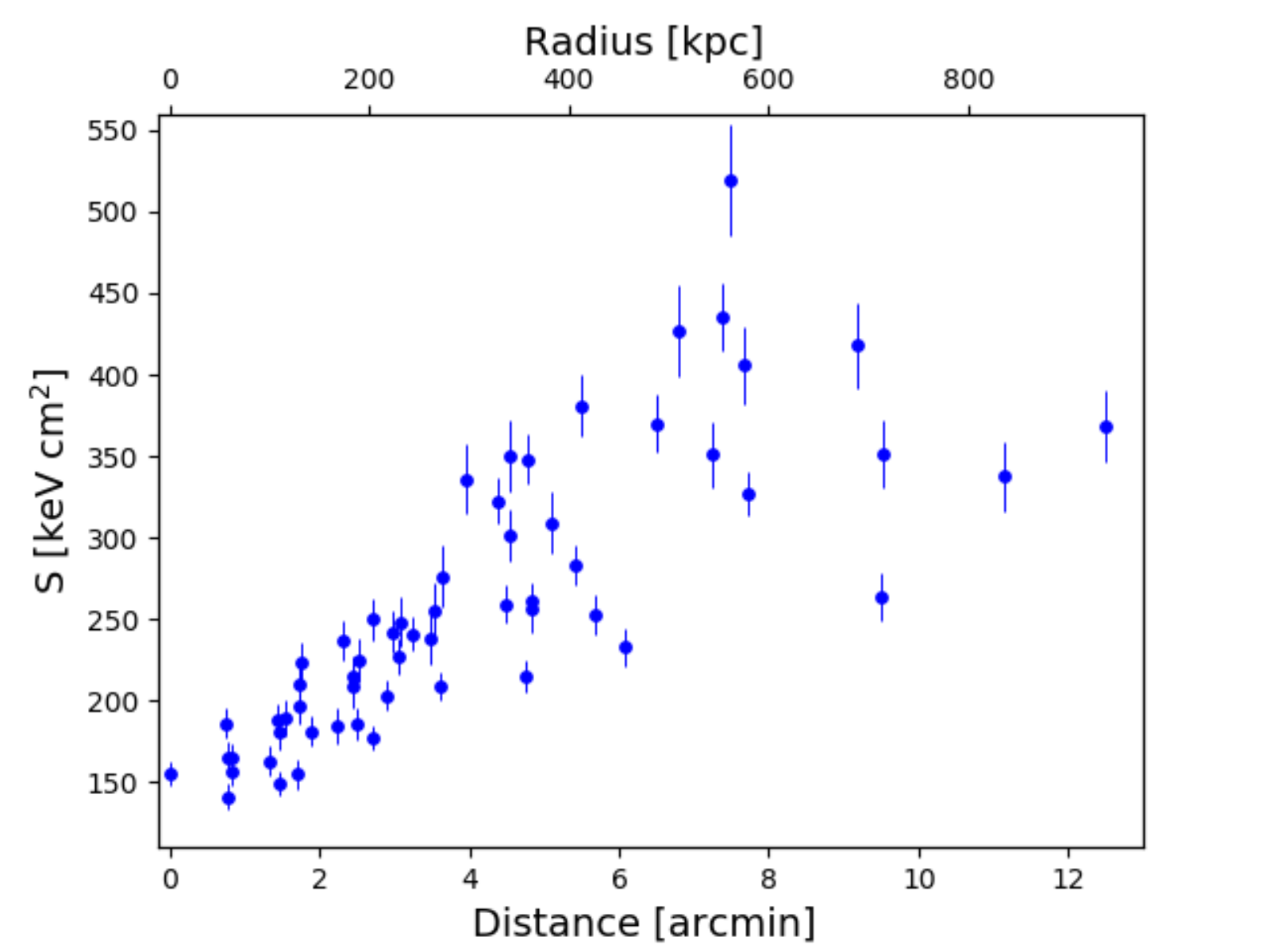}
  \includegraphics[angle=0,width=9.1cm]{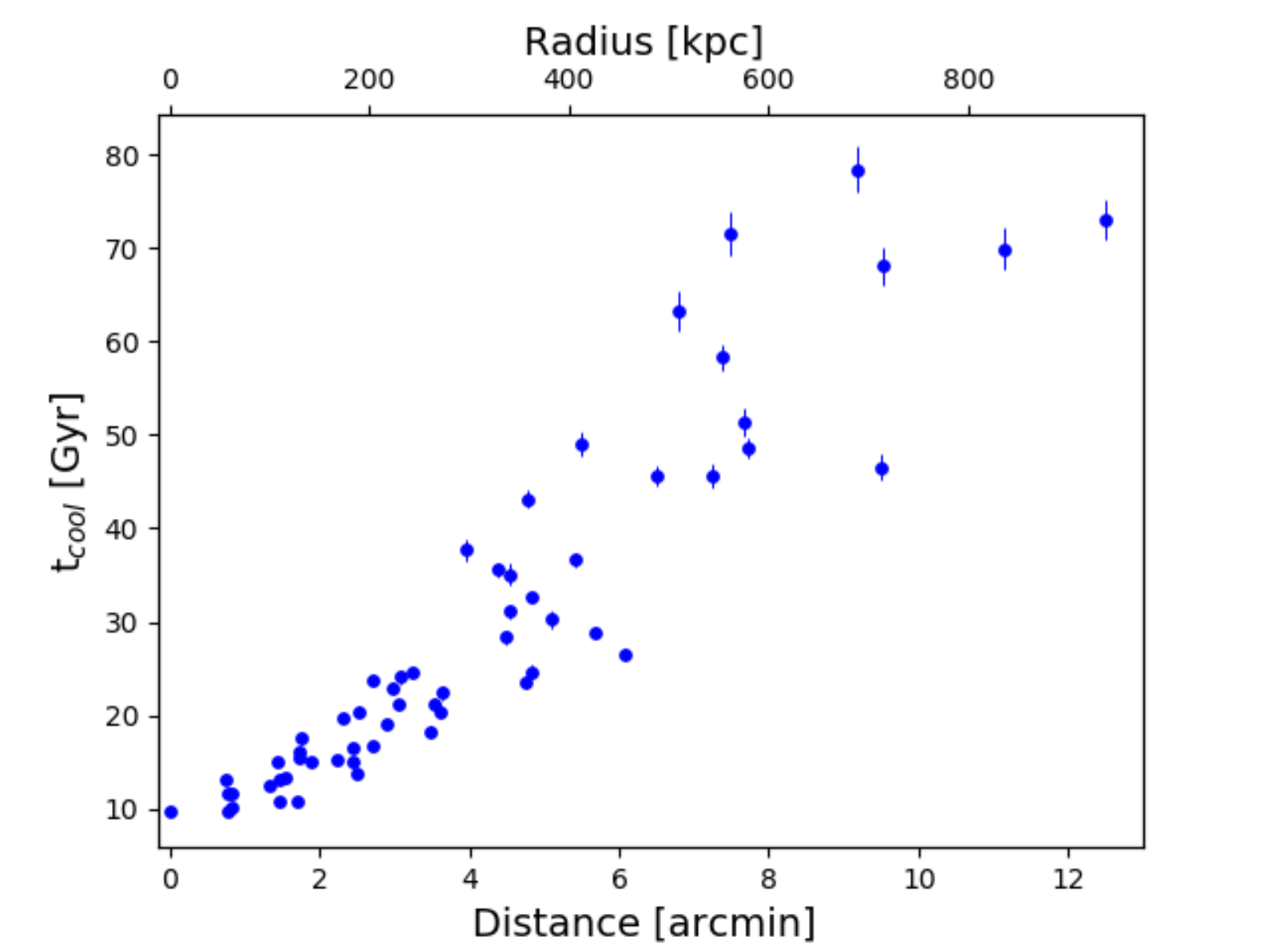}
  }
\caption{Temperature (keV), pressure  (keV cm$^{-3}$), entropy  (keV cm$^2$), and  cooling time (Gyr) shown in Fig.~\ref{fig:map20rxc}, plotted  as a function of the distances from the RXCJ1825 centroid.}
\label{fig:profmaprxc}
\end{figure*}

\subsection{Thermodynamic maps}
\label{sec:mapx}

\subsubsection{The central regions of RXCJ1825}
\label{sec:maprxc}
We selected the regions for the 2D spectral analysis of the central part of RXCJ1825
using the Weighted Voronoi Tessellation (WVT) adaptive binning algorithm by \citet{diehl06}.
We set a minimum value of 20 for the $S/N$ applied to the background subtracted MOS2 image computed in the 0.4-2 keV energy band. 
The extraction regions of the spectra overlapped to the X-ray residual map and the sampled core region of RXCJ1825 in the optical band are shown in the left and right panels of Fig.~\ref{fig:map20}, respectively. As already discussed in Sect. \ref{sec:specx}, we limit this analysis to the central regions of RXCJ1825 where the cluster emission outshines the background.

As is conventional in X-ray astronomy, we quantified the pressure assuming that the ICM is an ideal gas and the entropy using the adiabatic constant $K = k T n^{-2/3}$ ($T$ and $n$ being the gas temperature and density, and $k$ the Boltzmann constant). 
We computed the pressure and the entropy of the ICM in the cluster core in physical units without a proper deprojection of the data. We simply assumed that the volume, $V$, from which the emission is produced is related to the area, $A$, from which we observe it through the equation: $V \propto A^{3/2}$. This is a rather crude way of estimating physical quantities; however in the central regions of RXCJ1825 where the gas distribution is far from being spherically symmetric, it is probably not much less accurate than more traditional methods. This approximation is adequate to estimate the order of magnitude of the physical quantities at the center of the cluster.

The analysis is done with NH fixed at the \citep{willingale13} value and we checked that leaving NH free did not produce appreciable differences in the maps.

Figure~\ref{fig:map20rxc} shows the derived maps for temperature, pressure, entropy, and cooling time in the core of RXCJ1825.
The uncertainties on the thermodynamic parameters shown in the maps can be seen in Fig.~\ref{fig:profmaprxc}, where we plot their profiles as a function of distance from the X-ray centroid of the cluster. 
We also produced a metal abundance map; however the large uncertainties on this quantity did not allow us to draw statistically significant information on the variations of this quantity.

\begin{figure*}
\centering{
\hskip -0.3cm
     \includegraphics[angle=0,width=6.0cm]{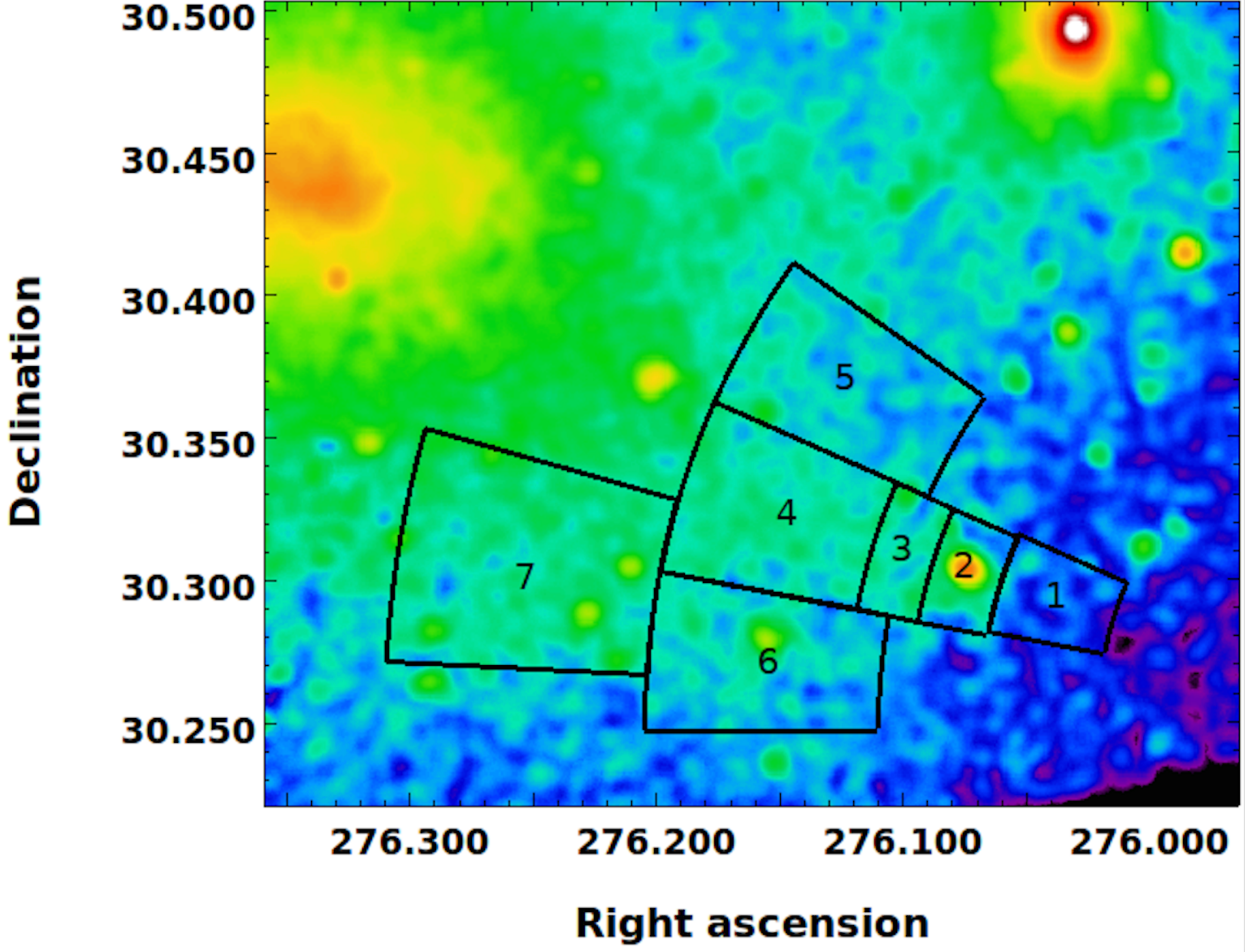}
     \includegraphics[angle=0,width=6.5cm]{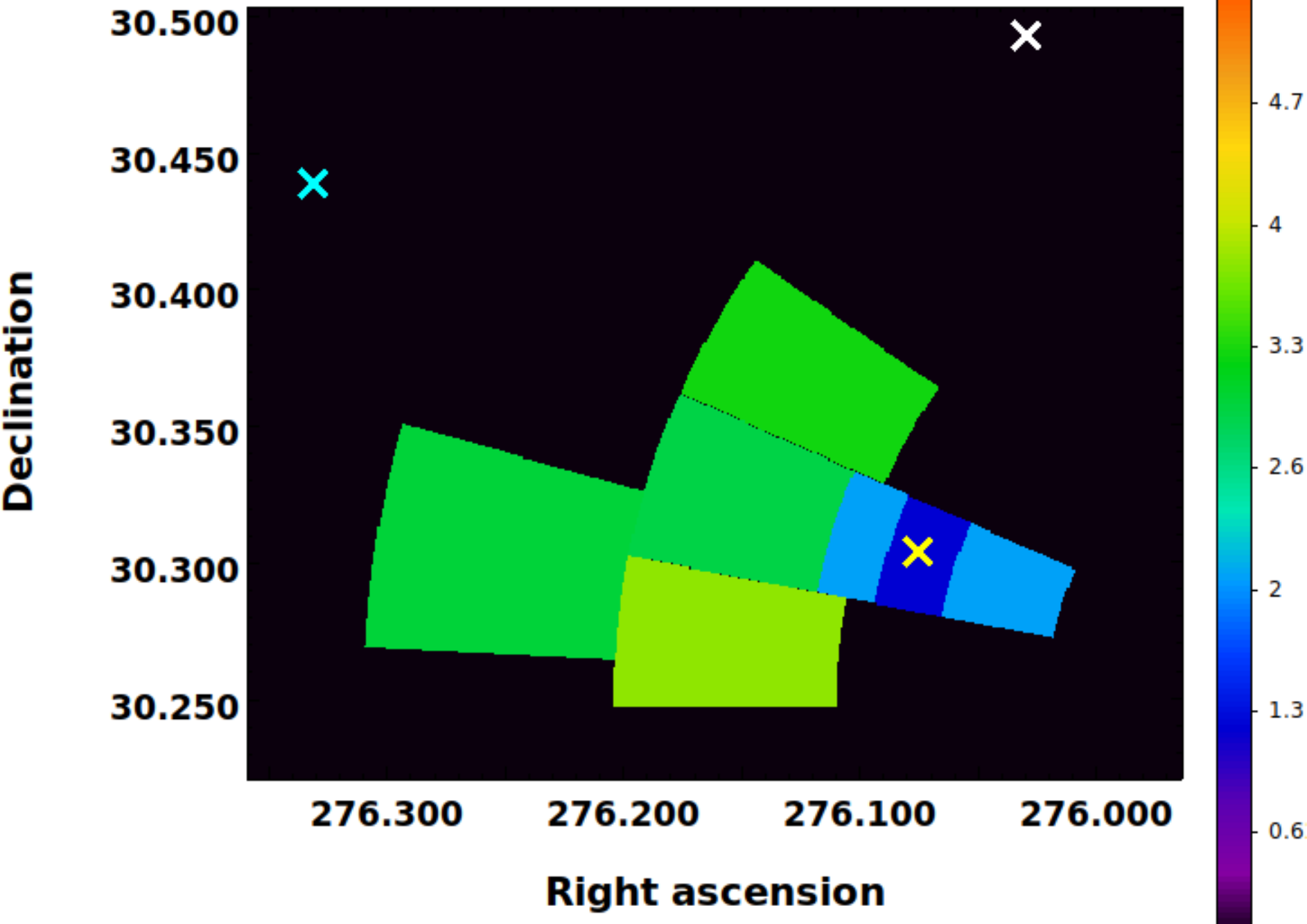}
     \includegraphics[angle=0,width=5.8cm]{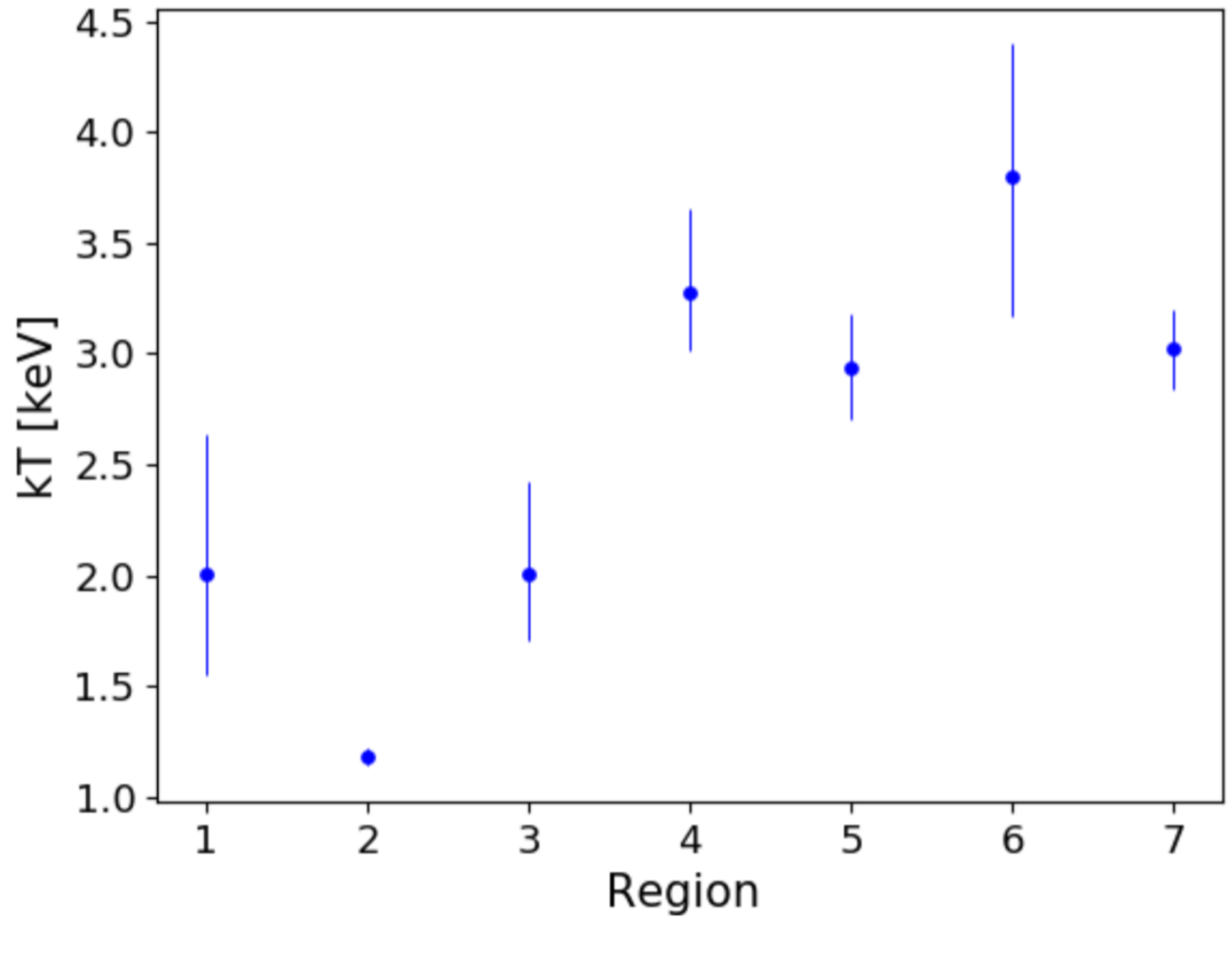}}
\caption{{\it Left panel}: Regions chosen to perform the spectral analysis around the SG and along the emission bridge towards RXCJ1825. {\it Middle panel}: Temperature map in keV. The yellow, cyan, and white crosses are the positions of the SG, the X-ray centroid of RXCJ1825, and the BCG-CC, respectively. {\it Right panel}: Temperature of the regions, with their uncertainties.}
\label{fig:grps}
\end{figure*}

\subsection{The Southern Galaxy}
\label{sec:grps}
We analyzed the spectra extracted from ad-hoc selected regions around the SG, shown in Fig.~\ref{fig:grps}, with the X-COP pipeline described in Sect.\ref{sec:specx}.
Gas temperature, iron abundance, and normalization of the {\it apec} model are free parameters in the spectral fit, whereas the redshift is fixed at the optical value of the SG (z=0.0708, \citealt{girardi19}) and the gas column density is fixed at the value of \cite{willingale13}.
Figure~\ref{fig:grps} shows the temperature distribution in these regions.

\begin{figure}[ht]
\centering{
\hskip 1cm    
\includegraphics[angle=0,width=8.5cm]{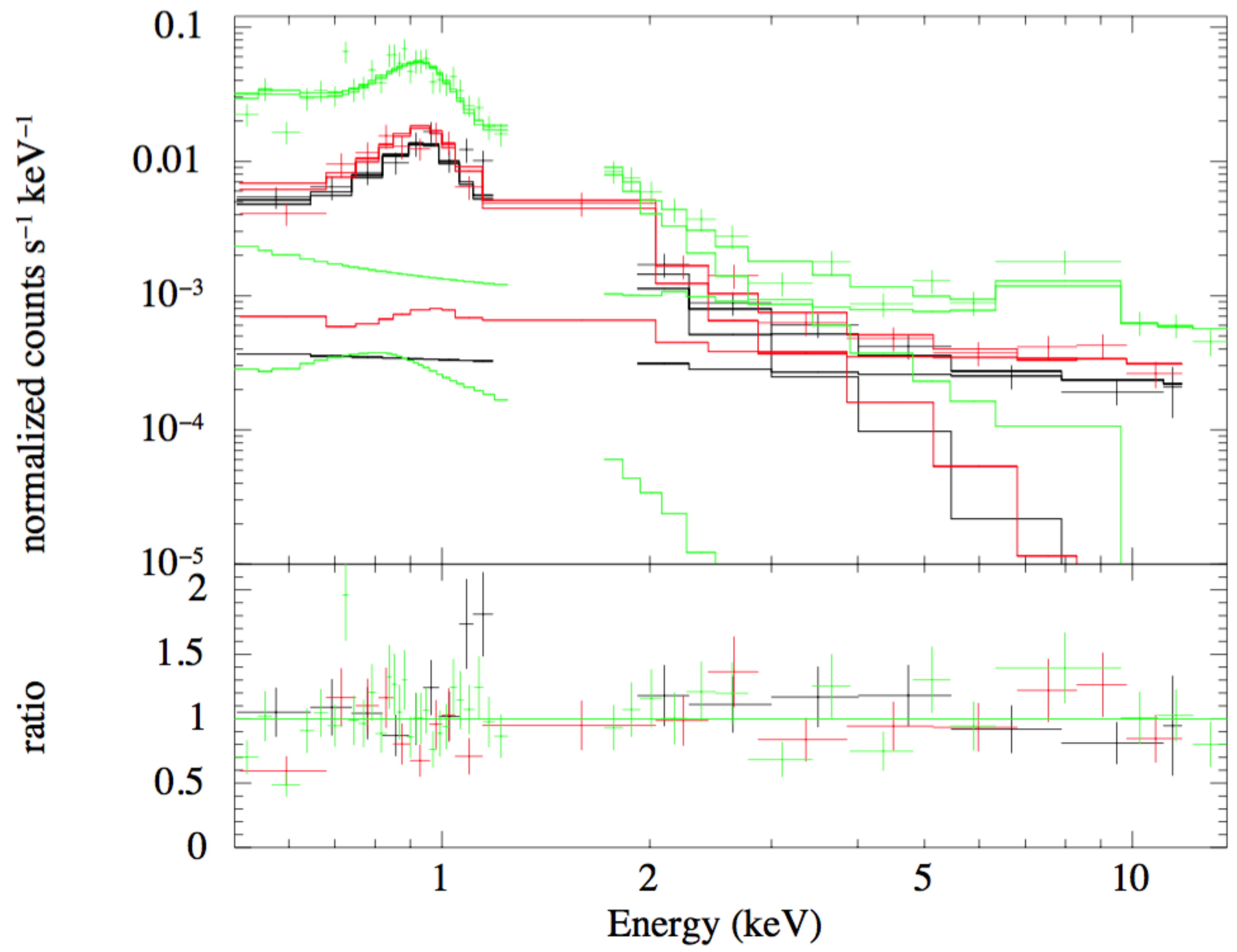}}
\caption{Spectrum of Region 2 in Fig.~\ref{fig:grps} that contains the SG (re-binned for clarity). Crosses are the EPIC data, continuous lines the best fit models (see  \citealt{ghirardini19}). The L-shell blend at $\sim 0.8-1.0$ keV is evident and the shape of the 1-T thermal spectrum is well determined with a reduced $\chi^2$ = 1.1.}
\label{fig:grpsspe}
\end{figure}

Region 2 centered on the bright elliptical galaxy is the coolest one with a temperature of $1.21\pm0.04$ keV measured with very small uncertainties. 
Indeed its spectrum (Fig.~\ref{fig:grpsspe}) shows a noticeable iron L-shell line blend and is undoubtedly thermal. No such feature is found in the spectra of the other regions indicating that in all likelihood this component is associated to the point source coincident with the SG and discussed in Sect.~\ref{sec:sbsg}. 
To better discriminate the emission from the point source from that of the ambient gas we described the source emission with a two-temperature model, where we fixed the temperature, iron abundance, and normalization of the first component to those of the surrounding regions (e.g., Regions 1 and 3 in Fig.~\ref{fig:grps}). The best fit returns a temperature of $1.12\pm 0.05$ keV and an metal abundance of $0.54\pm 0.21$. Verification of these values using the spectral analysis of a spectrum extracted from a 30 arcsec radius circle on the SG position shows that they agree to within $1\sigma$.

In the regions surrounding Region 2 (i.e., Regions 1 and 3 in Fig.~\ref{fig:grps}), the temperature is $\sim 2$ keV, which is consistent with the temperature of gas that once belonged to a group around the elliptical galaxy and is now spread out by the gravitational interaction with RXCJ1825, and also with the value of the temperature of RXCJ1825 at this distance (see Fig.~\ref{fig:profrxc}). We therefore cannot decipher the origin of this gas with our data.

\begin{figure*}
\centering{
\includegraphics[angle=0,width=16.cm]{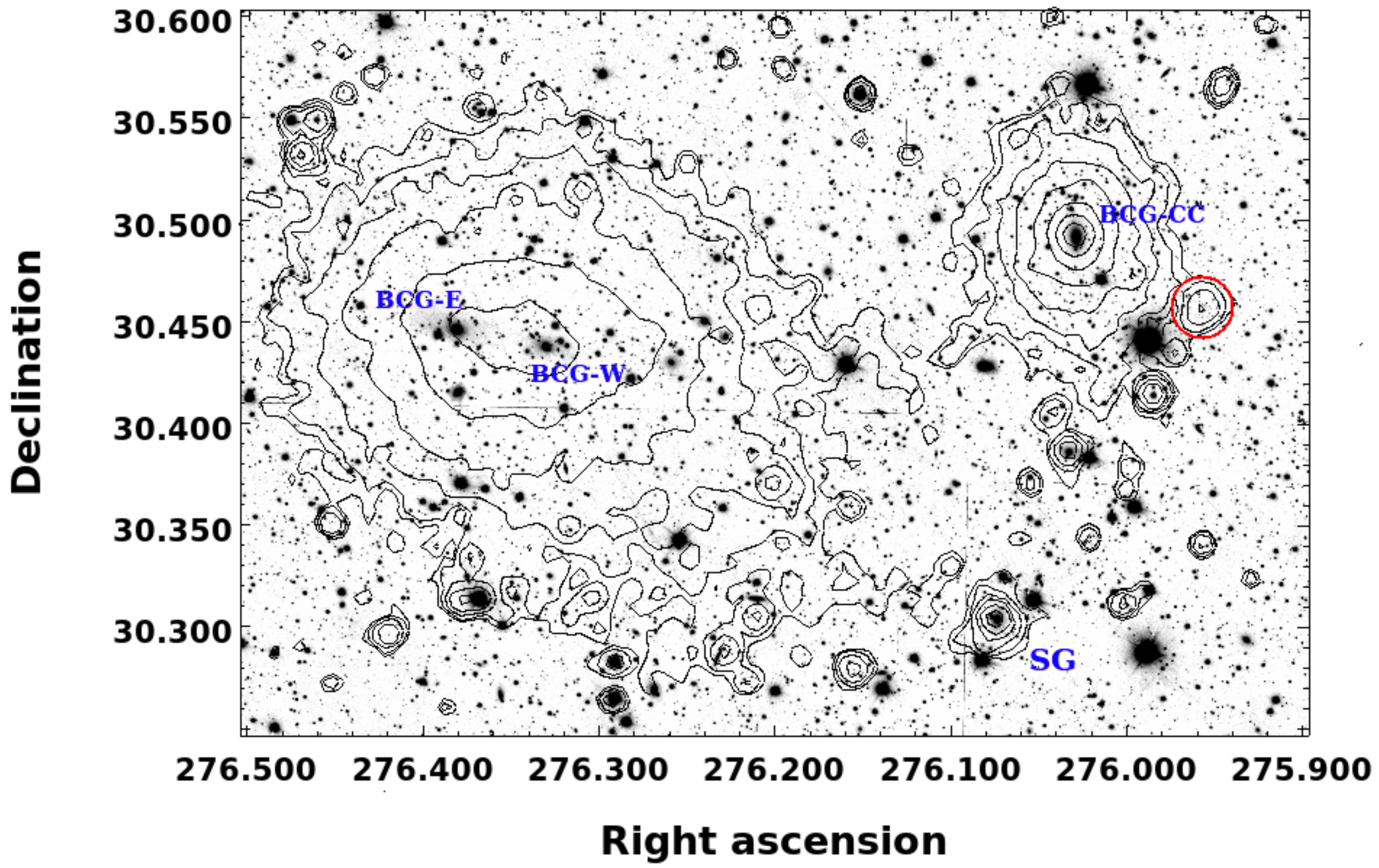}}
\caption{Pan-STARRs r-band image of  RXCJ1825 and CIZAJ1824 complex with the contour levels of the XMM X-ray emission superimposed. The red circle is an unidentified extended source discussed in Sect.\ref{sec:ciza_d}. In the figure, north is at the top and east to the left.}
\label{fig:optx}
\end{figure*}

\section{Discussion}
\label{sec:disc}
In this section we discuss the results of our X-ray analysis and their implications on the dynamical history of the structures in the Lyra complex. 
We complement our results by comparing them with those from recent work in the optical \citep{girardi19} and radio \citep{botteon19} bands on this system.

\subsection{The overall picture of the Lyra complex}
\label{sec:overall_d}
The XMM-Newton image shown in Fig.~\ref{fig:sbx} reveals that the Lyra complex environment is exceptional in many respects. The main structure, namely cluster RXCJ1825, is clearly far from being dynamically relaxed: the distribution of the surface brightness emission is irregular with a flat core,  elongated in the east-west direction, the same direction that connects its two BCGs. This is a characteristic often observed in unrelaxed clusters such as the Coma cluster.
The other prominent structure in the field northwest of RXCJ1825 is CIZAJ1824, which shows all the X-ray characteristics typically observed in a relaxed and unperturbed cool-core cluster.  
In the southwest regions of RXCJ1825, there are patchy, diffuse X-ray, low-surface-brightness structures that indicate the presence of gas which could be connected to the main cluster or to in-falling substructures at different destruction (or post-merger) stages. 

The dynamical analysis of the galaxies in the Lyra complex \citep{girardi19} confirms that all these structures are very close in redshift space, that is, that the whole system discussed here is gravitationally connected and bound. 
Figure~\ref{fig:optx} shows the optical Pan-STARRs r-band image \citep{panstarrs16_database} of the whole Lyra system with the X-ray contours taken from Fig.~\ref{fig:sbx}. In the figure, the prominent and brightest cluster galaxies are labeled.
The dynamical analysis of the member galaxies \citep{girardi19} confirms the unrelaxed nature of RXCJ1825, as well as the membership of the SG ($z=0.0746$, \citealt{girardi19}) to the Lyra system.

Intriguingly, we do not find any evidence of jumps in the surface brightness, which are typically observed in merging systems like the Lyra complex. 
One possibility is that these are transient features which have already disappeared in an advanced merger. This could be supported by the fact that on a large scale RXCJ1825 appears regular (apart from the southwest region). Another possibility is that such features are below the XMM-Newton spatial resolution or that they are at unfavorable projection angles. This last option is discussed further in Sect.~\ref{sec:hunting_d}.
In the following we take into consideration the prominent structures of the Lyra system individually.

\subsection{The central regions of RXCJ1825}
\label{sec:rxc_d}
The center of RXCJ1825 is a particularly complex region, where we clearly find evidence for an ongoing merger. The main features in the core are the east-west elongation of the X-ray morphology on the same line connecting the two BCGs, and the high temperatures ($>5.5$ keV) measured everywhere with no signs of decrements typical of cool-cores (Fig.~\ref{fig:map20rxc} and Fig.~\ref{fig:profmaprxc}). 

The merger scenario is also strongly supported by recent observations of this cluster at optical and radio wavelengths.
The dynamical analysis of the member galaxies in \citep{girardi19} shows that the two BCGs (i.e., BCG-W and BCG-E) are close to being coplanar on the sky plane (their velocities are coincident within the errors, \citealt{girardi19}) and aligned, together with their intra-cluster light (ICL) envelopes, elongated in the same direction as the X-ray emission.  
Moreover, \cite{botteon19}  discovered a 1.0 Mpc $\times$ 0.8 Mpc radio halo in RXCJ1825. Giant radio halos like this are observed only in merging systems \citep[e.g.,][]{vanweeren19}.

The maps of the thermodynamic properties of the ICM shown in Fig.~\ref{fig:map20rxc} can provide some indications to unravel the state of the merger.
Inspecting the pressure map (Fig.~\ref{fig:profmaprxc}), we find that its peak and the centroid of the X-ray emission (determined by the shape of the isophotes on large scales) are coincident. 
They can both be considered indicators of the bottom of the potential well of the system and the fact that they coincide supports this inference. 
The pressure map also shows an area of general maximum in the region that connects the two BCGs and hence indicates that the potential well is elongated in that direction.

If the cluster were dynamically relaxed, we would expect a situation in which the gas would be stratified around the center of the potential well, with the lower entropy coinciding with the pressure peak.
Instead, we observe in Fig.~\ref{fig:map20} that the surface brightness peak and the lowest gas entropy coincide with the BCG-W, $\sim 40$ kpc west of the pressure peak. 
This tells us that the gas is not relaxed and that it is likely still moving, but it also suggests  that the motions involved are slow, otherwise the collisional gas would have decoupled from the BCG-W.

Similar considerations can be applied to the other brightest galaxy, BCG-E, which is $\sim 150$ kpc from the X-ray centroid, albeit with less compelling evidence.
Also in this case BCG-E is close to a small surface brightness peak (see the residual map in Fig. \ref{fig:residual}) in a region where the entropy is still low, indicating denser gas still retained by the galaxy and suggesting  relatively slow motions.

The two BCGs have very similar radial velocities, suggesting that they are almost at rest with respect to each other very close to the plane of the sky (high tangential components of the velocity tend to be excluded by the absence of indications of shocks and other features in the ICM; see Sect.~\ref{sec:search}). This suggests that they are in the final stages of the merger, where the virialization process is already in an advanced state. 

Typically, in the centers of cool-core clusters the entropy of the gas is less than $\sim 20-30$ keV cm$^2$, while in interacting systems it exceeds several 100 keV cm$^2$ \citep[e.g.,][]{cavagnolo09}.
Our entropy map (Fig.~\ref{fig:map20rxc}) shows values of  140-180 keV cm$^2$ in a region of $\sim 100$ kpc radius around the centroid, typical of intermediate systems (see Fig.~1 in \citealt{pratt10}), that is, neither relaxed cool cores nor mergers. This suggests that 
the core of RXCJ1825 could be in a late or post-merger phase during which the entropy stratification found in the relaxed system is in the process of being reconstituted.  

\subsection{CIZAJ1824}
\label{sec:ciza_d}
The picture emerging from our X-ray analysis of the CIZAJ1824 cluster is clearly that of a very relaxed structure consistent with a typical cool core (Fig.~\ref{fig:profciza}). 
Temperature and abundance azimuthally averaged profiles in Fig.~\ref{fig:profciza} show the usual temperature decrements in the core associated to a increasing metal abundance, always observed in cool-core clusters \citep[e.g.,][]{degrandi04}.

The relaxed dynamical state of the ICM is evident from the regular peaked surface brightness profile (Fig.~\ref{fig:allsectors}), and from the residual and unsharp-mask maps (Fig.~\ref{fig:residual} and Fig.~\ref{fig:unsharp}) that show no signs of interactions, such as a sloshing spiral, or distortions, such as strong deviation from the radial symmetry or emission bridges towards RXCJ1825 or the SG (see also boxes C and D in Fig.~\ref{fig:sector-box}). 

The analysis of the galaxy distribution and dynamics in \cite{girardi19} suggests that the  pair RXCJ1825-CIZAJ1824 is bound (probability $\sim 80\%$), with the two clusters well detected and separated. A simple bimodal dynamical model shows that CIZAJ1825 is most likely located in front of RXCJ1825, moving towards it with a projection angle between the plane of the sky and the collision axis of 30-70 deg. 
This geometry implies that the physical distance between the two clusters could be anywhere between 1.5 Mpc ($\alpha=30$ deg) and 3.8 Mpc ($\alpha=70$ deg). In the first case the two systems are within their respective R$_{500}$, while in the latter they lie outside their respective R$_{200}$.
The absence of any evidence of perturbation in CIZAJ1824 favors a large value for the physical distance ($\simg 2$ Mpc) and therefore for the angle ($\simg 50$ deg).

From the gravitational masses measured by our analysis within R$_{200}$ of both clusters (Sect.~\ref{sec:profrxc} and \ref{sec:profciza}), $\sim 7\times 10^{14}$ M$_{\sun}$ for RXCJ1825 and $\sim 4\times 10^{14}$ M$_{\sun}$ for CIZAJ1824, we can infer that a future merger of this cluster pair would have a mass ratio of about 1:2.

\begin{figure}
    \centering{
    \includegraphics[angle=0,width=6.5cm]{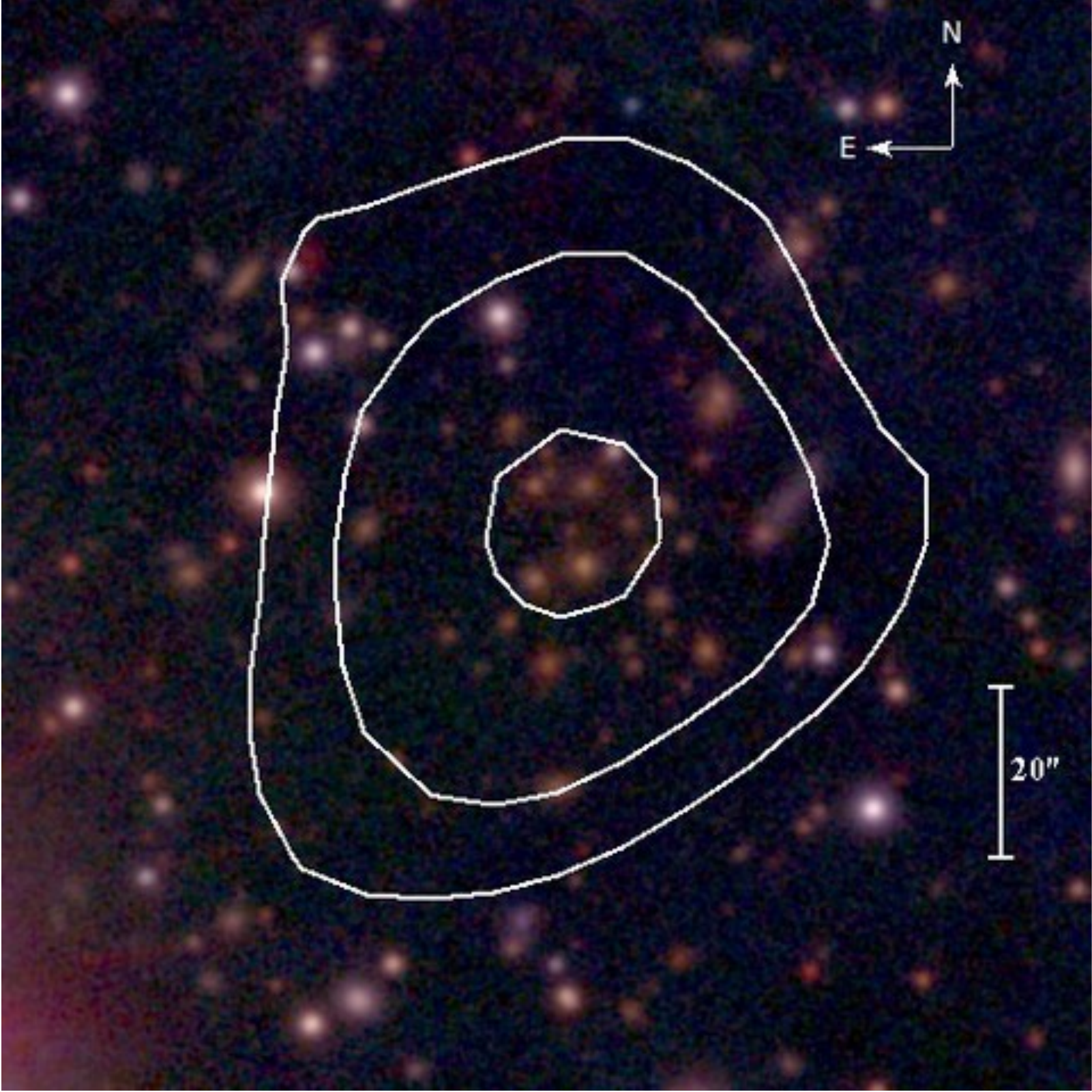}}
 \caption{Pan-STARRs r-band image of the extended X-ray source SW of CIZAJ1824 (red circle in Fig.~\ref{fig:optx}). X-ray contours are overplotted in white.}
        \label{fig:Irs}
\end{figure}

We conclude this section by noting that $\sim 4.5$ arcmin ($\sim$ 350 kpc) southwest of the central galaxy BCG-CC of CIZAJ1824 there is a small extended source ($\sim 1-2$ arcmin radius) at R.A.=18h:23m:49.4s DEC=+30d:27m:23.4s (J2000.0); see red circle in Fig.~\ref{fig:optx}. The Pan-STARRs r-band image at this position (Fig.~\ref{fig:Irs}) suggests the presence of a group of galaxies. We find that the colors of the brightest galaxies ($r_{PS} \lesssim +21$) on the line of sight of this source are consistent with a redshift of $\sim 0.2-0.3$ and therefore we conclude that this system is a probable background galaxy group, unrelated to the Lyra complex.
    
\subsection{The Southern Galaxy}  
\label{sec:sg_d}
Both the surface brightness image in Fig.~\ref{fig:sbx} and the residual map in Fig.~\ref{fig:residual} show that south and southwest of the main RXCJ1825 there is a wide region with weak and diffuse X-ray emission.
The shape of this emission is wider close to RXCJ1825 and then becomes narrower and elongated towards a bright elliptical galaxy that is the brightest object in this region, the SG (see also Fig.~\ref{fig:residual}).

In Sect.~\ref{sec:sbsg} we showed that the emission associated to the SG is point-like (implying that its extension must be smaller than $\sim 12-13$ kpc), and that it is embedded in diffuse emission.
From spectral analysis (Sect.~\ref{sec:grps}) we found that the point-like source is cooler ($\sim 1.$ keV) and metal richer ($Z_{Fe}\sim 0.5$) than the ambient gas (kT $\sim 2$ keV, $Z_{Fe}\sim 0.2$, see also Fig.~\ref{fig:grps}-\ref{fig:grpsspe}). 
The SG is among the five brightest ellipticals in the Lyra complex suggesting that it was once at the center of a galaxy group. 
We interpret the cool emission around this galaxy as a remnant of the gaseous atmosphere once surrounding it. As expected, this gas is highly enriched in metals \citep[e.g.,][and references therein]{sato09_a262}. 
Moreover, since this remnant has not evaporated through contact with the ambient medium, thermal conduction must be inhibited.
Such objects have been reported in the literature by several authors, who refer to them as galactic "coronae" \citep{sun09}. The discovery of a corona associated to the SG supports the idea that it once was the central galaxy of a group.
If the scenario we describe here is accurate, the diffuse surface brightness excess seen southwest of RXCJ1825 could be due to gas stripped from the group and/or to cluster gas dislocated by the passage of the group. 
The same processes are likely responsible for the low-surface-brightness extension of the radio halo towards the SG detected by \cite{botteon19} with LOFAR. 

\subsection{Hunting for infalling substructures}
\label{sec:hunting_d}
\begin{figure}
    \centering{
    \includegraphics[angle=0,width=9.cm]{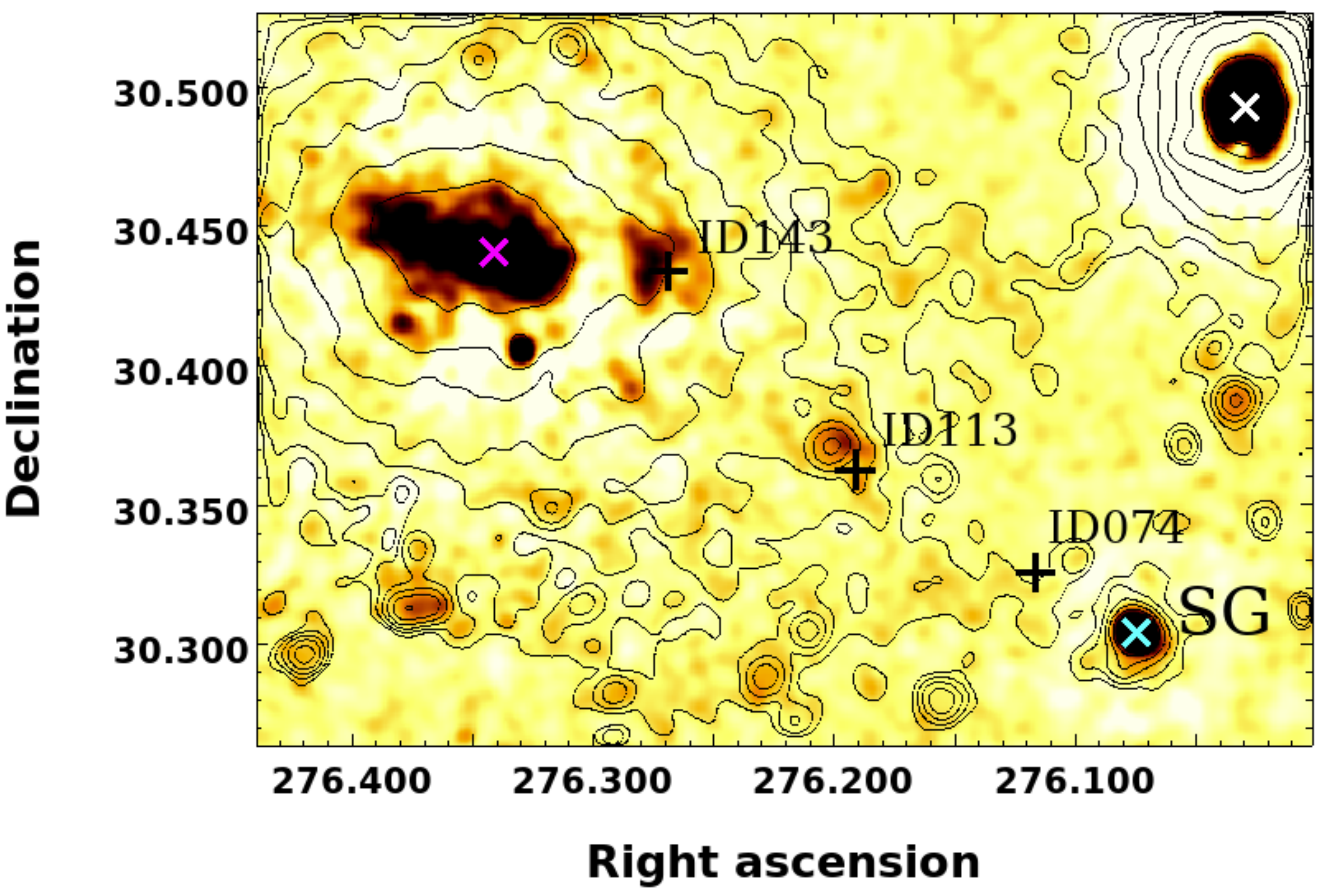}
}
 \caption{Unsharp-mask image of the Lyra complex with the high-velocity galaxies selected by \cite{girardi19} shown as black crosses (ID143, ID113, and ID074 in Table 6 of \citealt{girardi19}).
 The magenta, white, and cyan crosses are the X-ray centroid of RXCJ1825, the BCG-CC, and the SG,  respectively. Black contours are the surface brightness levels.}
        \label{fig:highv}
\end{figure}

\cite{girardi19} identified a high-velocity region south of CIZAJ1824 most likely not associated to either RXCJ1825 or CIZAJ1824 (see their Sect. 4.4 and Fig. 11), whose brightest galaxy is the SG.
In order to detect possible group members, the authors selected galaxies with velocities larger than that of the SG, finding three galaxies aligned between RXCJ1825 and the SG (ID074, ID113, and ID143 in their Table~6).
To search for possible X-ray counterparts of these galaxies we overlayed their positions on the unsharp-mask image which is more sensitive to small-scale structures (Fig.~\ref{fig:highv}, black crosses). 
Remarkably, we find that two of the three high-velocity galaxies have associated excesses (ID143 and ID113). Even more remarkably, the emission is offset from the galaxies in the same direction that connects all these high-velocity galaxies with the SG, suggesting that this alignment is not coincidental. 
Finally, from a spectral analysis of the excess emission associated to ID113, we find a temperature of kT$=2.0^{+0.5}_{-0.3}$ keV, lower at the $\sim 3\sigma$ c.l. than that of the surrounding ICM, kT$=4.0\pm0.2$ keV. 

From the optical analysis we know that these galaxies are moving away from us at a greater velocity than that of RXCJ1825. 
We interpret the X-ray excesses we observe as stripped gas from the high-velocity galaxies. This is supported by  our finding that one of the two excesses features a substantially lower temperature than the surrounding ICM. 
Ram pressure stripping typically occurs during infall, as it is proportional to $\rho v^2$ and both ICM density and velocity increase towards the cluster center, suggesting that the  galaxies are currently falling onto RXCJ1825.
While the displacement of the excesses with respect to two of these high-velocity systems in a specific direction strongly argues for a velocity component on the plane of the sky, the lack of elongation of the excesses along the presumed direction of motion, as typically seen in ram-pressure-stripped galaxies and groups \citep[e.g.,][and references therein]{jachym19,eckert14_a2142}, suggests that the velocity component on the plane of the sky is much smaller that the one on the line of sight.
As part of this hypothesis, the difference in measured redshift between the galaxies and RXCJ can be taken as a reasonable estimate of the infall velocities. By performing the computation we derive infall velocities, $v_{i}=\Delta v/(1+z)$, of $2942\pm76$, $3196\pm125$, $2891\pm48,$ and $2830\pm73$ km/s respectively for ID143, ID113, ID074 and SG. 
We note in passing that a high line-of-sight to plane-of-the-sky velocity ratio may also explain the remarkable lack of jumps in the surface brightness (see Sect.~\ref{sec:search}).
The large projected distance between ID143 and SG together with the small velocity on the plane of the sky suggests that these galaxies were not originally part of the same group; however the alignment  amongst themselves and the X-ray excesses suggests at least a common origin, perhaps a filament or, more likely, a plane roughly orthogonal to that of the sky.

\section{Summary}
\label{sec:concl}
We have presented results from the X-COP mosaic observation of the Lyra complex whose main structure is the galaxy clusters pair RXCJ1825-CIZAJ1824. While RXCJ1825 is dynamically disturbed, with evidence of an undergoing main merger in the east-west direction, CIZAJ1824 appears as a very regular undisturbed cluster.
The recent discovery of a giant radio halo, filling the same volume of the thermal plasma in the center of RXCJ1825, supports the merger hypothesis in this cluster.
Southwest of RXCJ1825 there is a bright elliptical galaxy, the SG, located at the very end of an elongated diffuse X-ray emission joining the main cluster and this galaxy, suggesting a physical connection between the two systems. The optical study of the velocity field of the Lyra complex shows that the clusters pair and the SG are gravitationally bound at a mean redshift of $z=0.067$. 

In the following we provide a summary of our main findings.

\begin{itemize}
    \item[$\bullet$] RXCJ1825 is the largest structure in the Lyra complex with a total mass within R$_{200}$ ($1.82\pm0.17$ Mpc) of M$_{200}= 7.3\pm1.9$ M$_{\sun}$ and an eastwest elongated emission in the same line connecting its two BCGs. From the analysis of the ICM entropy and pressure distributions we infer that the gas is not at rest at the bottom of the cluster potential well but is still moving, although with relatively slow motions since no shocks or density jumps have been detected anywhere, and it is tending towards complete virialization. This late stage or post-merger scenario in RXCJ1825 is also strongly supported by recent optical and radio observations.  
    \item[$\bullet$] CIZAJ1824, the companion cluster west of RXCJ1825, is a cool-core cluster with a relaxed X-ray morphology which suggests an equally relaxed dynamic state. The estimated gravitational mass within R$_{200}=1.51\pm0.18$ Mpc is M$_{200}=4.2\pm1.5$ M$_{\sun}$. 
    \item[$\bullet$] No surface-brightness bridges are found between the pair RXCJ1825-CIZAJ1824, suggesting that CIZAJ1824 is in a pre-merger state with RXCJ1825. A future merger of this cluster pair would have a mass ratio of about 1:2. 
    \item[$\bullet$] The SG hosts an X-ray-emitting gaseous corona with extension smaller than $\sim 13 $ kpc, unresolved by our XMM-Newton observation, with a temperature of $\sim 1$ keV and a metal abundance of $\sim 0.5$ $Z_\sun$. The discovery of the corona associated to the SG supports the idea that it once was the central galaxy of a group now almost completely destroyed by interaction with RXCJ1825. 
    \item[$\bullet$] In this scenario the diffuse surface-brightness excess seen southwest of RXCJ1825 could be due to gas stripped from the group and/or to cluster gas dislocated by the passage of the group. 
    \item[$\bullet$] We identified three high-velocity galaxies that are aligned between RXCJ1825 and the SG and that show (2 out of 3) associated X-ray excesses. The spectral analysis of one of these emission excesses shows the presence (at $\sim 3\sigma$ c.l.) of gas at a lower temperature than that of the surrounding ambient ICM. 
    We interpret these X-ray excesses as stripped gas from the high-velocity galaxies and, since stripping typically occurs during infall, we infer that the galaxies are currently falling onto RXCJ1825 with infall velocities of $\sim 3000$ km/s.
\end{itemize}

Finally, we would like to highlight the successful observational strategy of the X-COP project for the detection and characterization of infalling galaxy groups on clusters.
Thanks to the off-axis mosaicing, the careful technical analysis of cluster outskirts, and last but not least the optical follow up work, we have been able to discover and study various stages of the virialization process of groups during their infall onto the main cluster. 
In the case of the Hydra-A/A780 cluster \citep{degrandi16_hydraa} we detected a long gas tail, half of which has been ram pressure stripped while the other half is still gravitationally bound to the group.
In A2142  \citep{eckert14_a2142,eckert17_a2142}, we found a  group almost entirely stripped of its gas (only $\sim 10\%$ of it still retained).
In the last case, studied here, we find evidence of a BCG with a corona but no group surrounding it, as well as two galaxies being stripped of their gas as they fall at high velocity onto RXCJ1825.

\begin{acknowledgements}
We acknowledge financial contribution from the contracts ASI 2015-046-R.0 and ASI-INAF n.2017-14-H.0.
M.G. is supported by the Lyman Spitzer Jr. Fellowship (Princeton University) and by NASA Chandra grants GO7-18121X and GO8-19104X.
This research is based on observations obtained with {\it XMM-Newton}, an ESA science mission with instruments and contributions directly funded by ESA Member States and the USA (NASA).
\end{acknowledgements}

\bibliography{biblio_Feb2019}

\begin{thebibliography}{50}
\expandafter\ifx\csname natexlab\endcsname\relax\def\natexlab#1{#1}\fi

\bibitem[{{Asplund} {et~al.}(2009){Asplund}, {Grevesse}, {Sauval}, \&
  {Scott}}]{asplund09}
{Asplund}, M., {Grevesse}, N., {Sauval}, A.~J., \& {Scott}, P. 2009, \araa, 47,
  481

\bibitem[{{Berrier} {et~al.}(2009){Berrier}, {Stewart}, {Bullock}, {Purcell},
  {Barton}, \& {Wechsler}}]{berrier09}
{Berrier}, J.~C., {Stewart}, K.~R., {Bullock}, J.~S., {et~al.} 2009, \apj, 690,
  1292

\bibitem[{{Botteon et al.}(2019)}]{botteon19}
{Botteon et al.} 2019, \aap submitted

\bibitem[{{Boulanger} {et~al.}(1996){Boulanger}, {Abergel}, {Bernard},
  {Burton}, {Desert}, {Hartmann}, {Lagache}, \& {Puget}}]{boulanger96}
{Boulanger}, F., {Abergel}, A., {Bernard}, J.-P., {et~al.} 1996, \aap, 312, 256

\bibitem[{{Boylan-Kolchin} {et~al.}(2009){Boylan-Kolchin}, {Springel}, {White},
  {Jenkins}, \& {Lemson}}]{boylan09}
{Boylan-Kolchin}, M., {Springel}, V., {White}, S. D.~M., {Jenkins}, A., \&
  {Lemson}, G. 2009, \mnras, 398, 1150

\bibitem[{{Cavagnolo} {et~al.}(2009){Cavagnolo}, {Donahue}, {Voit}, \&
  {Sun}}]{cavagnolo09}
{Cavagnolo}, K.~W., {Donahue}, M., {Voit}, G.~M., \& {Sun}, M. 2009, \apjs,
  182, 12

\bibitem[{{Cavaliere} \& {Fusco-Femiano}(1976)}]{cavaliere76}
{Cavaliere}, A. \& {Fusco-Femiano}, R. 1976, \aap, 49, 137

\bibitem[{{De Grandi} {et~al.}(2016){De Grandi}, {Eckert}, {Molendi},
  {Girardi}, {Roediger}, {Gaspari}, {Gastaldello}, {Ghizzardi}, {Nonino}, \&
  {Rossetti}}]{degrandi16_hydraa}
{De Grandi}, S., {Eckert}, D., {Molendi}, S., {et~al.} 2016, \aap, 592, A154

\bibitem[{{De Grandi} {et~al.}(2004){De Grandi}, {Ettori}, {Longhetti}, \&
  {Molendi}}]{degrandi04}
{De Grandi}, S., {Ettori}, S., {Longhetti}, M., \& {Molendi}, S. 2004, \aap,
  419, 7

\bibitem[{{De Luca} \& {Molendi}(2004)}]{deluca04}
{De Luca}, A. \& {Molendi}, S. 2004, \aap, 419, 837

\bibitem[{{Diehl} \& {Statler}(2006)}]{diehl06}
{Diehl}, S. \& {Statler}, T.~S. 2006, \mnras, 368, 497

\bibitem[{{Ebeling} {et~al.}(2002){Ebeling}, {Mullis}, \& {Tully}}]{ebeling02}
{Ebeling}, H., {Mullis}, C.~R., \& {Tully}, R.~B. 2002, \apj, 580, 774

\bibitem[{{Eckert}(2016)}]{eckert16}
{Eckert}, D. 2016, {PROFFIT: Analysis of X-ray surface-brightness profiles},
  Astrophysics Source Code Library

\bibitem[{{Eckert} {et~al.}(2017{\natexlab{a}}){Eckert}, {Ettori},
  {Pointecouteau}, {Molendi}, {Paltani}, \& {Tchernin}}]{eckert17_xcop}
{Eckert}, D., {Ettori}, S., {Pointecouteau}, E., {et~al.} 2017{\natexlab{a}},
  Astronomische Nachrichten, 338, 293

\bibitem[{{Eckert} {et~al.}(2017{\natexlab{b}}){Eckert}, {Gaspari}, {Owers},
  {Roediger}, {Molendi}, {Gastaldello}, {Paltani}, {Ettori}, {Venturi},
  {Rossetti}, \& {Rudnick}}]{eckert17_a2142}
{Eckert}, D., {Gaspari}, M., {Owers}, M.~S., {et~al.} 2017{\natexlab{b}}, \aap,
  605, A25

\bibitem[{{Eckert} {et~al.}(2015){Eckert}, {Jauzac}, {Shan}, {Kneib}, {Erben},
  {Israel}, {Jullo}, {Klein}, {Massey}, {Richard}, \&
  {Tchernin}}]{eckert15_nature}
{Eckert}, D., {Jauzac}, M., {Shan}, H., {et~al.} 2015, \nat, 528, 105

\bibitem[{{Eckert} {et~al.}(2014){Eckert}, {Molendi}, {Owers}, {Gaspari},
  {Venturi}, {Rudnick}, {Ettori}, {Paltani}, {Gastaldello}, \&
  {Rossetti}}]{eckert14_a2142}
{Eckert}, D., {Molendi}, S., {Owers}, M., {et~al.} 2014, \aap, 570, A119

\bibitem[{{Ettori} {et~al.}(2002){Ettori}, {Fabian}, {Allen}, \&
  {Johnstone}}]{ettori02_a1795}
{Ettori}, S., {Fabian}, A.~C., {Allen}, S.~W., \& {Johnstone}, R.~M. 2002,
  \mnras, 331, 635

\bibitem[{{Ettori} {et~al.}(2010){Ettori}, {Gastaldello}, {Leccardi},
  {Molendi}, {Rossetti}, {Buote}, \& {Meneghetti}}]{ettori10}
{Ettori}, S., {Gastaldello}, F., {Leccardi}, A., {et~al.} 2010, \aap, 524, A68

\bibitem[{{Ettori} {et~al.}(2019){Ettori}, {Ghirardini}, {Eckert},
  {Pointecouteau}, {Gastaldello}, {Sereno}, {Gaspari}, {Ghizzardi},
  {Roncarelli}, \& {Rossetti}}]{ettori19_xcop_mass}
{Ettori}, S., {Ghirardini}, V., {Eckert}, D., {et~al.} 2019, \aap, 621, A39

\bibitem[{{Ettori} \& {Molendi}(2011)}]{ettori11}
{Ettori}, S. \& {Molendi}, S. 2011, Memorie della Societa Astronomica Italiana
  Supplementi, 17, 47

\bibitem[{{Fakhouri} {et~al.}(2010){Fakhouri}, {Ma}, \&
  {Boylan-Kolchin}}]{fakhouri10}
{Fakhouri}, O., {Ma}, C.-P., \& {Boylan-Kolchin}, M. 2010, \mnras, 406, 2267

\bibitem[{{Flewelling} {et~al.}(2016){Flewelling}, {Magnier}, {Chambers},
  {Heasley}, {Holmberg}, {Huber}, {Sweeney}, {Waters}, {Calamida}, {Casertano},
  {Chen}, {Farrow}, {Hasinger}, {Henderson}, {Long}, {Metcalfe}, {Narayan},
  {Nieto-Santisteban}, {Norberg}, {Rest}, {Saglia}, {Szalay}, {Thakar},
  {Tonry}, {Valenti}, {Werner}, {White}, {Denneau}, {Draper}, {Hodapp},
  {Jedicke}, {Kaiser}, {Kudritzki}, {Price}, {Wainscoat}, {Builders},
  {Chastel}, {McLean}, {Postman}, \& {Shiao}}]{panstarrs16_database}
{Flewelling}, H.~A., {Magnier}, E.~A., {Chambers}, K.~C., {et~al.} 2016, arXiv
  e-prints, arXiv:1612.05243

\bibitem[{{Gao} {et~al.}(2012){Gao}, {Navarro}, {Frenk}, {Jenkins}, {Springel},
  \& {White}}]{gao12}
{Gao}, L., {Navarro}, J.~F., {Frenk}, C.~S., {et~al.} 2012, \mnras, 425, 2169

\bibitem[{{Ghirardini} {et~al.}(2019){Ghirardini}, {Eckert}, {Ettori},
  {Pointecouteau}, {Molendi}, {Gaspari}, {Rossetti}, {De Grandi}, {Roncarelli},
  {Bourdin}, {Mazzotta}, {Rasia}, \& {Vazza}}]{ghirardini19}
{Ghirardini}, V., {Eckert}, D., {Ettori}, S., {et~al.} 2019, \aap, 621, A41

\bibitem[{{Ghirardini} {et~al.}(2018){Ghirardini}, {Ettori}, {Eckert},
  {Molendi}, {Gastaldello}, {Pointecouteau}, {Hurier}, \&
  {Bourdin}}]{ghirardini18}
{Ghirardini}, V., {Ettori}, S., {Eckert}, D., {et~al.} 2018, \aap, 614, A7

\bibitem[{{Ghizzardi} {et~al.}(2014){Ghizzardi}, {De Grandi}, \&
  {Molendi}}]{ghizzardi14}
{Ghizzardi}, S., {De Grandi}, S., \& {Molendi}, S. 2014, \aap, 570, A117

\bibitem[{{Ghizzardi} {et~al.}(2004){Ghizzardi}, {Molendi}, {Pizzolato}, \& {De
  Grandi}}]{ghizzardi04}
{Ghizzardi}, S., {Molendi}, S., {Pizzolato}, F., \& {De Grandi}, S. 2004, \apj,
  609, 638

\bibitem[{{Girardi et al.}(2019)}]{girardi19}
{Girardi et al.} 2019, \aap submitted

\bibitem[{{Haines} {et~al.}(2018){Haines}, {Finoguenov}, {Smith}, {Babul},
  {Egami}, {Mazzotta}, {Okabe}, {Pereira}, {Bianconi}, {McGee}, {Ziparo},
  {Campusano}, \& {Loyola}}]{haines18_locuss}
{Haines}, C.~P., {Finoguenov}, A., {Smith}, G.~P., {et~al.} 2018, \mnras, 477,
  4931

\bibitem[{{Jachym} {et~al.}(2019){Jachym}, {Kenney}, {Sun}, {Combes},
  {Cortese}, {Scott}, {Sivanand am}, {Brinks}, {Roediger}, {Palous}, \&
  {Fumagalli}}]{jachym19}
{Jachym}, P., {Kenney}, J. D.~P., {Sun}, M., {et~al.} 2019, arXiv e-prints,
  arXiv:1905.13249

\bibitem[{{Kalberla} {et~al.}(2005){Kalberla}, {Burton}, {Hartmann}, {Arnal},
  {Bajaja}, {Morras}, \& {P{\"o}ppel}}]{kaberla05}
{Kalberla}, P.~M.~W., {Burton}, W.~B., {Hartmann}, D., {et~al.} 2005, \aap,
  440, 775

\bibitem[{{Kocevski} {et~al.}(2007){Kocevski}, {Ebeling}, {Mullis}, \&
  {Tully}}]{kocevski07_ciza}
{Kocevski}, D.~D., {Ebeling}, H., {Mullis}, C.~R., \& {Tully}, R.~B. 2007,
  \apj, 662, 224

\bibitem[{{Kravtsov} \& {Borgani}(2012)}]{kravtsov12_rev}
{Kravtsov}, A.~V. \& {Borgani}, S. 2012, \araa, 50, 353

\bibitem[{{Kriss} {et~al.}(1983){Kriss}, {Cioffi}, \& {Canizares}}]{kriss83}
{Kriss}, G.~A., {Cioffi}, D.~F., \& {Canizares}, C.~R. 1983, \apj, 272, 439

\bibitem[{{Leccardi} \& {Molendi}(2008)}]{leccardi08_t}
{Leccardi}, A. \& {Molendi}, S. 2008, \aap, 486, 359

\bibitem[{{Lovisari} {et~al.}(2017){Lovisari}, {Forman}, {Jones}, {Ettori},
  {Andrade-Santos}, {Arnaud}, {D{\'e}mocl{\`e}s}, {Pratt}, {Rand all}, \&
  {Kraft}}]{lovisari17}
{Lovisari}, L., {Forman}, W.~R., {Jones}, C., {et~al.} 2017, \apj, 846, 51

\bibitem[{{McCammon} {et~al.}(2002){McCammon}, {Almy}, {Apodaca}, {Bergmann
  Tiest}, {Cui}, {Deiker}, {Galeazzi}, {Juda}, {Lesser}, {Mihara},
  {Morgenthaler}, {Sanders}, {Zhang}, {Figueroa-Feliciano}, {Kelley},
  {Moseley}, {Mushotzky}, {Porter}, {Stahle}, \& {Szymkowiak}}]{mccammon02}
{McCammon}, D., {Almy}, R., {Apodaca}, E., {et~al.} 2002, \apj, 576, 188

\bibitem[{{McGee} {et~al.}(2009){McGee}, {Balogh}, {Bower}, {Font}, \&
  {McCarthy}}]{mcgee09}
{McGee}, S.~L., {Balogh}, M.~L., {Bower}, R.~G., {Font}, A.~S., \& {McCarthy},
  I.~G. 2009, \mnras, 400, 937

\bibitem[{{Planck Collaboration} {et~al.}(2014){Planck Collaboration}, {Ade},
  {Aghanim}, {Armitage-Caplan}, {Arnaud}, {Ashdown}, {Atrio-Barand ela},
  {Aumont}, {Aussel}, {Baccigalupi}, \& et~al.}]{planck14}
{Planck Collaboration}, {Ade}, P.~A.~R., {Aghanim}, N., {et~al.} 2014, \aap,
  571, A29

\bibitem[{{Pratt} {et~al.}(2010){Pratt}, {Arnaud}, {Piffaretti},
  {B{\"o}hringer}, {Ponman}, {Croston}, {Voit}, {Borgani}, \&
  {Bower}}]{pratt10}
{Pratt}, G.~W., {Arnaud}, M., {Piffaretti}, R., {et~al.} 2010, \aap, 511, A85

\bibitem[{{Rossetti} {et~al.}(2016){Rossetti}, {Gastaldello}, {Ferioli},
  {Bersanelli}, {De Grandi}, {Eckert}, {Ghizzardi}, {Maino}, \&
  {Molendi}}]{rossetti16}
{Rossetti}, M., {Gastaldello}, F., {Ferioli}, G., {et~al.} 2016, \mnras, 457,
  4515

\bibitem[{{Sato} {et~al.}(2009){Sato}, {Matsushita}, \&
  {Gastaldello}}]{sato09_a262}
{Sato}, K., {Matsushita}, K., \& {Gastaldello}, F. 2009, \pasj, 61, 365

\bibitem[{{Schlegel} {et~al.}(1998){Schlegel}, {Finkbeiner}, \&
  {Davis}}]{schlegel98}
{Schlegel}, D.~J., {Finkbeiner}, D.~P., \& {Davis}, M. 1998, \apj, 500, 525

\bibitem[{{Smith} {et~al.}(2001){Smith}, {Brickhouse}, {Liedahl}, \&
  {Raymond}}]{smith01_apec}
{Smith}, R.~K., {Brickhouse}, N.~S., {Liedahl}, D.~A., \& {Raymond}, J.~C.
  2001, \apjl, 556, L91

\bibitem[{{Snowden} {et~al.}(2008){Snowden}, {Mushotzky}, {Kuntz}, \&
  {Davis}}]{snowden08_esas}
{Snowden}, S.~L., {Mushotzky}, R.~F., {Kuntz}, K.~D., \& {Davis}, D.~S. 2008,
  \aap, 478, 615

\bibitem[{{Sun} {et~al.}(2009){Sun}, {Voit}, {Donahue}, {Jones}, {Forman}, \&
  {Vikhlinin}}]{sun09}
{Sun}, M., {Voit}, G.~M., {Donahue}, M., {et~al.} 2009, \apj, 693, 1142

\bibitem[{{van Weeren} {et~al.}(2019){van Weeren}, {de Gasperin}, {Akamatsu},
  {Br{\"u}ggen}, {Feretti}, {Kang}, {Stroe}, \& {Zandanel}}]{vanweeren19}
{van Weeren}, R.~J., {de Gasperin}, F., {Akamatsu}, H., {et~al.} 2019, \ssr,
  215, 16

\bibitem[{{Walker} {et~al.}(2019){Walker}, {Simionescu}, {Nagai}, {Okabe},
  {Eckert}, {Mroczkowski}, {Akamatsu}, {Ettori}, \&
  {Ghirardini}}]{walker19_rev}
{Walker}, S., {Simionescu}, A., {Nagai}, D., {et~al.} 2019, \ssr, 215, 7

\bibitem[{{Willingale} {et~al.}(2013){Willingale}, {Starling}, {Beardmore},
  {Tanvir}, \& {O'Brien}}]{willingale13}
{Willingale}, R., {Starling}, R.~L.~C., {Beardmore}, A.~P., {Tanvir}, N.~R., \&
  {O'Brien}, P.~T. 2013, \mnras, 431, 394

\end{thebibliography}
%\newpage
\section*{Appendix A}
\begin{table*}
        \caption[]{Best-fit $\beta$-model parameters for RXCJ1825 and CIZAJ1824 clusters. Columns are: cluster name, X-ray centroid coordinates in degrees [J2000], beta parameter, core radius in arcmin, central surface brightness in cts s$^{-1}$ arcmin$^{-2}$, constant sky background in units of $10^{-4}$ cts s$^{-1}$ arcmin$^{-2}$, chi-square and degrees of freedom. Neither RXCJ1825 nor CIZAJ1824 show a significant best-fit improvement with a double-$\beta$ model, as evident from the F-test.}
         \label{tab:bestfit}
              $$ 
           \begin{array}{l c c c c c c c}
            \hline
            \noalign{\smallskip}
            \hline
            \noalign{\smallskip}
\mathrm{Cluster} & \mathrm{RA}~\mathrm{DEC}~(\mathrm{J}2000) & \beta & r_c & S_0 & const & \chi^2 & dof  \\ 
            \hline
            \noalign {\smallskip}  
\text{RXCJ1825}  & 18:25:21.77, +30:26:25.3 & 0.562\pm0.008 & 3.08\pm0.06 &  0.0146\pm0.0002 & 1.9\pm0.1 & 318.4 & 140 \\      
\text{CIZAJ1824} & 18:24:07.11, +30:29:34.7  & 0.445\pm0.003 & 0.18\pm0.01 & 0.2082\pm0.0078 & 1.5\pm0.2 & 126.5 & 56  \\      
                        \noalign{\smallskip}                        
            \hline                                          
            \noalign{\smallskip}                            
            \hline                                          
         \end{array}
        $$
\end{table*}
\label{sec:appendix}

\begin{figure*}
    \centering{
    \includegraphics[angle=0,width=8.5cm]{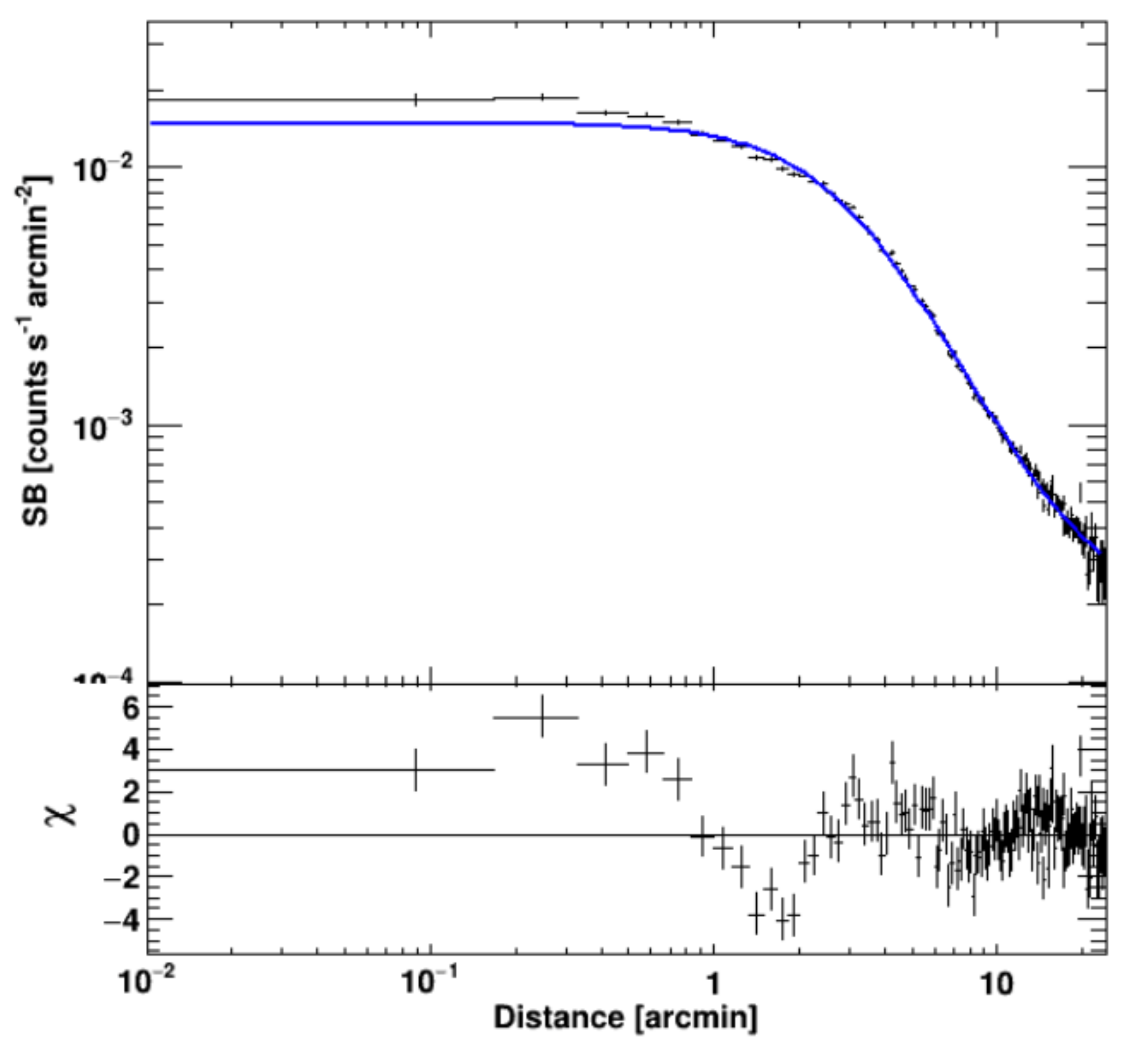}
    \includegraphics[angle=0,width=8.5cm]{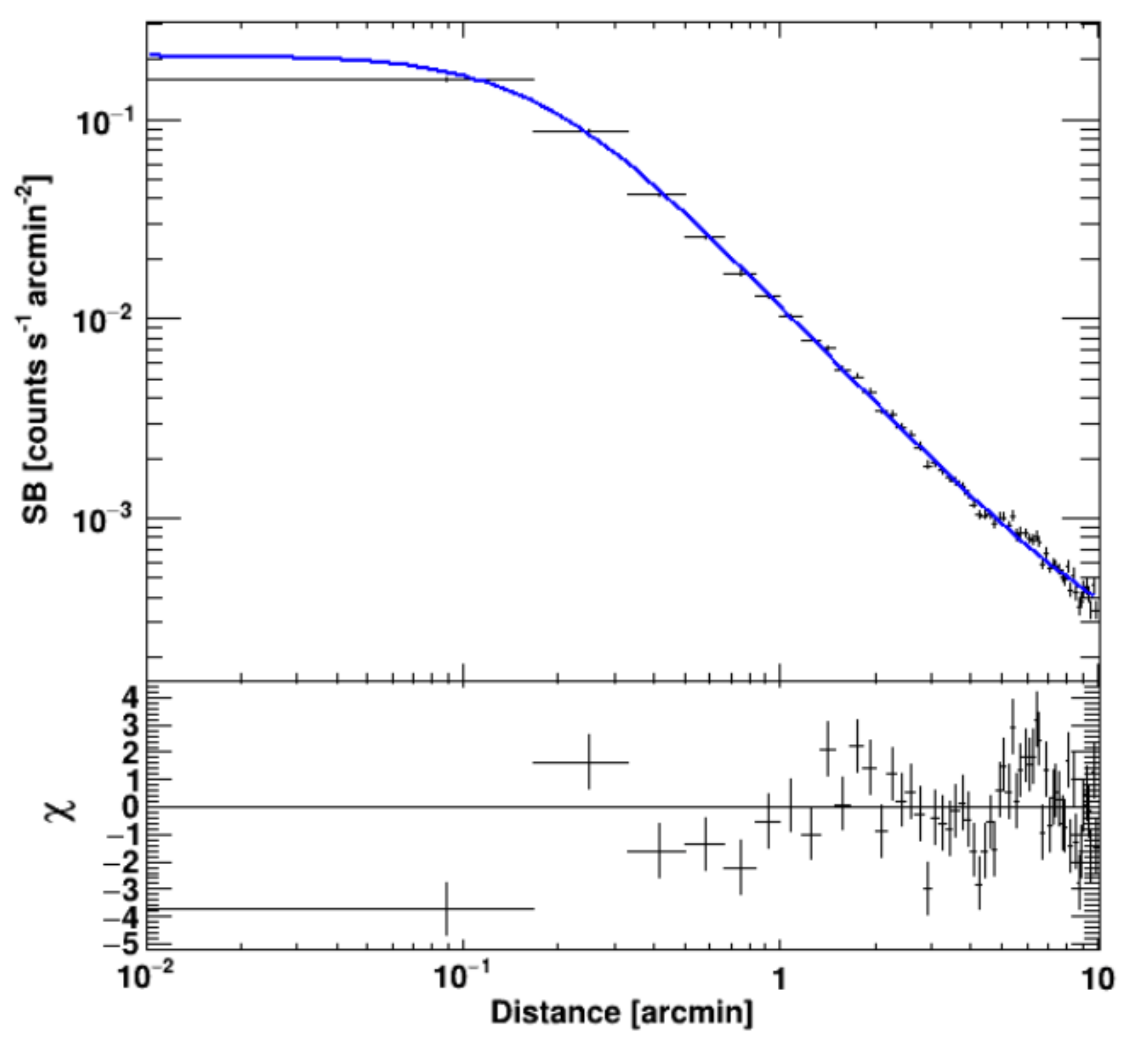}
}
 \caption{Surface brightness profile and best-fit $\beta$-model (blue line) of RXCJ1825 ({\it left panel}) and CIZAJ1824 ({\it right panel}). The profiles are centered on the cluster centroid, the elliptical annuli have ellipticity and P.A. of 0.25 and 165 deg for RXCJ1825, and 0.24 and 94 deg for CIZAJ1824. The distance is computed along the major axis.}
        \label{fig:sbprof}
\end{figure*}

\end{document}